\documentclass{article}

\usepackage{PRIMEarxiv}
\usepackage{caption}
\usepackage{subcaption}
\usepackage[utf8]{inputenc}
\usepackage[T1]{fontenc}
\usepackage{hyperref}
\usepackage{url}
\usepackage{booktabs}
\usepackage{amsfonts}
\usepackage{amsmath}
\usepackage{nicefrac}
\usepackage{microtype}
\usepackage{fancyhdr}
\usepackage{graphicx}
\usepackage{braket}
\usepackage{float}
\usepackage{placeins}

\usepackage{multirow}
\graphicspath{{media/}}

\providecommand{\isAPAandChicago}[2]{#2}

\title{Characterizing Entanglement in Combinations of Bell States through Superposition and Mixing: An Increase in Entanglement on Introducing Depolarizing, Phase Damping, and Amplitude Damping Noise}

\author{
  Nishant Chaudhari \\
  Department of Physics and Astronomy \\
  University of Southern Mississippi \\
  Hattiesburg, MS\\
  \texttt{nishant.chaudhari@usm.edu} \\
   \And
  Jean-Fran\c{c}ois Van Huele \\
  Department of Physics and Astronomy \\
  Brigham Young University \\
  Provo, UT\\
  \texttt{vanhuele@byu.edu} \\
}

\begin{document}
\maketitle

\begin{abstract}
We systematically characterize quantum entanglement in all sixteen combinations of 
two-qubit Bell states across three operational regimes: pure superposition, noiseless 
mixing, and asymmetric mixing of one noisy and one pure Bell state. For pure 
superpositions, we calculate the von Neumann entropy and show that entanglement depends critically on the relative phase $\phi$ between superposed states. For mixed states, 
we use concurrence C under three physically motivated 
decoherence channels. The Depolarizing (DP) and Phase Damping (PD) channels are 
modeled as global, collective operations on the two-qubit Hilbert space, while 
Amplitude Damping (AD) is treated as independent local decay on each qubit, 
reflecting the asymmetric, energy-dissipating nature of spontaneous emission. 
A central and counterintuitive result is that increasing noise can \textit{raise} 
concurrence in specific parameter regimes across all three channels. We further map 
the boundary between mathematical entanglement ($C > 0$) and operational quantum 
non-locality using the Horodecki criterion, finding this 
boundary to be strongly channel-dependent. Under Phase Damping, the Bell-local 
entanglement regime vanishes entirely for identical and same bases mixtures, 
meaning any surviving entanglement guarantees a CHSH violation. Under Depolarizing 
noise, a large Bell-local region requires a mixing probability  $r > 1/\sqrt{2}$ to recover 
non-locality at maximum noise. For Amplitude Damping, certain Bell state mixtures exhibit a non-monotonic Horodecki parameter $M(T)$: while concurrence decreases monotonically with noise, the capacity for CHSH violation is lost at intermediate noise levels and then restored at high noise, as maximal damping repurifies the noisy branch of the mixture. These results provide a unified analytical 
reference for entanglement and non-locality management across Bell-state combinations 
and decoherence channels relevant to Noisy Intermediate-Scale Quantum (NISQ) 
architectures.
\end{abstract}


\section{Introduction}
\label{sec:intro}
Noise is an irreducible feature in quantum information~\cite{ref-nielsen}. In many 
instances, we try to suppress noise to avoid or postpone the occurrence of decoherence 
of the quantum state. Different types of noise, such as Depolarizing (DP), Amplitude 
Damping (AD), and Phase Damping (PD), decrease the purity of the state, typically 
decreasing entanglement, and, to the extent entanglement is a quantum resource, thereby 
reducing this quantum resource. A significant research effort has been conducted to 
mitigate the noise in quantum computers~\cite{ref-kandala}. Our motivation is to study 
these types of noise and give a quantitative analysis of how they affect combinations of 
Bell states, which are maximally entangled states for two-qubit systems and a fundamental 
ingredient for quantum information applications.

The four Bell states form a complete orthonormal basis for the entire two-qubit Hilbert 
space, meaning any valid two-qubit state can be expressed as a linear superposition of 
these basis states. This makes them the foundational building blocks of quantum information 
science, central to applications ranging from quantum teleportation to quantum error 
correction. This work provides a systematic characterization of entanglement in different 
combinations of Bell states under superposition and mixing with noise. While multi-qubit systems such as three-qubit GHZ and W states are also of significant interest, the two-qubit Bell basis offers a setting where every combination can be treated exactly and analytically, allowing the constructive role of noise to be isolated cleanly. Extending this methodology to three or more qubits would introduce additional degrees of freedom that obscure these physical insights and is beyond the scope of the current study.

We consider three separate cases: superposition of Bell states, mixing of Bell states without 
noise, and mixing of pure Bell states with noisy Bell states. We systematically examine 
the possibility of subjecting Bell states to DP, PD, and AD noise. For DP and PD, we 
model collective global decoherence acting on the two-qubit system as a whole, while 
for AD we treat each qubit as undergoing independent local decay, reflecting the 
asymmetric, energy-dissipating nature of spontaneous emission. In some cases, we find 
that noise can enhance entanglement depending on the mixture and noise parameter. We 
then take this a step further to see whether the entanglement increase in the presence 
of noise can also produce a CHSH violation. Since noise mitigation is an active area of 
research in quantum information, we want to establish a framework for characterizing and 
understanding the fundamentals of noise. Our selection of DP, PD, and AD is motivated 
by the fact that they significantly decrease the purity of the state and form the basis 
for most types of noise leading to decoherence.

\section{Methods for characterizing entanglement and noise}
\label{sec:methods}

\subsection{Entanglement}
\label{sec:entanglement}
Entanglement is a quantum correlation between two or more quantum systems that 
transcends classical correlation. In an entangled state, the joint states of two 
qubits cannot be factorized, i.e., written as the tensor product of individual 
states from each subsystem. Mathematically, a two-qubit state
\[
\ket{\Psi} \in \mathcal{H}_A \otimes \mathcal{H}_B
\]
is \textbf{entangled} if it cannot be written as
\[
\ket{\Psi} \neq \ket{\psi}_A \otimes \ket{\phi}_B,
\]
where \( \ket{\psi}_A \in \mathcal{H}_A \) and \( \ket{\phi}_B \in \mathcal{H}_B \) 
are single-qubit states.

Whereas a tenet of reductionism in science posits that the whole is explained through 
its parts, in an entangled state, the joint system is fully known, whereas the 
individual systems are not~\cite{schrodinger1935}. We can have different degrees of 
entanglement, from maximally entangled states to completely unentangled or separable 
states, and we can use quantitative entanglement measures to quantify the degree of 
entanglement. In this paper, we will focus on two-qubit states, and we will work with 
particular maximally entangled two-qubit states: Bell states. There are exactly four 
of them, and they form an orthonormal basis for the two-qubit Hilbert space
\begin{align*}
\ket{\phi^+}&=\frac{\ket{00}+\ket{11}}{\sqrt{2}}, &
\ket{\phi^-}&=\frac{\ket{00}-\ket{11}}{\sqrt{2}}, \\
\ket{\psi^+}&=\frac{\ket{01}+\ket{10}}{\sqrt{2}}, &
\ket{\psi^-}&=\frac{\ket{01}-\ket{10}}{\sqrt{2}},
\end{align*}
where $\ket{0}$ and $\ket{1}$ are the basis states in the computational basis and 
where we have adopted the simplified notation $\ket{00}=\ket{0}\otimes \ket{0}$, 
$\ket{01}=\ket{0}\otimes \ket{1}$, etc.

They serve as standard examples to test and understand entanglement. Notably, any maximally entangled two-qubit state can be converted into a Bell state, confirming that all four Bell states are equivalent as entanglement resources~\cite{ref-nielsen}. Bell states are central to both 
the theoretical foundation and practical applications of quantum information. They 
define what it means for two qubits to be entangled and serve as a key resource in 
quantum information science, in teleportation, quantum error 
correction~\cite{ref-nielsen} and quantum random number generation~\cite{pironio2010}.

There are several ways to characterize the degree of entanglement. In this work, we 
will use two specific measures: von Neumann Entropy $S$ and Concurrence $C$. For 
these measures, we only need the density matrix $\rho$ of the state to characterize 
entanglement. A general state is given as
\[
\rho=\sum_i p_i\ket{\psi_i}\bra{\psi_i},
\]
where $p_i$ is the probability of the system being in the pure quantum states 
$\ket{\psi_i}$.

For pure states, we use the von Neumann entropy~\cite{ref-vonneumann} 
as an entanglement measure, following Bennett et al.~\cite{bennett1996},
\[
S = -\sum_i \lambda_i \log_2 \lambda_i,
\]
where $\lambda_i$ are the eigenvalues of the partial-traced $2\times 2$ matrix of 
the original state's density matrix $\rho$. To quantify the degree of entanglement 
in the pure states, we partially trace~\cite{ref-steeb} one of the two qubits 
of the resultant states. The resultant entropy does not depend on which qubit is 
partially traced over. Partial tracing is a method to focus on one of the subsystems 
in a composite quantum system by mathematically removing the degrees of freedom 
of the other subsystem.

Concurrence, introduced by Wootters~\cite{wootters1998,ref-wootters}, is a method of 
characterizing the degree of entanglement for mixed two-qubit states, given as
\[
C=\max[0,\lambda_1-\lambda_2-\lambda_3-\lambda_4],
\]
where $\{\lambda_1,\lambda_2,\lambda_3,\lambda_4\}$ are the eigenvalues (in 
descending order such that $\lambda_1\ge\lambda_2\ge\lambda_3\ge\lambda_4$) of the 
matrix $R$ such that
\[
R=\sqrt{\sqrt{\rho}\, \tilde{\rho}\, \sqrt{\rho}},
\]
where $\tilde{\rho}=(\sigma_y\otimes \sigma_y)\rho^*( \sigma_y\otimes \sigma_y)$ 
and $\sigma_y$ is the Pauli-Y matrix, and * indicates complex conjugation. The transition from $\rho$ to $\tilde{\rho}$ is called the spin flip operation. 

In all cases, we model the noise in the Bell states using density operators. Density 
operators correspond to Hermitian, positive semi-definite matrices with unit trace. 
They provide a complete description of both pure and mixed quantum states. This makes 
them especially suitable for modeling noise, decoherence, and entanglement 
degradation~\cite{ref-nielsen}.

Furthermore, because the maximally entangled Bell states inherently possess an 
``X-state'' structure, where non-zero elements are restricted to the 
main diagonal and anti-diagonal of the $4\times4$ matrix, the noise channels considered in this work 
(Depolarizing, Phase Damping, and Amplitude Damping) preserve this 
structure~\cite{yu2007evolution}, and the calculation of concurrence can be significantly 
simplified. Instead of computing the eigenvalues of the complex matrix $R$ as 
required in Wootters' general formulation, we can determine the concurrence directly 
from the density matrix elements~\cite{yu2007evolution}. For a general two-qubit 
X-state, the concurrence reduces to the expression:
\begin{equation}
    C(\rho_X) = 2 \max \left\{ 0,\, |\rho_{14}| - \sqrt{\rho_{22}\rho_{33}},\, 
    |\rho_{23}| - \sqrt{\rho_{11}\rho_{44}} \right\}.
    \label{eq:x_state_concurrence}
\end{equation}
Throughout the subsequent sections, we leverage 
Equation~(\ref{eq:x_state_concurrence}) to analytically evaluate the entanglement 
dynamics under varying noise conditions. In our analysis, we represent each Bell 
state as a $4\times4$ density matrix in the computational basis. This formalism 
makes it possible to incorporate noise channels directly through Kraus operators, as 
defined in Section~\ref{sec:noise}.
\subsection{Noise}
\label{sec:noise}
Whereas superposition and mixing are essentially mathematical operations on quantum states, noise is a physical process arising from uncontrolled interactions between the quantum system and its environment. It can be defined as any unintended interaction between a quantum system and its 
environment that causes the system to be degraded or lose information. One of the 
methods to mathematically describe a state evolving under noise is by implementing 
the \textbf{Kraus operator} formalism~\cite{ref-nielsen}. A noisy quantum operation 
acting on a density matrix $\rho$ can be written as
\[
\rho'=\sum_{i} K_i(p)\rho K_i(p)^\dagger,
\]
where $K_i(p)$ is the set of Kraus operators specific to each noise type and $\rho'$ is the 
resulting noisy density matrix parameterized by the noise probability $p$.

The noise channels studied in this work fall into two physical categories. For 
Depolarizing (DP) and Phase Damping (PD) noise, we adopt a model of collective 
decoherence. In NISQ architectures~\cite{preskill2018}, noise is an unavoidable 
feature of current hardware. For two-qubit systems in particular, 
modeling collective global decoherence is a natural idealization 
when the qubits interact with a shared environment. Therefore, rather than modeling these 
channels as independent, local events acting on single qubits (which would require 
taking the tensor product of single-qubit Kraus operators), we treat the two-qubit 
Bell states as a single $d=4$ quantum system undergoing global, correlated noise 
operations. For Amplitude Damping (AD), however, we model each qubit as undergoing 
independent local decay. This reflects the asymmetric, energy-dissipating nature of spontaneous emission, where each qubit independently loses its excitation to 
its own environmental mode. The two-qubit AD Kraus operators are therefore the 
tensor products of single-qubit AD operators, as derived explicitly in 
Appendix~\ref{app:ad_kraus}. The noise channels we use in this paper are the following:

\begin{itemize}
    \item \textbf{Depolarizing Noise:}
    Depolarizing noise uniformly reduces the purity of the state, with equal 
    probability of applying any Pauli error. The Kraus operators for depolarizing 
    noise acting on a single qubit are
    \[
    K_0=\sqrt{1-\frac{3p}{4}}\,I,\quad K_1=\sqrt{\frac{p}{4}}\,X,\quad 
    K_2=\sqrt{\frac{p}{4}}\,Y, \quad K_3=\sqrt{\frac{p}{4}}\,Z,
    \]
    where $\{I,X,Y,Z\}$ are the Pauli operators and $p$ is the noise probability. 
    We model global depolarizing noise by treating the two-qubit pair as a single 
    entity in a $d=4$ Hilbert space. Following the standard generalized 
    $d$-dimensional depolarizing channel~\cite{ref-nielsen}, the state remains 
    perfectly intact with probability $1-p$, and with probability $p$ it is replaced 
    by the completely mixed state $\frac{I}{4}$. This is physically equivalent to 
    applying the full set of 16 collective $4\times4$ Pauli Kraus operators to the 
    joint state. The resulting channel is given by:
    \begin{equation}
    \rho'=(1-p)\rho+p\,\frac{I}{4}.
    \label{eq:dp_channel}
    \end{equation}
    If $p=1$ the state becomes maximally mixed, and if $p=0$ the purity is 
    unchanged.

    \item \textbf{Phase Damping Noise:}
    Phase damping noise affects only the coherence of a quantum state, destroying 
    off-diagonal elements without exchanging energy with the environment. In a 
    density matrix, a complete PD error zeroes all off-diagonal elements while 
    leaving the diagonal populations intact. The Kraus operators for 
    single-qubit PD noise are
    \[
    K_0=\sqrt{1-p}\;I, \quad K_1=\begin{bmatrix}
        \sqrt{p}&0\\
        0&0
    \end{bmatrix}, \quad K_2=\begin{bmatrix}
        0&0\\
        0&\sqrt{p}
    \end{bmatrix}.
    \]
    For phase damping, we model a global dephasing channel that uniformly suppresses all off-diagonal coherences simultaneously without exchanging energy. We generalize 
    this as a statistical mixture:
    \begin{equation}
    \rho'=(1-p)\rho+p\,\eta,
    \label{eq:pd_channel}
    \end{equation}
    where $\eta$ is the fully dephased density matrix (all off-diagonal elements 
    exactly zero), and $p$ dictates the strength of this collective coherence loss.

    \item \textbf{Amplitude Damping Noise:}
    Unlike Depolarizing and Phase Damping, Amplitude Damping is an asymmetric, 
    energy-dissipating process modeling the decay of an excited qubit into its 
    ground state~\cite{ref-nielsen}. Because 
    this decay is a local, single-qubit process (each qubit independently loses 
    its excitation to its own environmental mode), we model the two-qubit AD 
    channel as the tensor product of independent single-qubit operators rather than 
    as a collective global channel. A complete amplitude-damping error corresponds to 
    a non-unitary decay from $\ket{1}$ to $\ket{0}$, permanently destroying 
    coherence. The Kraus operators for single-qubit AD noise are:
        \[
        K_0=\begin{bmatrix}
            1&0\\
            0&\sqrt{1-p}
        \end{bmatrix}, \quad K_1=\begin{bmatrix}
            0&\sqrt{p}\\
            0&0
        \end{bmatrix}.
        \]
        Taking the tensor product for the two-qubit system yields the effective 
        Kraus operators (see Appendix~\ref{app:ad_kraus} for the full derivation):
        \[
        K_0=\begin{bmatrix}
            1&0&0&0\\
            0&\sqrt{1-p}&0&0\\
            0&0&\sqrt{1-p}&0\\
            0&0&0&(1-p)
        \end{bmatrix},
        \quad K_1=\begin{bmatrix}
            0&\sqrt{p}&0&0\\
            0&0&0&0\\
            0&0&0&\sqrt{p(1-p)}\\
            0&0&0&0
        \end{bmatrix},
        \]
        \[
        K_2=\begin{bmatrix}
            0&0&\sqrt{p}&0\\
            0&0&0&\sqrt{p(1-p)}\\
            0&0&0&0\\
            0&0&0&0
        \end{bmatrix}, \quad K_3=\begin{bmatrix}
            0&0&0&p\\
            0&0&0&0\\
            0&0&0&0\\
            0&0&0&0
        \end{bmatrix}.
        \]
\end{itemize}

\subsection{Quantum Non-Locality and the Horodecki Criterion}
\label{sec:horodecki_method}
While a non-zero concurrence $C$ provides a quantification of 
entanglement in the given state, it is well-established that not 
all entangled states are capable of demonstrating quantum 
non-locality~\cite{werner1989}. In mixed entangled states, the 
state can be in the regime of \textbf{Bell-Local Entanglement}, 
where the state is entangled but fails to violate the CHSH 
inequality~\cite{clauser1969,werner1989} via projective measurements.

To determine whether a two-qubit state can violate the CHSH inequality and quantify the maximum achievable violation, we use 
the Horodecki criterion~\cite{horodecki1995}. Any two-qubit density matrix $\rho$ 
can be mapped to a $3\times3$ correlation tensor, the $T$-matrix, whose elements 
are defined by $T_{jk} = \mathrm{Tr}[\rho(\sigma_j \otimes \sigma_k)]$. The maximum 
CHSH violation $S_{\max}$ is bounded by
\begin{equation}
    S_{\max} = 2\sqrt{M(T)}\;,
\end{equation}
where $M(T)$ is the sum of the two largest eigenvalues of the $T^\top T$ matrix where $\top $ indicates transposition.
The condition $M(T) > 1$ is necessary and sufficient for violation of the CHSH 
inequality specifically.

The evaluation of $M(T)$ requires different computational approaches depending on 
the operational regime. For arbitrary pure superpositions, the $T$-matrix is 
generally non-diagonal due to off-diagonal correlations that depend on the relative 
phase $\phi$. Therefore, to evaluate $M(T)$ in the superposition section, we compute 
the full general correlation matrix $T$, construct $T^\top T$, and explicitly 
calculate its two largest eigenvalues. Conversely, in our mixed and noisy state 
regimes, the mixing methodology strictly preserves the X-state structure of the 
density matrices across all evaluated noise channels. When $\rho$ is an X-state, the corresponding $T$-matrix is strictly diagonal, 
meaning the off-diagonal $T_{jk}$ terms are exactly zero. The three diagonal 
entries $t_1$, $t_2$, and $t_3$ can be extracted directly from the density 
matrix elements as:
\begin{align}
    t_{1} &= 2(\rho_{14}+\rho_{23}), \label{eq:t1}\\
    t_{2} &= 2(\rho_{23}-\rho_{14}), \label{eq:t2}\\
    t_{3} &= \rho_{11}+\rho_{44}-\rho_{22}-\rho_{33}. \label{eq:t3}
\end{align}
Since $T^\top T$ is then also diagonal, its eigenvalues are simply $t_1^2$, 
$t_2^2$, and $t_3^2$. The Horodecki parameter therefore reduces to the sum 
of the two largest among these three squared values, allowing us to bypass 
full matrix diagonalization and evaluate $M(T)$ directly from the density 
matrix elements:
\begin{equation}
    M(T) = \text{sum of the two largest values among } \{t_1^2,\, t_2^2,\, t_3^2\}.
\end{equation}
We utilize these metrics to generate phase diagrams that cleanly classify the 
parameter space into three distinct physical regimes:
\begin{itemize}
    \item \textbf{Blue (Quantum Non-Local):} $M(T) > 1$. The state is entangled 
    and robust enough to violate the CHSH inequality.
    \item \textbf{Orange (Bell-Local Entanglement):} Entangled but $M(T) \leq 1$. 
    The state remains mathematically inseparable, but the noise has degraded the quantum correlations such that it does not violate the CHSH inequality under projective measurements~\cite{werner1989}. Entanglement is measured by $S > 0$ 
    for pure superpositions and $C > 0$ for mixed and noisy states.
    \item \textbf{Gray (Separable / Classical):} The state has undergone 
    entanglement sudden death~\cite{yu2004sudden,yu2007evolution} and is entirely classical, 
    with $S = 0$ for pure superpositions and $C = 0$ for mixed and noisy states.
\end{itemize}

Finally, to cleanly map the phase transitions between these regimes in the noise 
sections, we define an effective noise parameter:
\begin{equation}
    x = p(1-r).
    \label{eq:effective_noise}
\end{equation}
This parameter elegantly accounts for the interplay between the channel noise 
probability $p$ and the pure state mixing probability $r$, simplifying the boundary 
equations in our subsequent analysis.

\section{Results}
\label{sec:results}

\subsection{Superposition of two Bell states without noise}
\label{sec:superposition}
We superpose two Bell states with arbitrary real coefficients $s$ and a relative phase 
angle $e^{i\phi}$, and use von Neumann entropy to characterize the entanglement. 
We do this by partial tracing the resulting density matrix and then calculating its 
von Neumann entropy. If the entropy is one, then the state is maximally entangled. 
Since superposition does not decrease the purity of the resultant state, using the 
entropy method is the most suitable way to characterize the entanglement. The general normalized
superposition state takes the form
\[
\ket{\delta} = \sqrt{1-s}\,\ket{B_1} + e^{i\phi}\sqrt{s}\,\ket{B_2},
\]
where $\ket{B_1}$ and $\ket{B_2}$ are any two Bell states, $s \in [0,1]$ controls 
the relative weight, and $\phi \in [0, 2\pi]$ is the relative phase. In total, we 
have sixteen combinations, as we have four individual Bell states to work with. We 
start with $\ket{\phi^+}$ and combine it with  all the Bell states.

\begin{itemize}

\item \textbf{For $\ket{\phi^+}$ with $\ket{\phi^+}$:}
On superposing two identical Bell states, the result after normalization is always 
proportional to $\ket{\phi^+}$ itself, regardless of $s$ and $\phi$. Since 
$\ket{\phi^+}$ is a maximally entangled Bell state, the superposition remains 
maximally entangled with entropy $S = 1$ for all values of $s$ and $\phi$. The same reasoning applies to the four combinations for identical Bell states.

\item \textbf{For $\ket{\phi^+}$ with $\ket{\phi^-}$:}
Let $\ket{\delta}$ denote the superposition state,
\[
\ket{\delta}=\sqrt{1-s}\,\ket{\phi^+}+e^{i\phi}\sqrt{s}\,\ket{\phi^-}.
\]
We construct the density operator $\rho=\ket{\delta}\bra{\delta}$,
\[
\rho=\begin{bmatrix}
    \frac{1}{2}+\sqrt{s(1-s)}\cos\phi & 0 & 0 & 
    \frac{1}{2}-s+\sqrt{s(1-s)}\,i\sin\phi\\
    0 & 0 & 0 & 0\\
    0 & 0 & 0 & 0\\
    \frac{1}{2}-s-\sqrt{s(1-s)}\,i\sin\phi & 0 & 0 & 
    \frac{1}{2}-\sqrt{s(1-s)}\cos\phi
\end{bmatrix}.
\]
The partial trace of this state with respect to one of the qubits, $\rho_t$, is
\[
\rho_t=\begin{bmatrix}
    \frac{1}{2}+\sqrt{s(1-s)}\cos\phi & 0\\
    0 & \frac{1}{2}-\sqrt{s(1-s)}\cos\phi
\end{bmatrix}.
\]
Here the diagonal elements are the eigenvalues, such that
\[S=-[\lambda_1\log_2\lambda_1+\lambda_2\log_2\lambda_2],\]
\begin{equation}
\begin{split}
S = -\Bigg[&\frac{1+2\sqrt{s(1-s)}\cos\phi}{2}
    \log_2\!\left(\frac{1+2\sqrt{s(1-s)}\cos\phi}{2}\right) \\
   &+\frac{1-2\sqrt{s(1-s)}\cos\phi}{2}
    \log_2\!\left(\frac{1-2\sqrt{s(1-s)}\cos\phi}{2}\right)\Bigg].
\end{split}
\label{eq:entropy_phiplus_phiminus}
\end{equation}
The von Neumann entropy is plotted in Figure~\ref{fig:vn1} for different values 
of the phase angle $\phi$.

\begin{figure}[htbp]
    \centering
    \includegraphics[width=0.7\textwidth]{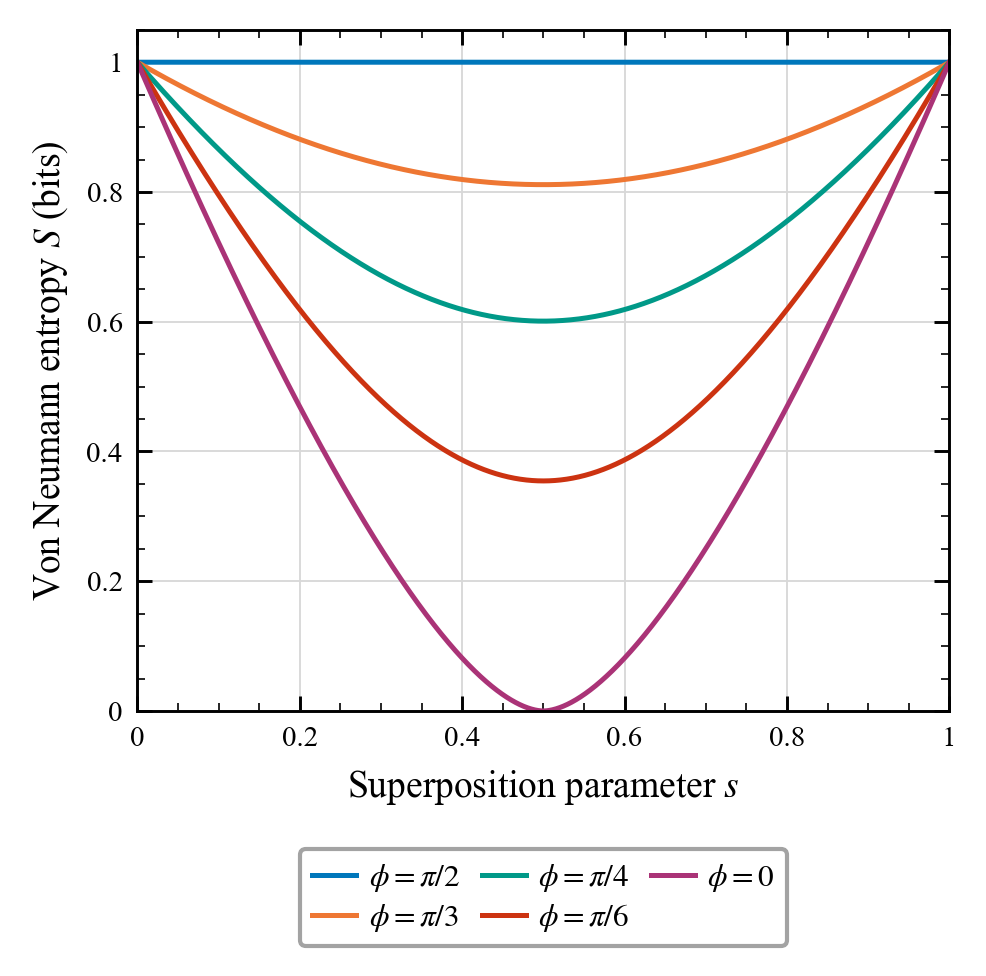}
    \caption{Von Neumann entropy $S$ as a function of the relative weight $s$ in 
    a superposition of $\ket{\phi^+}$ and $\ket{\phi^-}$ for different values of 
    the phase angle $\phi$.}
    \label{fig:vn1}
\end{figure}

The graph is symmetric for supplementary angle pairs: $0$ and $\pi$, 
$\frac{\pi}{4}$ and $\frac{3\pi}{4}$, $\frac{\pi}{6}$ and $\frac{5\pi}{6}$, 
$\frac{\pi}{3}$ and $\frac{2\pi}{3}$, etc. For $\phi=0$, the entropy goes to $0$ 
at $s=0.5$, implying that the state is completely separable. At any value of 
$\phi$, the entropy reaches its minimum at $s=0.5$. The resulting state becomes 
completely separable only when $\phi=0$ or $\phi=\pi$ and remains maximally entangled for $\phi=\frac{\pi}{2}$ for all $s$ values.

\item \textbf{For $\ket{\phi^+}$ with $\ket{\psi^+}$:}
\[
\ket{\delta}=\sqrt{1-s}\,\ket{\phi^+}+e^{i\phi}\sqrt{s}\,\ket{\psi^+},
\]
\[
\rho=\begin{bmatrix}
    \frac{1-s}{2} & \frac{e^{-i\phi}\sqrt{s(1-s)}}{2} & 
    \frac{e^{-i\phi}\sqrt{s(1-s)}}{2} & \frac{1-s}{2}\\
    \frac{e^{i\phi}\sqrt{s(1-s)}}{2} & \frac{s}{2} & \frac{s}{2} & 
    \frac{e^{i\phi}\sqrt{s(1-s)}}{2}\\
    \frac{e^{i\phi}\sqrt{s(1-s)}}{2} & \frac{s}{2} & \frac{s}{2} & 
    \frac{e^{i\phi}\sqrt{s(1-s)}}{2}\\
    \frac{1-s}{2} & \frac{e^{-i\phi}\sqrt{s(1-s)}}{2} & 
    \frac{e^{-i\phi}\sqrt{s(1-s)}}{2} & \frac{1-s}{2}
\end{bmatrix}.
\]
Again, the partial trace is
\[
\rho_t=\begin{bmatrix}
    \frac{1}{2} & \sqrt{s(1-s)}\cos\phi\\
    \sqrt{s(1-s)}\cos\phi & \frac{1}{2}
\end{bmatrix}.
\]
The von Neumann entropy is
\begin{equation*}
\begin{split}
S = -\Bigg[&\frac{1+2\sqrt{s(1-s)}\cos\phi}{2}
    \log_2\!\left(\frac{1+2\sqrt{s(1-s)}\cos\phi}{2}\right) \\
   &+\frac{1-2\sqrt{s(1-s)}\cos\phi}{2}
    \log_2\!\left(\frac{1-2\sqrt{s(1-s)}\cos\phi}{2}\right)\Bigg].
\end{split}
\end{equation*}
The entropy is identical to that found in the case of $\ket{\phi^+}$ and 
$\ket{\phi^-}$, and the results from Figure~\ref{fig:vn1} also apply. 
This follows from the fact that $\ket{\psi^+}$ can be obtained from 
$\ket{\phi^+}$ by applying a Pauli $X$ gate to a single qubit 
($X \otimes I\,\ket{\phi^+} = \ket{\psi^+}$). Since local unitaries 
act on only one subsystem, they leave all entanglement measures invariant, 
and both the von Neumann entropy and the Horodecki parameter $M$ are 
therefore identical for any superposition of $\ket{\phi^+}$ with 
$\ket{\psi^+}$ and the corresponding superposition with $\ket{\phi^-}$.

\item \textbf{For $\ket{\phi^+}$ with $\ket{\psi^-}$:}
\[
\ket{\delta}=\sqrt{1-s}\,\ket{\phi^+}+e^{i\phi}\sqrt{s}\,\ket{\psi^-},
\]
\[
\rho=\begin{bmatrix}
    \frac{1-s}{2} & \frac{e^{-i\phi}\sqrt{s(1-s)}}{2} & 
    -\frac{e^{-i\phi}\sqrt{s(1-s)}}{2} & \frac{1-s}{2}\\
    \frac{e^{i\phi}\sqrt{s(1-s)}}{2} & \frac{s}{2} & -\frac{s}{2} & 
    \frac{e^{i\phi}\sqrt{s(1-s)}}{2}\\
    -\frac{e^{i\phi}\sqrt{s(1-s)}}{2} & -\frac{s}{2} & \frac{s}{2} & 
    -\frac{e^{i\phi}\sqrt{s(1-s)}}{2}\\
    \frac{1-s}{2} & \frac{e^{-i\phi}\sqrt{s(1-s)}}{2} & 
    -\frac{e^{-i\phi}\sqrt{s(1-s)}}{2} & \frac{1-s}{2}
\end{bmatrix},
\]
giving the partial-traced density matrix as
\[
\rho_t=\begin{bmatrix}
    \frac{1}{2} & \sqrt{s(1-s)}\,i\sin\phi\\
    -\sqrt{s(1-s)}\,i\sin\phi & \frac{1}{2}
\end{bmatrix}.
\]
The off-diagonal elements now carry $i\sin\phi$ rather than $\cos\phi$, 
The von Neumann entropy is
\begin{equation*}
\begin{split}
S = -\Bigg[&\frac{1+2\sqrt{s(1-s)}\sin\phi}{2}
    \log_2\!\left(\frac{1+2\sqrt{s(1-s)}\sin\phi}{2}\right) \\
   &+ \frac{1-2\sqrt{s(1-s)}\sin\phi}{2}
    \log_2\!\left(\frac{1-2\sqrt{s(1-s)}\sin\phi}{2}\right)\Bigg].
\end{split}
\end{equation*}
The result is plotted in Figure~\ref{fig:vn2}.\\

We repeat the same analysis for all remaining superpositions of Bell states 
starting from $\ket{\phi^-}$, $\ket{\psi^+}$, and $\ket{\psi^-}$. In each 
case, the partial-traced density matrix takes the same $2 \times 2$ form, with 
off-diagonal elements proportional to either $\cos\phi$ or $\sin\phi$ depending 
on the superposed states bases($\phi$ vs $\psi)$ and phases ($+$ vs $-$). The von Neumann 
entropy therefore takes one of the two functional forms already derived: 
Eq.~(\ref{eq:entropy_phiplus_phiminus}) when the superposition couples states 
of either same bases but opposite phases or different bases but same phases (e.g.\ $\ket{\phi^+}$ with $\ket{\phi^-}$, or $\ket{\phi^+}$ with $\ket{\psi^+}$), and the $\sin\phi$ counterpart above when it couples 
states of different bases and opposite phases (e.g.\ $\ket{\phi^+}$ with $\ket{\psi^-}$).

Notably, in the special case of $\phi=0$, the $\ket{\phi^+}$/$\ket{\psi^-}$ 
reduced density matrix diagonalizes completely,
\[
\rho_t=\begin{bmatrix}
    \frac{1}{2} & 0\\
    0 & \frac{1}{2}
\end{bmatrix}.
\]
The parameter $s$ cancels out entirely, leaving a maximally mixed reduced 
state with entropy $S = 1$ for all $s$. This means the superposition of 
$\ket{\phi^+}$ and $\ket{\psi^-}$ is maximally entangled regardless of 
the superposition weight, a property that holds for no other non-identical 
pairing with $\ket{\phi^+}$. Each Bell state therefore has exactly two 
partner states that guarantee maximal entanglement under superposition: 
itself, and a different basis and opposite phase state.

Because the phase dependence is rotated by $\pi/2$, the entropy now reaches its minimum at 
$\phi = \frac{\pi}{2}$ rather than $\phi = 0$, as seen in 
Figure~\ref{fig:vn2}.

\begin{figure}[htbp]
    \centering
    \includegraphics[width=0.7\textwidth]{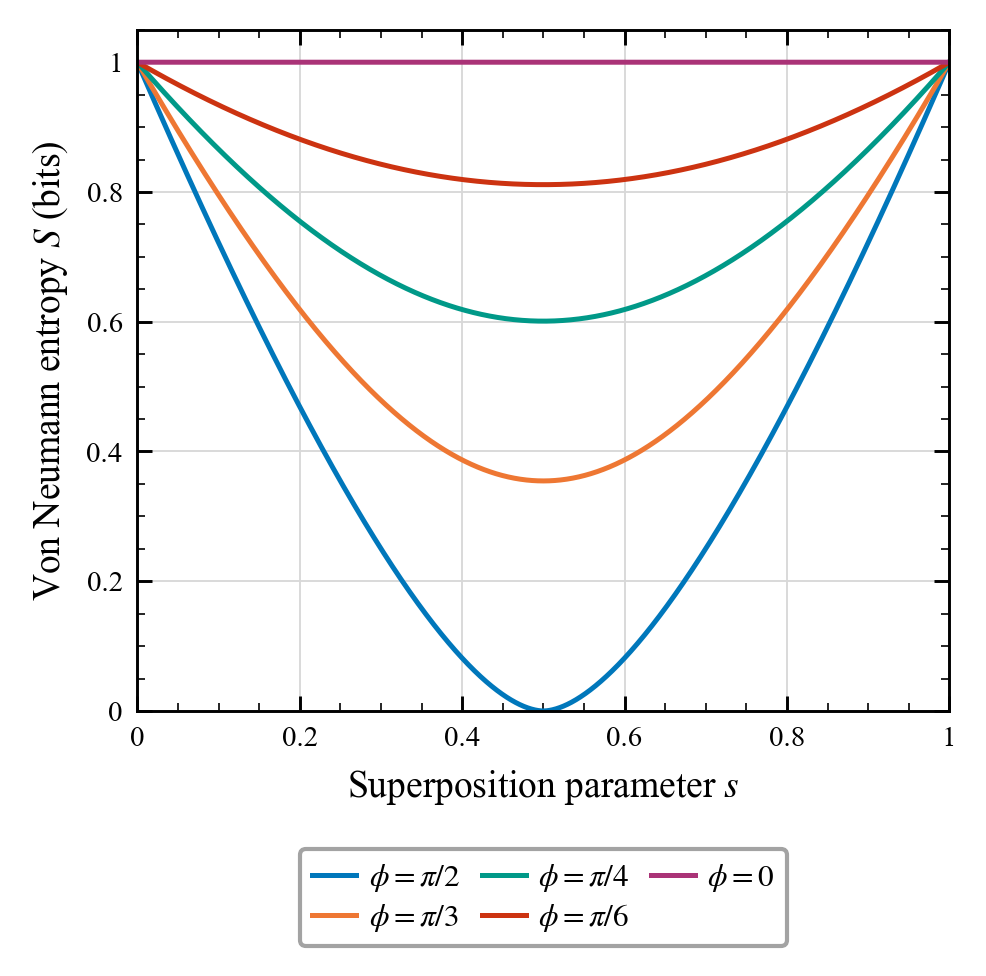}
    \caption{Von Neumann entropy $S$ as a function of the relative weight $s$ in 
    a superposition of $\ket{\phi^+}$ and $\ket{\psi^-}$.}
    \label{fig:vn2}
\end{figure}

Since the $\cos$ and $\sin$ functions coincide at $\phi=\frac{\pi}{4}$, 
the entropy curves from Figures~\ref{fig:vn1} and \ref{fig:vn2} match 
exactly at that angle. The full entanglement structure across all four 
Bell states is synthesized in Figures~\ref{fig:diagram0} 
and~\ref{fig:diagram90} for $\phi=0$ and $\phi=\frac{\pi}{2}$ respectively, 
and the pattern generalizes by symmetry to all Bell states.

\begin{figure}[htbp]
    \centering
    \includegraphics[width=0.7\textwidth]{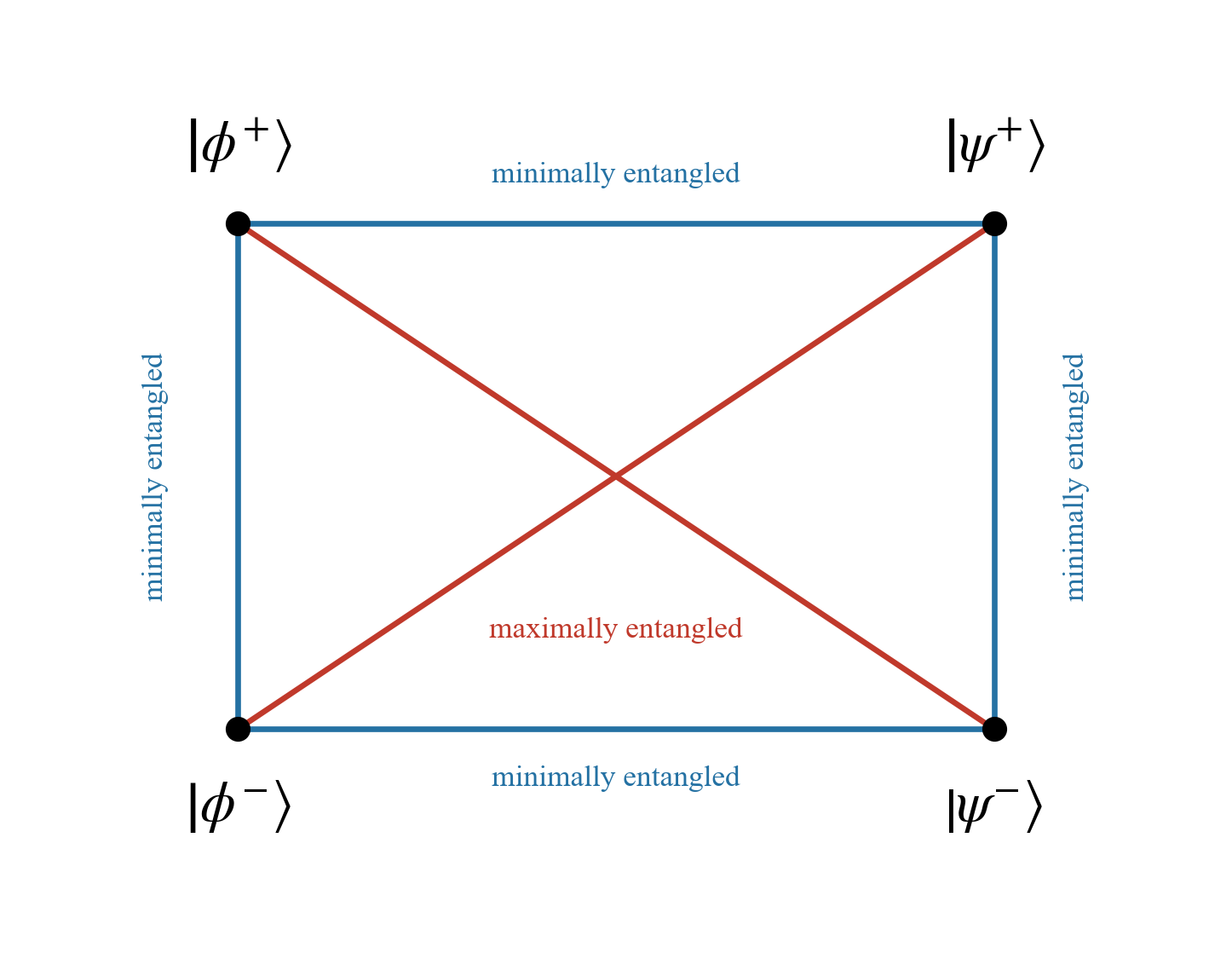}
    \caption{Degree of entanglement for $\ket{\phi^+}$, $\ket{\phi^-}$, 
    $\ket{\psi^+}$, $\ket{\psi^-}$ when $\phi=0$.}
    \label{fig:diagram0}
\end{figure}

\begin{figure}[htbp]
    \centering
    \includegraphics[width=0.7\textwidth]{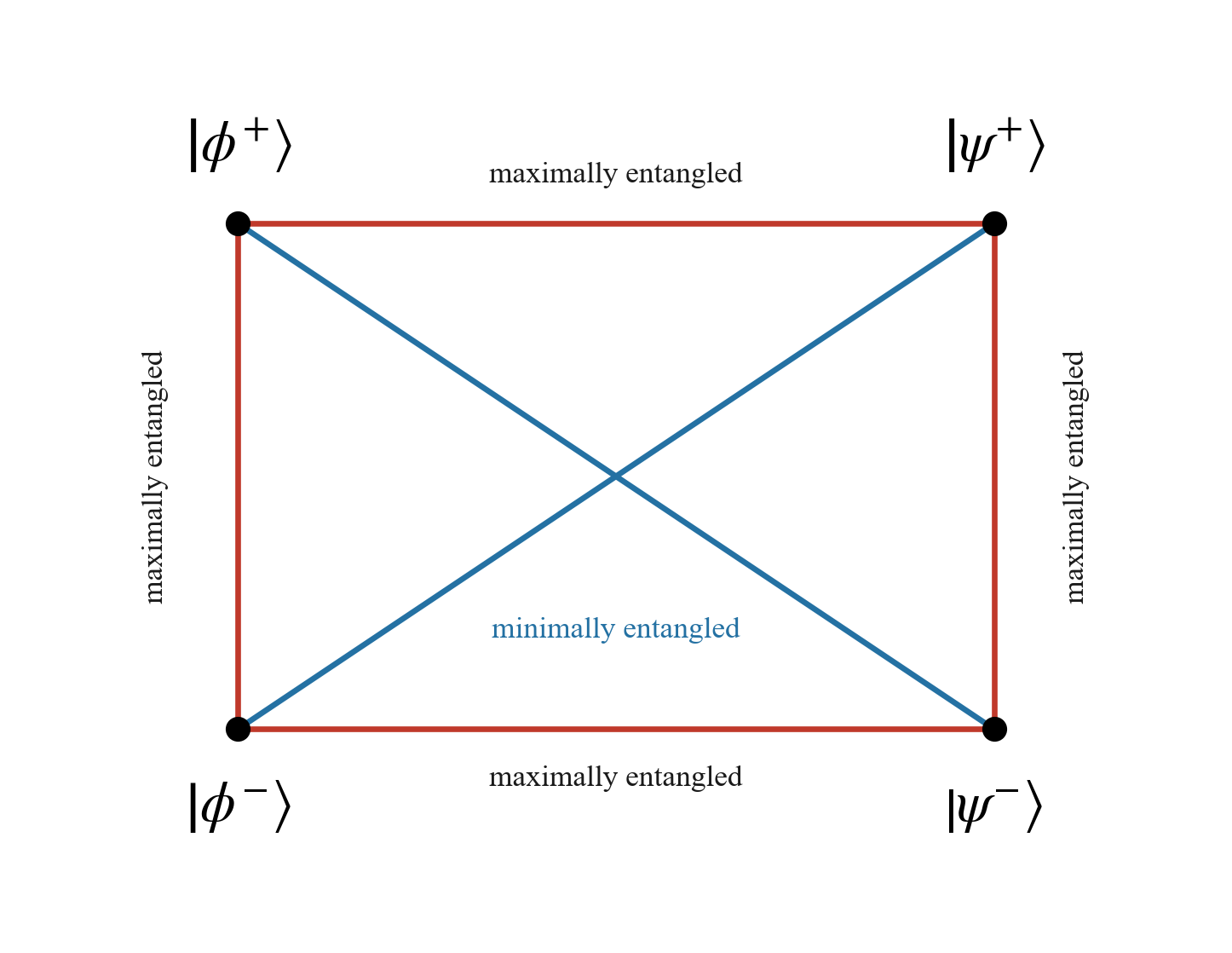}
    \caption{Degree of entanglement for $\ket{\phi^+}$, $\ket{\phi^-}$, 
    $\ket{\psi^+}$, $\ket{\psi^-}$ when $\phi=\frac{\pi}{2}$.}
    \label{fig:diagram90}
\end{figure}

\FloatBarrier

Considering the symmetry in all the calculations, we can generalize the 
entropy for $\phi=0$ into a table. Let
\[
\cup \;=\; -\left[\frac{1+2\sqrt{s(1-s)}}{2}\log_2\!\left(\frac{1+2\sqrt{s(1-s)}}{2}\right)
+\frac{1-2\sqrt{s(1-s)}}{2}\log_2\!\left(\frac{1-2\sqrt{s(1-s)}}{2}\right)\right],
\]
where $\cup$ denotes an $s$-dependent entropy and $1$ corresponds to the maximally 
entangled state. Table~\ref{tab:superposition} collects the entropy for all sixteen 
superpositions at $\phi=0$. The block structure of the table directly reflects the same classification above: same bases but opposite phases or different bases but same phases (off-diagonal blocks) yield $\cup$, 
while identical or different bases and opposite phases that preserve maximal entanglement yield $1$.

\begin{table}[htbp]
\centering
\begin{tabular}{lcccc}
\toprule
States & $\ket{\phi^+}$ & $\ket{\phi^-}$ & $\ket{\psi^+}$ & $\ket{\psi^-}$ \\
\midrule
$\ket{\phi^+}$ & $1$ & $\cup$ & $\cup$ & $1$ \\
$\ket{\phi^-}$ & $\cup$ & $1$ & $1$ & $\cup$ \\
$\ket{\psi^+}$ & $\cup$ & $1$ & $1$ & $\cup$ \\
$\ket{\psi^-}$ & $1$ & $\cup$ & $\cup$ & $1$ \\
\bottomrule
\end{tabular}
\caption{Entropy as a function of the relative weight $s$ for superpositions of 
all Bell states when $\phi=0$.}
\label{tab:superposition}
\end{table}
For $\ket{\phi}=\frac{\pi}{2}$, the symbols 1 and $\cup$ are reversed in the table.

\end{itemize}
\subsection{Horodecki Criterion for Superpositions of Bell States}
\label{sec:sup_horodecki}

We apply the Horodecki criterion~\cite{horodecki1995} to each Bell superposition state 
$\rho(s,\phi)$ to determine regions of CHSH non-locality in the $(s,\phi)$ 
parameter space. For each state we compute the correlation matrix 
$T_{jk} = \mathrm{Tr}[\rho(\sigma_j \otimes \sigma_k)]$, form $U = T^\top T$, 
and evaluate $M(T) = u_1 + u_2$ (the sum of the two largest eigenvalues of 
$U$). The state violates the CHSH inequality and is therefore quantum non-local 
if and only if $M(\rho) > 1$. Entanglement is quantified via the von Neumann 
entropy of the reduced state. 
The phase diagrams below follow the color convention of Section~\ref{sec:horodecki_method}: 
\textit{blue} (quantum non-local, $M>1$), \textit{orange} (Bell-local entanglement, 
$M \leq 1$ but $S>0$), and \textit{gray} (separable, $S=0$). A key result is that 
the orange region is absent for all three superpositions considered: entanglement 
and CHSH non-locality share an identical boundary for pure Bell state superpositions. 
This is a concrete manifestation of Gisin's theorem, that every entangled pure 
two-qubit state violates the CHSH inequality~\cite{gisin1991}.

\begin{figure}[htbp]
    \centering
    \begin{minipage}[t]{0.48\textwidth}
        \centering
        \includegraphics[width=\linewidth]{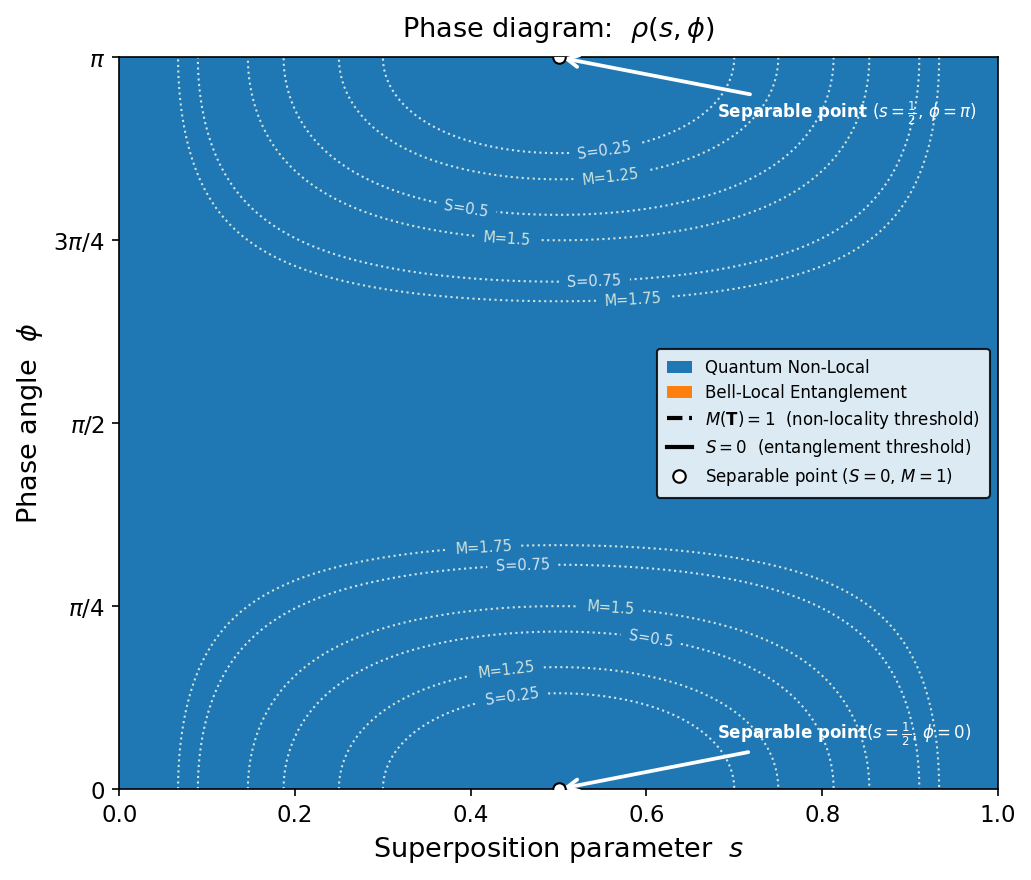}
        \caption{Phase diagram for $\ket{\phi^+}$--$\ket{\phi^-}$ (and 
        $\ket{\phi^+}$--$\ket{\psi^+}$). Separable points at 
        $(s,\phi) = (\tfrac{1}{2}, 0)$ and $(\tfrac{1}{2},\pi)$.}
        \label{fig:phase1}
    \end{minipage}
    \hfill
    \begin{minipage}[t]{0.48\textwidth}
        \centering
        \includegraphics[width=\linewidth]{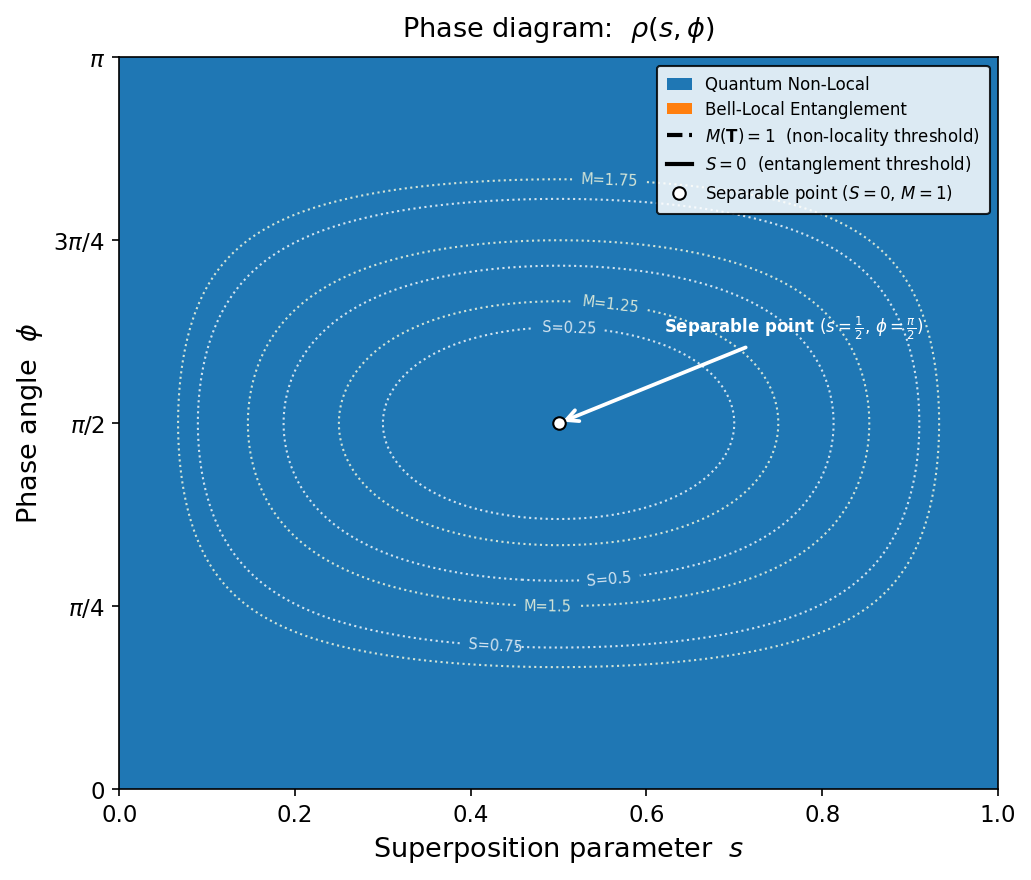}
        \caption{Phase diagram for $\ket{\phi^+}$--$\ket{\psi^-}$. The phase 
        structure rotates by $\pi/2$; the separable point shifts to 
        $(s,\phi) = (\tfrac{1}{2}, \tfrac{\pi}{2})$.}
        \label{fig:phase2}
    \end{minipage}
\end{figure}

\begin{itemize}

    \item \textbf{$\ket{\phi^+}$ with $\ket{\phi^-}$:} The Horodecki parameter is
    \begin{equation}
        M(s,\phi) = 2 - 4s(1-s)\cos^2\!\phi.
        \label{eq:M_phiplus_phiminus}
    \end{equation}
    Since $4s(1-s) \leq 1$ for all $s \in (0,1)$, with equality only at 
    $s = \tfrac{1}{2}$, the condition $M = 1$ reduces to the isolated points 
    $(s, \phi) = (\tfrac{1}{2}, 0)$ and $(\tfrac{1}{2}, \pi)$. The state is 
    non-local and entangled everywhere else in the parameter space. The 
    boundaries of non-locality and entanglement coincide exactly, confirming 
    the absence of any Bell-local entanglement region.

    \item \textbf{$\ket{\phi^+}$ with $\ket{\psi^+}$:} The analysis yields the 
    same expression,
    \begin{equation}
        M(s,\phi) = 2 - 4s(1-s)\cos^2\!\phi,
    \end{equation}
    with identical separable points and phase structure. This reflects that 
    $\ket{\phi^-}$ and $\ket{\psi^+}$ are related by a local unitary, leaving 
    both $S$ and $M$ invariant.

    \item \textbf{$\ket{\phi^+}$ with $\ket{\psi^-}$:} Here the parameter is
    \begin{equation}
        M(s,\phi) = 2\!\left(2s^2\sin^2\!\phi - 2s\sin^2\!\phi + 1\right) 
        = 2 - 4s(1-s)\sin^2\!\phi.
        \label{eq:M_phiplus_psiminus}
    \end{equation}
    The dependence shifts from $\cos^2\phi$ to $\sin^2\phi$, rotating the phase 
    structure by $\pi/2$. The separable point moves to 
    $(s,\phi) = (\tfrac{1}{2}, \tfrac{\pi}{2})$, and the state is maximally 
    entangled and non-local along $\phi = 0$ and $\phi = \pi$ for any $s$. Notice that Fig~\ref{fig:phase1} and Fig~\ref{fig:phase2} are identical under phase translation of $\frac{\pi}{2}.$

\end{itemize}

\FloatBarrier

In both diagrams the entire parameter space is quantum non-local (blue), with 
the orange Bell-local entanglement region absent. The white dotted contours show 
isolines of constant $S$ and $M$, confirming that both quantities increase 
monotonically as the state moves away from the separable point. The innermost 
contours (lowest $S$, $M$ closest to 1) encircle the unique separable point, 
while the outermost contours approach the maximally entangled boundary $S = 1$, 
$M = 2$. This behavior is fully consistent with the von Neumann entropy analysis 
presented earlier: the separable boundary $S = 0$ and the non-locality threshold 
$M = 1$ are governed by the same condition $4s(1-s)f(\phi) = 1$ (where 
$f = \cos^2\phi$ or $\sin^2\phi$ depending on the superposition), confirming 
that for these pure Bell superpositions, entanglement is equivalent to CHSH 
non-locality.
\subsection{Mixing two pure Bell states without noise}
\label{sec:mixing}
We now mix two pure Bell states and find the degree of entanglement of the resulting 
state. Since mixing two pure states yields a mixed state, we use 
concurrence to evaluate the entanglement. The general method we use to mix two pure 
Bell states $\ket{B_1}$ and $\ket{B_2}$ is to combine the corresponding density matrices with a mixing probability 
$r$:
\[
\rho_{\mathrm{mixed}} = (1-r)\rho_{\ket{B_1}} + r\rho_{\ket{B_2}}.
\]
The resulting state is a valid mixed density matrix. We note that all mixed Bell state 
density matrices in this section retain the X-state structure, so the concurrence can 
be evaluated directly via Eq.~(\ref{eq:x_state_concurrence}). For cases where the matrix $R$ differs from $\rho$ (as occurs in the $\ket{\phi^-}$ mixtures 
below), we verified explicitly that the general Wootters formula and the X-state 
shortcut yield identical eigenvalues and hence identical concurrence.

We start with $\ket{\phi^+}$,
\[
\rho=\begin{bmatrix}
    \frac{1}{2}&0&0&\frac{1}{2}\\
    0&0&0&0\\
    0&0&0&0\\
    \frac{1}{2}&0&0&\frac{1}{2}
\end{bmatrix}.
\]
The combination of $\ket{\phi^+}$ with any other Bell state is given by
\[
\rho_{\mathrm{mixed}} = (1-r)\rho_{\ket{\phi^+}} + r
\left[\rho_{\ket{\phi^+}},\,\rho_{\ket{\phi^-}},\,\rho_{\ket{\psi^+}},\,
\rho_{\ket{\psi^-}}\right].
\]

\begin{itemize}

\item \textbf{For $\ket{\phi^+}$ and $\ket{\phi^+}$:}
\[
\rho_{\mathrm{mixed}}=\begin{bmatrix}
    \frac{1}{2}&0&0&\frac{1}{2}\\
    0&0&0&0\\
    0&0&0&0\\
    \frac{1}{2}&0&0&\frac{1}{2}
\end{bmatrix}.
\]
Since we are mixing the same state, the resulting density is unchanged. The 
concurrence is the same as that of the pure $\ket{\phi^+}$, which is one. Hence 
we can neglect the four possible mixtures of identical states, leaving twelve 
distinct combinations.

\item \textbf{For $\ket{\phi^+}$ and $\ket{\phi^-}$:}
\[
\rho_{\mathrm{mixed}}=(1-r)\rho_{\ket{\phi^+}}+r\rho_{\ket{\phi^-}} = 
\begin{bmatrix}
    \frac{1}{2}&0&0&\frac{1}{2}-r\\
    0&0&0&0\\
    0&0&0&0\\
    \frac{1}{2}-r&0&0&\frac{1}{2}
\end{bmatrix}.
\]
We calculate $R = \sqrt{\sqrt{\rho}\,\tilde{\rho}\,\sqrt{\rho}}$, where 
$\tilde{\rho}=(\sigma_y\otimes\sigma_y)\rho^*(\sigma_y\otimes\sigma_y)$. Since 
$\rho$ is purely real, $\rho^*=\rho$. Furthermore, the X-state symmetry of this 
matrix makes it invariant under the spin-flip operation, so $\tilde{\rho} = \rho$ 
and hence $R = \rho$.

The eigenvalues of $R$ are $\lambda_1=r$, $\lambda_2=1-r$, $\lambda_3=0$, 
$\lambda_4=0$. Their ordering depends on $r$, giving two cases:
\begin{itemize}
    \item Case-I $(0 \leq r \leq \tfrac{1}{2})$: $C = \max[0,\,1-2r].$
    \item Case-II $(\tfrac{1}{2} \leq r \leq 1)$: $C = \max[0,\,2r-1].$
\end{itemize}
Both cases are captured by $C = |2r-1|$. At $r=0$ or $r=1$ (no mixing), the 
concurrence equals one, corresponding to a maximally entangled state.

\item \textbf{For $\ket{\phi^+}$ and $\ket{\psi^+}$:}
\[
\rho_{\mathrm{mixed}}=\begin{bmatrix}
    \frac{1-r}{2}&0&0&\frac{1-r}{2}\\
    0&\frac{r}{2}&\frac{r}{2}&0\\
    0&\frac{r}{2}&\frac{r}{2}&0\\
    \frac{1-r}{2}&0&0&\frac{1-r}{2}
\end{bmatrix}.
\]
The four eigenvalues of $R$ are again $\lambda_1=r$, $\lambda_2=1-r$, 
$\lambda_3=0$, $\lambda_4=0$, yielding the same concurrence as in the previous case.

\item \textbf{For $\ket{\phi^+}$ and $\ket{\psi^-}$:}
\[
\rho_{\mathrm{mixed}}=\begin{bmatrix}
    \frac{1-r}{2}&0&0&\frac{1-r}{2}\\
    0&\frac{r}{2}&-\frac{r}{2}&0\\
    0&-\frac{r}{2}&\frac{r}{2}&0\\
    \frac{1-r}{2}&0&0&\frac{1-r}{2}
\end{bmatrix}.
\]
The eigenvalues of $R$ are again identical, giving the same concurrence.

We see that mixing $\ket{\phi^+}$ with any of the other Bell states yields the 
same concurrence for all three non-identical pairings. We expect the same result 
for the other three Bell states. This contrasts with the superposition case, where 
the specific choice of Bell states affected the entanglement.

To verify, we repeat the analysis starting from $\ket{\phi^-}$,
\[
\rho_{\ket{\phi^-}}=\begin{bmatrix}
    \frac{1}{2}&0&0&-\frac{1}{2}\\
    0&0&0&0\\
    0&0&0&0\\
    -\frac{1}{2}&0&0&\frac{1}{2}
\end{bmatrix}.
\]

\item \textbf{For $\ket{\phi^-}$ with $\ket{\psi^+}$:}
\[
\rho_{\mathrm{mixed}}=\begin{bmatrix}
    \frac{1-r}{2}&0&0&\frac{-1+r}{2}\\
    0&\frac{r}{2}&\frac{r}{2}&0\\
    0&\frac{r}{2}&\frac{r}{2}&0\\
    \frac{-1+r}{2}&0&0&\frac{1-r}{2}
\end{bmatrix}.
\]
Here $R \neq \rho$; however, the eigenvalues of $R$ are $\{0,\,0,\,1-r,\,r\}$, 
identical to the previous cases, so the concurrence is the same.

\item \textbf{For $\ket{\phi^-}$ and $\ket{\psi^-}$:}
\[
\rho_{\mathrm{mixed}}=\begin{bmatrix}
    \frac{1-r}{2}&0&0&\frac{-1+r}{2}\\
    0&\frac{r}{2}&-\frac{r}{2}&0\\
    0&-\frac{r}{2}&\frac{r}{2}&0\\
    \frac{-1+r}{2}&0&0&\frac{1-r}{2}
\end{bmatrix}.
\]
Once again $R \neq \rho$, but the eigenvalues of $R$ are identical to all 
previous non-identical cases, giving the same concurrence. We continued the 
same method for the remaining Bell state pairings and find the same eigenvalues 
throughout.

\end{itemize}

\begin{figure}[htbp]
    \centering
    \begin{minipage}[t]{0.48\textwidth}
        \centering
        \includegraphics[width=\linewidth]{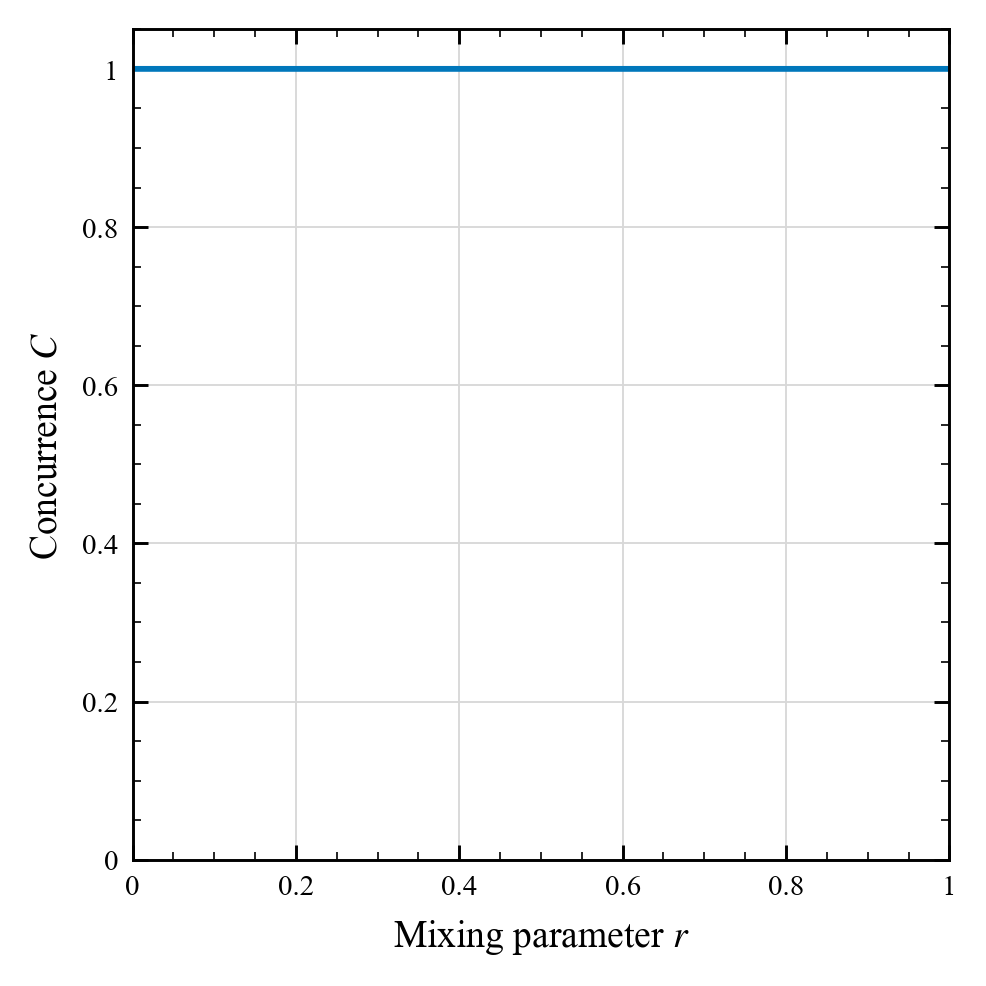}
        \caption{Concurrence as a function of the mixing parameter $r$ for identical 
        Bell states.}
        \label{fig:mix_id}
    \end{minipage}
    \hfill
    \begin{minipage}[t]{0.48\textwidth}
        \centering
        \includegraphics[width=\linewidth]{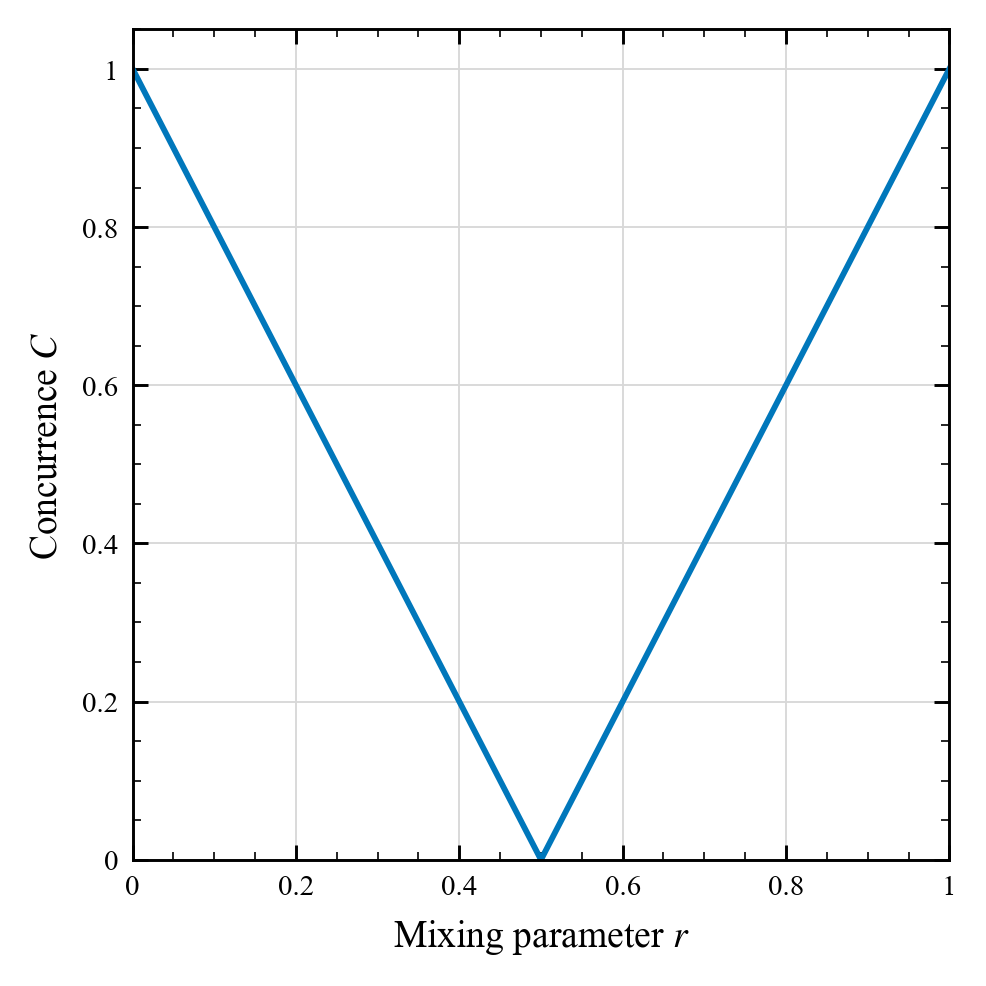}
        \caption{Concurrence as a function of the mixing parameter $r$ for 
        non-identical Bell states.}
        \label{fig:mix_nonid}
    \end{minipage}
\end{figure}

We present two graphs showing all mixing combinations. Figure~\ref{fig:mix_id} 
shows concurrence $C$ versus mixing parameter $r$ for identical Bell states, 
and Figure~\ref{fig:mix_nonid} shows the same for non-identical Bell states.
\FloatBarrier
In conclusion, mixing a Bell state with itself leaves the state unchanged and 
hence maximally entangled ($C=1$), whereas mixing two non-identical Bell states 
produces a characteristic V-shaped concurrence varying linearly from $C=1$ at $r=0$ or 
$r=1$ to $C=0$ at equal mixing weights $r=0.5$.

\subsection{Horodecki Criterion for Mixing of Non-Identical Bell States}
\label{sec:mix_horodecki}
\begin{figure}[H]
    \centering
    \includegraphics[width=0.65\linewidth]{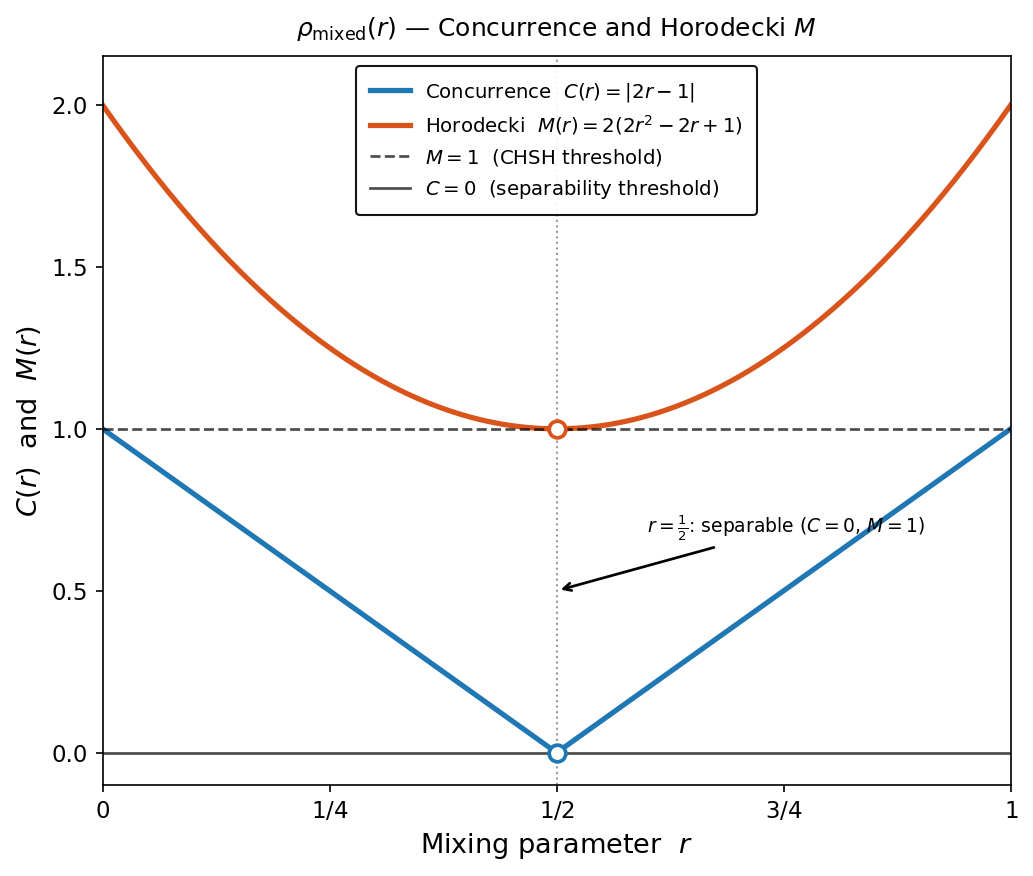}
    \caption{Concurrence $C(r)$ and Horodecki parameter $M(r)$ for the mixture 
    of any two non-identical Bell states. At $r = \tfrac{1}{2}$, the concurrence 
    vanishes ($C = 0$) and the Horodecki parameter reaches its minimum value 
    $M = 1$, marking the unique separable point where entanglement and CHSH 
    non-locality are simultaneously lost. The result is identical for all six 
    non-identical pairings.}
    \label{fig:mixed}
\end{figure}

A key result is that all six non-identical Bell state pairings ($\ket{\phi^\pm}$, 
$\ket{\psi^\pm}$ taken in pairs) yield \textit{identical} concurrence and 
Horodecki parameter:
\begin{equation}
    C(r) = |2r - 1|,
    \label{eq:C_mix}
\end{equation}
\begin{equation}
    M(r) = 2\left(2r^2 - 2r + 1\right).
    \label{eq:M_mix}
\end{equation}
This universality was verified numerically for all six combinations and contrasts 
with the superposition case, where the specific choice of Bell states affected the 
result. The two quantities satisfy the exact relation $M(r) = 1 + C^2(r)$, so 
$M > 1$ whenever $C > 0$, confirming that entanglement and CHSH non-locality share 
the same boundary at $r = \tfrac{1}{2}$. This shared boundary is displayed in 
Figure~\ref{fig:mixed}.

The state is entangled and CHSH non-local for all $r \neq \tfrac{1}{2}$, with 
no Bell-local entanglement region. At $r = \tfrac{1}{2}$ the mixture is maximally 
disordered between the two Bell states, yielding a separable state with $C = 0$ 
and $M = 1$.
\subsection{Introducing DP noise in one Bell state and mixing with a pure Bell state}
\label{sec:dp}
From this section on, we incorporate noise into the Bell states. In this section, 
we first introduce DP noise to a pure Bell state under the noise probability $p$ 
and then mix it with a pure Bell state under the mixing probability $r$. To 
effectively parametrize the concurrence, we use the effective noise parameter 
$x = p(1-r)$ defined in Eq.~(\ref{eq:effective_noise}). The reconstructed density 
matrix takes the form
\[
\rho_{\mathrm{reconstructed}} = (1-r)\rho_{\mathrm{noisy}} + r\rho_{\mathrm{original}},
\]
where
\[
\rho_{\mathrm{noisy}} = (1-p)\rho_{\ket{\phi^+}} + p\,\frac{I}{4}.
\]
All reconstructed matrices in this section retain the X-state structure, so 
concurrence is evaluated directly via Eq.~(\ref{eq:x_state_concurrence}). We 
start with $\ket{\phi^+}$ and derive the reconstructed density matrix for each 
mixing combination.

\begin{figure}[htbp]
    \centering
    \begin{minipage}[t]{0.48\textwidth}
        \centering
        \includegraphics[width=\linewidth]{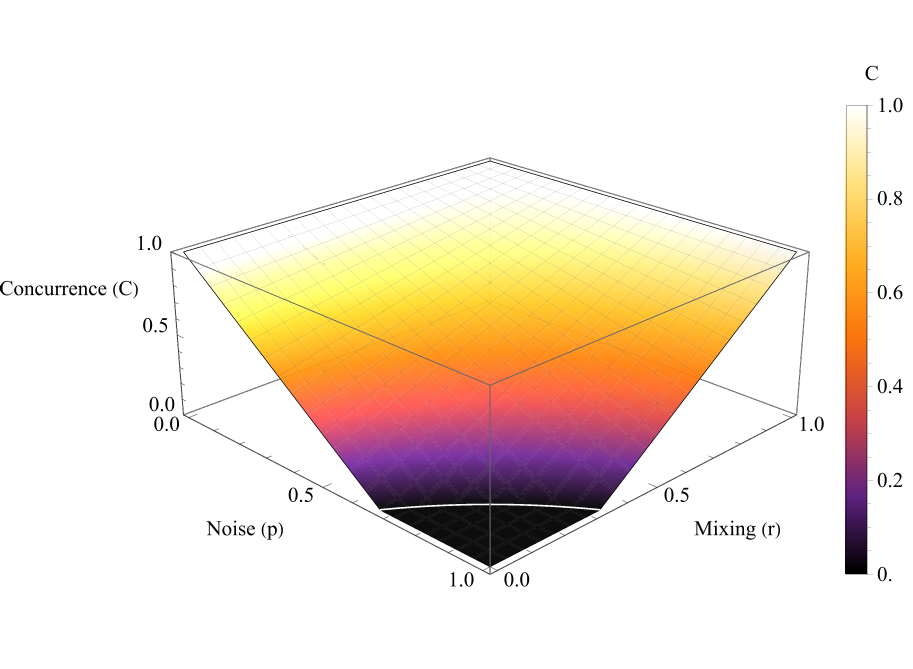}
        \caption{Concurrence $C$ as a function of mixing parameter $r$ and DP noise 
        $p$ for identical Bell states.}
        \label{fig:dp_identical}
    \end{minipage}
    \hfill
    \begin{minipage}[t]{0.48\textwidth}
        \centering
        \includegraphics[width=\linewidth]{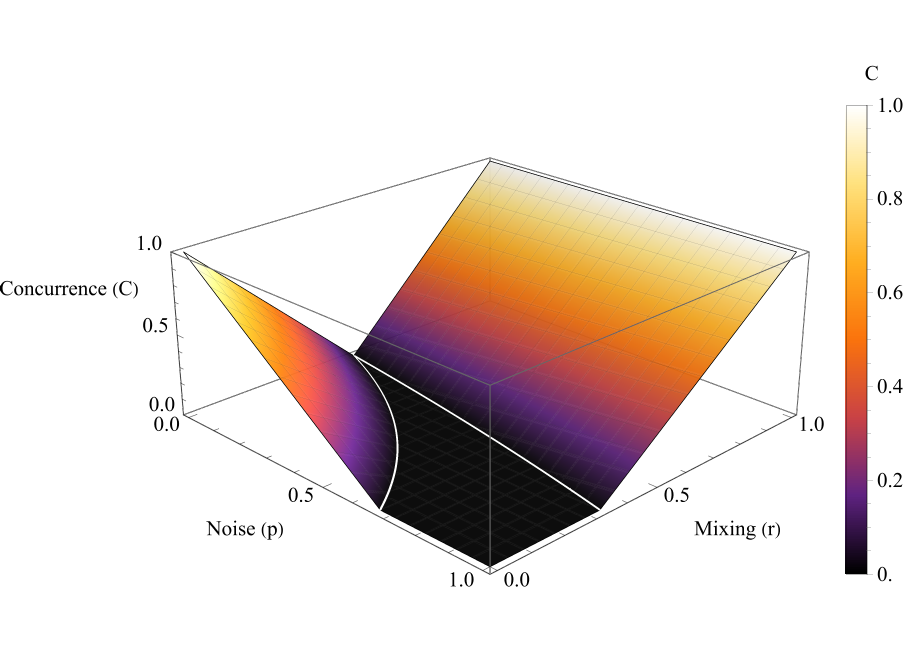}
        \caption{Concurrence $C$ as a function of mixing parameter $r$ and DP noise 
        $p$ for non-identical Bell states.}
        \label{fig:dp_non_identical}
    \end{minipage}
\end{figure}

\begin{itemize}

\item \textbf{For $\ket{\phi^+}$ with $\ket{\phi^+}$:}
\[
\rho_{\mathrm{reconstructed}} = \begin{bmatrix}
    \frac{2-p+pr}{4} & 0 & 0 & \frac{1-p+pr}{2}\\
    0 & \frac{p(1-r)}{4} & 0 & 0\\
    0 & 0 & \frac{p(1-r)}{4} & 0\\
    \frac{1-p+pr}{2} & 0 & 0 & \frac{2-p+pr}{4}
\end{bmatrix}.
\]
The concurrence is
\begin{equation}
    C = \max\!\left[0,\;1-\frac{3x}{2}\right].
    \label{eq:dp_identical_C}
\end{equation}
We plot the resulting concurrence as a function of $p$ and $r$ in 
Fig.~\ref{fig:dp_identical}.
We show that the concurrence monotonically decreases with noise reducing
exactly to the Werner state $\rho = (1-x)\ket{\phi^+}\bra{\phi^+} + xI/4$ 
with $x = p(1-r)$, and our thresholds recover the known separability ($x < 2/3$) 
and CHSH ($x < 1-1/\sqrt{2}$) boundaries~\cite{werner1989,horodecki1995} as a 
consistency check.
\item \textbf{For $\ket{\phi^+}$ with $\ket{\phi^-}$:}
\[
\rho_{\mathrm{reconstructed}} = \begin{bmatrix}
    \frac{2-p+pr}{4} & 0 & 0 & \frac{1-p-2r+pr}{2}\\
    0 & \frac{p(1-r)}{4} & 0 & 0\\
    0 & 0 & \frac{p(1-r)}{4} & 0\\
    \frac{1-p-2r+pr}{2} & 0 & 0 & \frac{2-p+pr}{4}
\end{bmatrix}.
\]
The concurrence is
\begin{equation}
    C = \max\!\left[0,\;|1-2r-x| - \frac{x}{2}\right].
    \label{eq:dp_nonidentical_C}
\end{equation}

\item \textbf{For $\ket{\phi^+}$ with $\ket{\psi^-}$:}
\[
\rho_{\mathrm{reconstructed}} = \begin{bmatrix}
    \frac{2-p-2r+pr}{4} & 0 & 0 & \frac{1-p-r+pr}{2}\\
    0 & \frac{p-pr+2r}{4} & -\frac{r}{2} & 0\\
    0 & -\frac{r}{2} & \frac{p-pr+2r}{4} & 0\\
    \frac{1-p-r+pr}{2} & 0 & 0 & \frac{2-p-2r+pr}{4}
\end{bmatrix}.
\]
Evaluating the concurrence via Eq.~(\ref{eq:x_state_concurrence}) and 
substituting $x = p(1-r)$ yields the same universal non-identical expression 
as Eq.~(\ref{eq:dp_nonidentical_C}).

\item \textbf{For $\ket{\phi^+}$ with $\ket{\psi^+}$:}
\[
\rho_{\mathrm{reconstructed}} = \begin{bmatrix}
    \frac{2-p-2r+pr}{4} & 0 & 0 & \frac{1-p-r+pr}{2}\\
    0 & \frac{p-pr+2r}{4} & \frac{r}{2} & 0\\
    0 & \frac{r}{2} & \frac{p-pr+2r}{4} & 0\\
    \frac{1-p-r+pr}{2} & 0 & 0 & \frac{2-p-2r+pr}{4}
\end{bmatrix}.
\]
Again, the concurrence reduces to Eq.~(\ref{eq:dp_nonidentical_C}).

\end{itemize}

We repeat the process with all combinations of noisy and noiseless Bell states 
and obtain identical results throughout. The universality collapses all sixteen 
combinations into just two closed-form expressions, now written in terms of $r$ and $p$ explicitly 
\begin{itemize}
    \item For identical Bell states:
    \begin{equation}
        C = \max\!\left[0,\;1 - \frac{3p(1-r)}{2}\right],
        \label{eq:dp_id_final}
    \end{equation}
    \item For non-identical Bell states:
    \begin{equation}
        C = \max\!\left[0,\;|1-2r-p(1-r)| - \frac{p(1-r)}{2}\right].
        \label{eq:dp_nonid_final}
    \end{equation}
\end{itemize}

\FloatBarrier
Figure~\ref{fig:dp_identical} shows concurrence versus noise for identical Bell 
states, while Figure~\ref{fig:dp_non_identical} covers non-identical Bell states. 
Both figures recover the noiseless results (Figures~\ref{fig:mix_id} and 
\ref{fig:mix_nonid}) at $p=0$: $C=1$ for identical states and the characteristic 
V-shaped profile for non-identical states.

The behavior in Figure~\ref{fig:dp_identical} is straightforward: as noise 
increases from $p=0$ to $p=1$, the concurrence decreases monotonically, 
reflecting the uniform purity loss of the DP channel.

Figure~\ref{fig:dp_non_identical} reveals a more striking result. The surface 
is divided by the white curve $p(1-r) = 1-2r$ into two qualitatively different 
regimes. To the left of this curve (low noise or low $r$), concurrence decreases 
with noise as expected. To the right (high $r$ or high noise), concurrence 
\textit{increases} with noise, reaching its maximum at $p = 1$ for any fixed 
$r > 0.5$.

The mechanism is a consequence of the signs of the off-diagonal coherences. 
Consider mixing a noisy $\ket{\phi^+}$ (coherence $+\frac{1}{2}$ in the $\rho_{14}$ 
position) with a pure $\ket{\phi^-}$ (coherence $-\frac{1}{2}$). When $r > 0.5$, 
the pure $\ket{\phi^-}$ branch dominates, but the noisy $\ket{\phi^+}$ branch is 
still present with its positive coherence partially canceling the negative coherence 
of the pure branch. The net off-diagonal element of the mixture is smaller in 
magnitude than it would be from the pure branch alone, and so is the concurrence. 
Depolarizing noise drives the noisy branch toward $I/4$, which has zero coherence. 
As $p$ increases, the noisy branch grows quieter and its canceling contribution 
shrinks. By $p = 1$ the noisy branch has zero coherence and the pure branch 
contributes without interference, recovering the maximum possible concurrence for 
that value of $r$.

This represents a regime of \textbf{noise-enhanced entanglement}: the noise is 
not creating entanglement, but destroying the destructive interference that was 
suppressing it. 

\subsection{Introducing PD noise in one Bell state and mixing with a pure Bell state}
\label{sec:pd}
We repeat the same process as in Section~\ref{sec:dp}, but with Phase Damping noise. 
A key physical distinction from the Depolarizing channel is that PD noise acts 
only on the off-diagonal coherences, leaving the diagonal populations entirely 
invariant. As a result, all reconstructed density matrices in this section retain 
a purely real, symmetric X-state structure with unpopulated inner diagonal elements 
, a feature that sharply distinguishes the PD channel from the DP channel.

\begin{itemize}

\item \textbf{For a noisy $\ket{\phi^+}$ with a pure $\ket{\phi^+}$:}
The reconstructed density operator is:
\[
\rho' = (1-r)\rho_{\mathrm{noisy}} + r\rho_{\ket{\phi^+}},
\]
where the phase-damped noisy state is:
\[
\rho_{\mathrm{noisy}} = (1-p)\rho_{\ket{\phi^+}} + p\begin{bmatrix}
    \frac{1}{2} & 0 & 0 & 0 \\
    0 & 0 & 0 & 0 \\
    0 & 0 & 0 & 0 \\
    0 & 0 & 0 & \frac{1}{2}
\end{bmatrix}.
\]
Carrying out the mixing procedure yields the reconstructed density matrix:
\begin{equation}
\rho' = \begin{bmatrix}
    \frac{1}{2} & 0 & 0 & \frac{1-p+pr}{2} \\
    0 & 0 & 0 & 0 \\
    0 & 0 & 0 & 0 \\
    \frac{1-p+pr}{2} & 0 & 0 & \frac{1}{2}
\end{bmatrix}.
\label{eq:pd_identical_rho}
\end{equation}
Since the density matrix is an X-state, we find the concurrence directly via 
Eq.~(\ref{eq:x_state_concurrence}):
\begin{equation}
    C = \max[0,\,1-x].
    \label{eq:pd_identical_C}
\end{equation}
The corresponding concurrence surface is shown in 
Figure~\ref{fig:pd_identical_surf}.

\begin{figure}[htbp]
    \centering
    \includegraphics[width=0.7\textwidth]{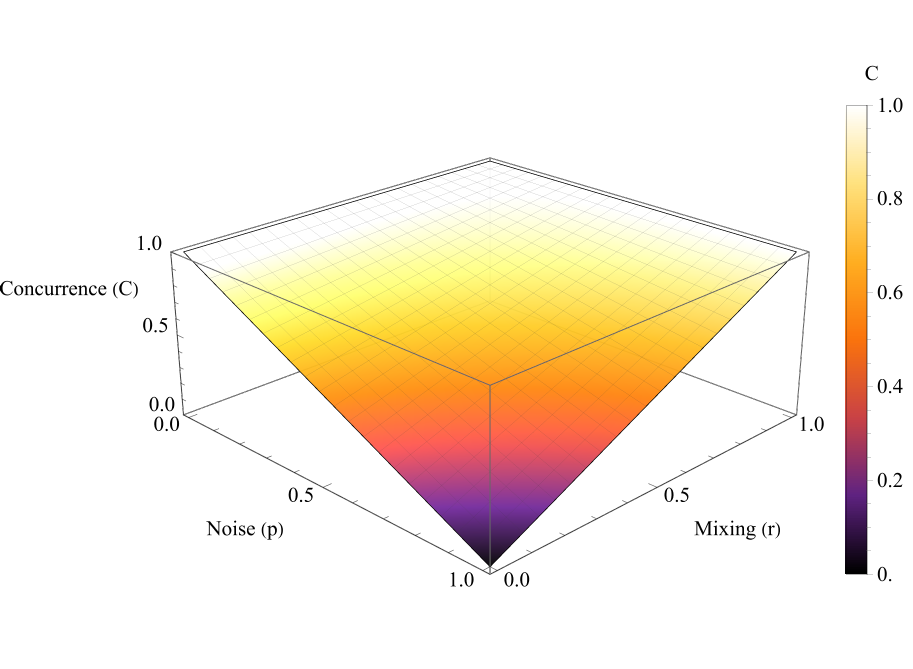}
    \caption{Concurrence $C$ as a function of mixing parameter $r$ and PD noise 
    parameter $p$ for identical Bell states.}
    \label{fig:pd_identical_surf}
\end{figure}

\item \textbf{For a noisy $\ket{\phi^+}$ with a pure $\ket{\phi^-}$:}
\[
\rho' = (1-r)\rho_{\mathrm{noisy}} + r\rho_{\ket{\phi^-}}.
\]
The reconstructed density matrix is:
\begin{equation}
\rho' = \begin{bmatrix}
    \frac{1}{2} & 0 & 0 & \frac{1-p-2r+pr}{2} \\
    0 & 0 & 0 & 0 \\
    0 & 0 & 0 & 0 \\
    \frac{1-p-2r+pr}{2} & 0 & 0 & \frac{1}{2}
\end{bmatrix}.
\label{eq:pd_sameparity_rho}
\end{equation}
The concurrence is:
\begin{equation}
    C = \max\left[0,\,|1-2r-x|\right].
    \label{eq:pd_sameparity_C}
\end{equation}
Due to the symmetry of the Bell basis, reversing the mixture (mixing a 
phase-damped $\ket{\phi^-}$ with a pure $\ket{\phi^+}$ yields an identical 
concurrence landscape, plotted in Figure~\ref{fig:pd_same_surf}. The resulting 
density matrices differ only by a global 
sign on the off-diagonal elements, which is eliminated by the absolute value 
in the concurrence formula.

\begin{figure}[htbp]
    \centering
    \includegraphics[width=0.7\textwidth]{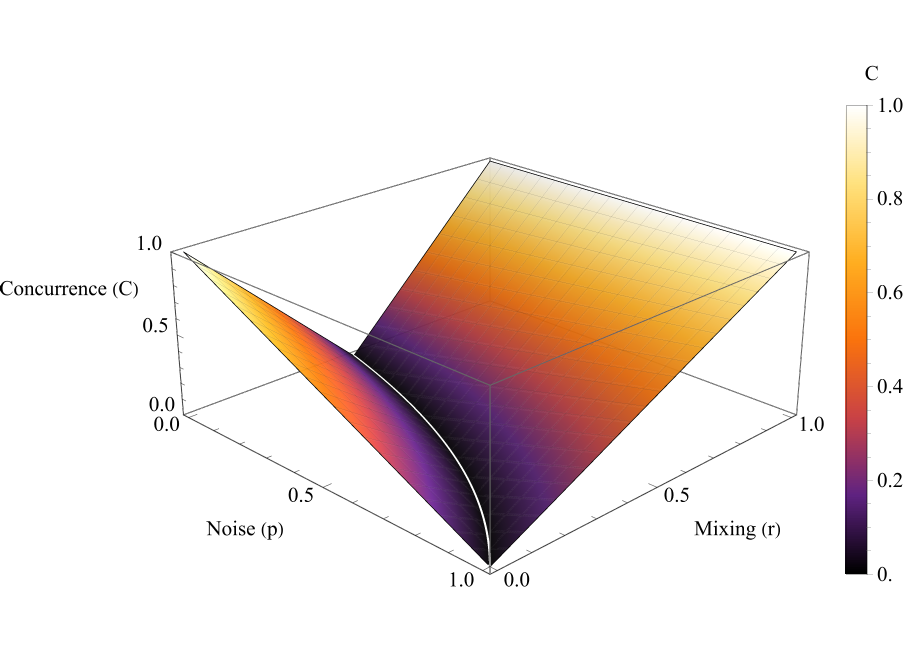}
    \caption{Concurrence $C$ as a function of mixing parameter $r$ for a noisy 
    $\ket{\phi^+}$ parametrized with $p$ mixed with a pure $\ket{\phi^-}$.}
    \label{fig:pd_same_surf}
\end{figure}

\item \textbf{For a noisy $\ket{\phi^+}$ mixed with $\ket{\psi^-}$ or $\ket{\psi^+}$:}
The reconstructed density operator for the mixture with a pure $\ket{\psi^-}$ is:
\[
\rho' = (1-r)\rho_{\mathrm{noisy}} + r\rho_{\ket{\psi^-}}.
\]
Substituting the phase-damped noisy state yields:
\begin{equation}
\rho' = \begin{bmatrix}
    \frac{1-r}{2} & 0 & 0 & \frac{(1-p)(1-r)}{2} \\
    0 & \frac{r}{2} & -\frac{r}{2} & 0 \\
    0 & -\frac{r}{2} & \frac{r}{2} & 0 \\
    \frac{(1-p)(1-r)}{2} & 0 & 0 & \frac{1-r}{2}
\end{bmatrix}.
\label{eq:pd_cross_rho}
\end{equation}
Evaluating the concurrence via Eq.~(\ref{eq:x_state_concurrence}) and substituting 
$x = p(1-r)$:
\begin{equation}
    C = \max\left[0,\,(1-p)(1-r)-r,\,r-(1-r)\right]
      = \max\left[0,\,1-x-2r,\,2r-1\right].
    \label{eq:pd_cross_C}
\end{equation}
Mixing the noisy $\ket{\phi^+}$ with a pure $\ket{\psi^+}$ instead yields a 
density matrix differing only by a positive sign on the inner off-diagonal 
elements ($\rho'_{23} = \rho'_{32} = +\frac{r}{2}$), giving the same concurrence 
by the absolute value in Eq.~(\ref{eq:x_state_concurrence}). The resulting 
concurrence surface is shown in Figure~\ref{fig:pd_cross_surf}.

\begin{figure}[htbp]
    \centering
    \includegraphics[width=0.7\textwidth]{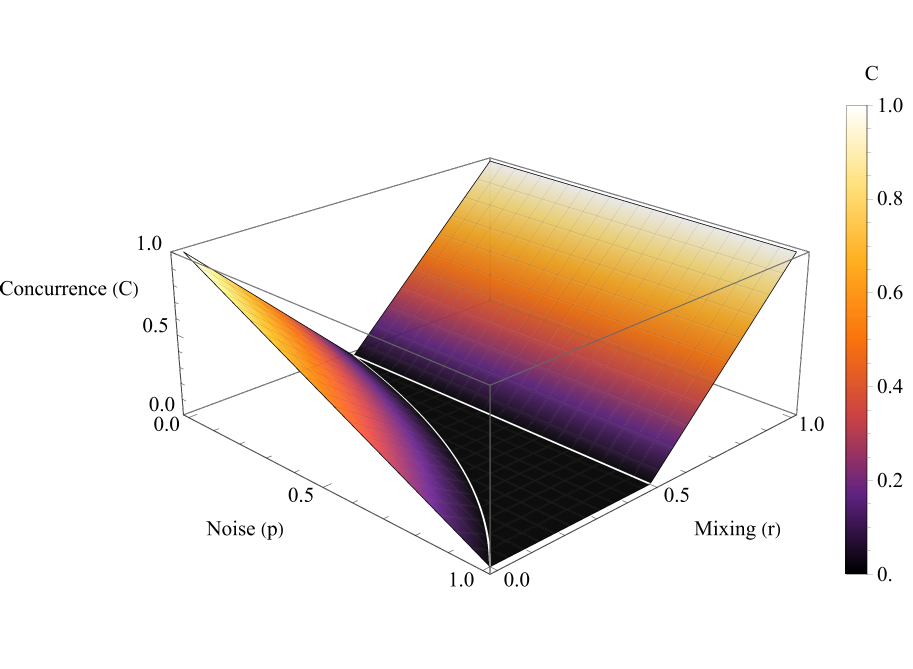}
    \caption{Concurrence $C$ as a function of mixing parameter $r$ for a noisy 
    $\ket{\phi^+}$ parametrized with $p$ mixed with a pure $\ket{\psi^-}$ or 
    $\ket{\psi^+}$.}
    \label{fig:pd_cross_surf}
\end{figure}

\end{itemize}

\FloatBarrier

Repeating the analysis for all remaining Bell state combinations yields identical 
results throughout. The three distinct mixing symmetries collapse into three 
closed-form concurrence expressions:
\begin{itemize}
    \item For identical Bell states:
    \begin{equation}
        C = \max\!\left[0,\;1 - p(1-r)\right],
        \label{eq:pd_id_final}
    \end{equation}
    \item For non-identical -- same basis but opposite phases (e.g.\ $\ket{\phi^+}$ with 
    $\ket{\phi^-}$):
    \begin{equation}
        C = \max\!\left[0,\;|1-2r-p(1-r)|\right],
        \label{eq:pd_sameparity_final}
    \end{equation}
    \item For non-identical -- different bases (e.g.\ $\ket{\phi^+}$ with $\ket{\psi^\pm}$):
    \begin{equation}
        C = \max\!\left[0,\;1-p(1-r)-2r,\;2r-1\right].
        \label{eq:pd_cross_final}
    \end{equation}
\end{itemize}
Figure~\ref{fig:pd_identical_surf} shows the identical Bell states case. Concurrence 
decreases monotonically with noise in all directions. This is because both 
branches have the same off-diagonal coherence with the same sign, so PD 
noise can only reduce the net coherence of the mixture. There is no 
noise-enhanced regime.

Figure~\ref{fig:pd_same_surf} shows the same-basis, opposite-phase case 
(e.g.\ noisy $\ket{\phi^+}$ with pure $\ket{\phi^-}$). Here the two 
branches have opposite signs on their off-diagonal coherence: $+\tfrac{1}{2}$ 
from $\ket{\phi^+}$ and $-\tfrac{1}{2}$ from $\ket{\phi^-}$. When $r > 0.5$, 
the noisy branch partially cancels the pure branch, reducing the net coherence. 
PD noise drives the noisy branch's coherence toward zero, silencing this 
cancellation. As a result, concurrence increases with noise for $r > 0.5$, 
with the white curve marking the separable boundary where the two branches 
cancel exactly.

Figure~\ref{fig:pd_cross_surf} shows the different-basis case (e.g.\ noisy 
$\ket{\phi^+}$ with pure $\ket{\psi^-}$). Here the two branches contribute 
to \textit{different} off-diagonal positions: $\ket{\phi^+}$ to $\rho_{14}$ 
and $\ket{\psi^-}$ to $\rho_{23}$. Since they occupy different positions, 
PD noise acting on the noisy branch has no effect on the pure branch's 
coherence. For $r > 0.5$ the surface is therefore completely flat in the 
noise direction: $C = 2r-1$ regardless of $p$.
\subsection{Adding Generalized AD noise on Bell states}
\label{sec:ad_bell}
We now introduce AD noise to the Bell states, using the two-qubit Kraus operators 
of Section~\ref{sec:noise} (derived in Appendix~\ref{app:ad_kraus}). The general 
equation for the evolution of a density matrix $\rho$ is:
\[
\rho' = \sum_{i=0}^{3} K_i\,\rho\,K_i^\dagger.
\]

\begin{itemize}

\item \textbf{For $\ket{\phi^+}$ and $\ket{\phi^-}$:}
Given the density matrix for the $\ket{\phi}$ states, we apply the AD Kraus 
operators and evaluate the concurrence via Eq.~(\ref{eq:x_state_concurrence}). 
The expression simplifies to:
\begin{equation}
    C = (1-p)^2.
    \label{eq:ad_phi_single_C}
\end{equation}
The graph showing concurrence versus noise probability $p$ for the $\ket{\phi}$ 
states is shown in Figure~\ref{fig:ad_phi_single}. Notice the non-linear decay, 
which arises because AD acts on the doubly-excited $\ket{11}$ 
component that both $\ket{\phi}$ states depopulate at rate proportional to $p^2$, since 
both qubits must decay.

\item \textbf{For $\ket{\psi^+}$ and $\ket{\psi^-}$:}
We introduce AD noise for the $\ket{\psi}$ states using the same Kraus operator 
formalism. The concurrence simplifies to:
\begin{equation}
    C = 1 - p.
    \label{eq:ad_psi_single_C}
\end{equation}
The graph is shown in Figure~\ref{fig:ad_psi_single}. In this case we get a 
linear decay. This makes sense: the $\ket{\psi}$ states are populated at 
$\ket{01}$ and $\ket{10}$ and do not populate the $\ket{11}$ component, so AD 
affects each excitation independently and symmetrically, giving the linear result. 
By contrast, the $\ket{\phi}$ states depopulate $\ket{11}$, producing the 
non-linear $(1-p)^2$ behavior.
\begin{figure}[htbp]
    \centering
    \begin{minipage}[t]{0.48\textwidth}
        \centering
        \includegraphics[width=\linewidth]{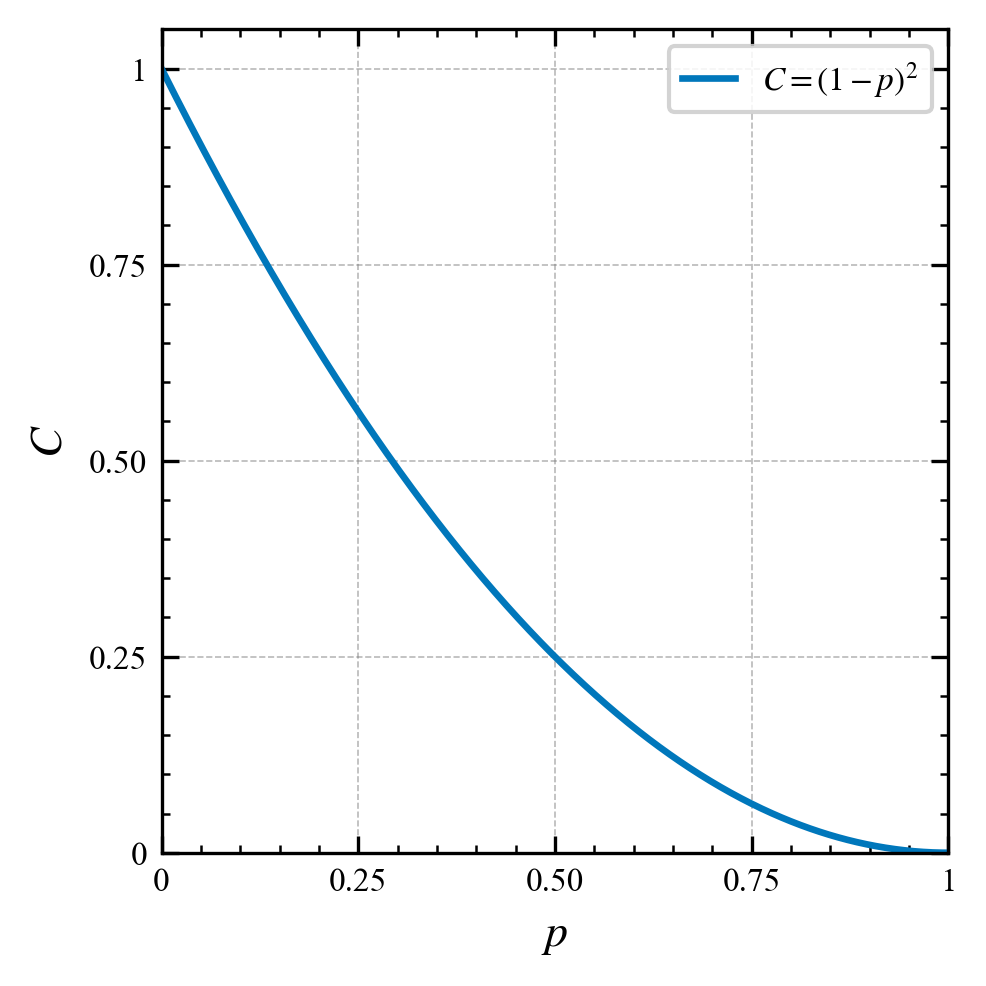}
        \caption{Concurrence $C$ as a function of the AD noise probability $p$ for 
        the $\ket{\phi^\pm}$ Bell states, $C = (1-p)^2$.}
        \label{fig:ad_phi_single}
    \end{minipage}
    \hfill
    \begin{minipage}[t]{0.48\textwidth}
        \centering
        \includegraphics[width=\linewidth]{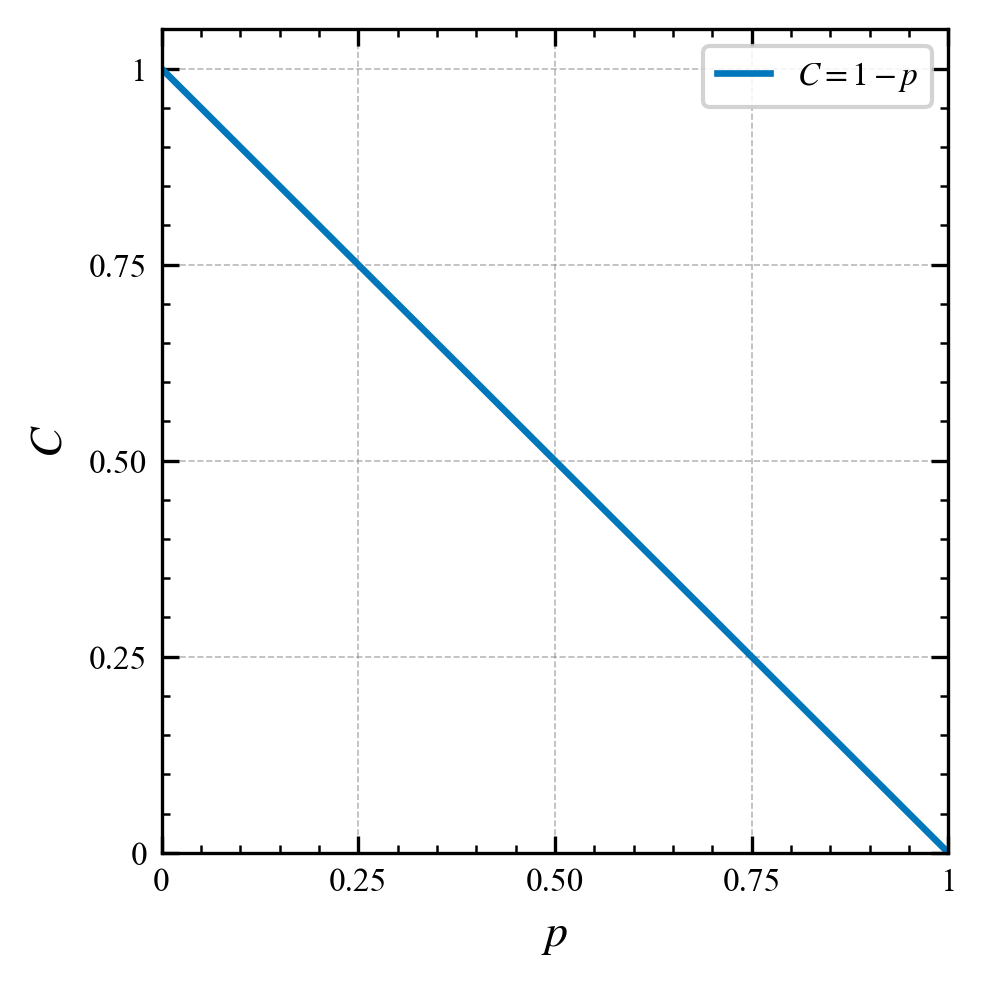}
        \caption{Concurrence $C$ as a function of the AD noise probability $p$ for 
        the $\ket{\psi^\pm}$ Bell states, $C = 1-p$.}
        \label{fig:ad_psi_single}
    \end{minipage}
\end{figure}
\end{itemize}

\FloatBarrier

\subsection{Mixing a pure Bell state with a noisy Amplitude Damping Bell state}
\label{sec:ad_mixing}
We now extend our method to the Amplitude Damping (AD) channel. We start by 
introducing AD noise to a pure Bell state with probability $p$ via the Kraus 
operators of Sections~\ref{sec:noise},~\ref{sec:ad_bell} and subsequently mix this noisy state with
a pure reference Bell state with probability $r$.

\textbf{In the identical Bell state cases:}
\begin{itemize}
\item \textbf{For a noisy $\ket{\phi^+}$ mixed with a pure $\ket{\phi^+}$:}

Applying the AD Kraus operators to $\ket{\phi^+}$, the noisy density matrix 
$\rho'$ is:
\[
\rho' = \sum_{i=0}^{3} K_i\rho_{\ket{\phi^+}}K_i^\dagger = \begin{bmatrix}
    \frac{1+p^2}{2} & 0 & 0 & \frac{1-p}{2} \\
    0 & \frac{p(1-p)}{2} & 0 & 0 \\
    0 & 0 & \frac{p(1-p)}{2} & 0 \\
    \frac{1-p}{2} & 0 & 0 & \frac{(1-p)^2}{2}
\end{bmatrix}.
\]
Unlike the Depolarizing and Phase Damping channels, Amplitude Damping represents 
an asymmetric energy loss, breaking the population symmetry such that 
$\rho'_{11} \neq \rho'_{44}$. We mix this noisy state with the pure 
$\ket{\phi^+}$ state:
\[
\rho'' = (1-r)\rho' + r\rho_{\ket{\phi^+}}.
\]
Expanding yields the reconstructed density matrix:
\begin{equation}
\rho'' = \begin{bmatrix}
    \frac{(1-r)(1+p^2)+r}{2} & 0 & 0 & \frac{1-p(1-r)}{2} \\
    0 & \frac{(1-r)p(1-p)}{2} & 0 & 0 \\
    0 & 0 & \frac{(1-r)p(1-p)}{2} & 0 \\
    \frac{1-p(1-r)}{2} & 0 & 0 & \frac{(1-r)(1-p)^2+r}{2}
\end{bmatrix}.
\label{eq:ad_phi_mix}
\end{equation}
Factoring the expression, the concurrence simplifies to:
\begin{equation}
    C = \max\!\left[0,\,(1-p)^2(1-r)+r\right].
    \label{eq:ad_identical_phi_C}
\end{equation}

\begin{figure}[htbp]
    \centering
    \includegraphics[width=0.7\textwidth]{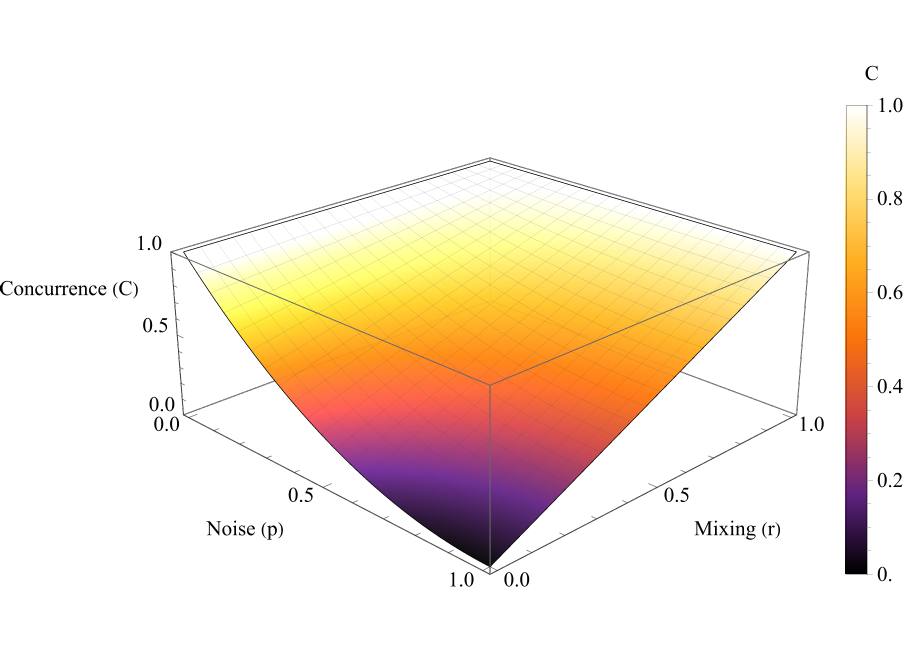}
    \caption{Concurrence $C$ as a function of mixing parameter $r$ for a noisy 
    $\ket{\phi^+}$ parametrized with $p$ mixed with a pure $\ket{\phi^+}$.}
    \label{fig:ad_id_surf}
\end{figure}

\item \textbf{For a noisy $\ket{\phi^-}$ mixed with a pure $\ket{\phi^-}$:}

Because the Amplitude Damping channel affects the populations of $\ket{00}$ and 
$\ket{11}$ irrespective of their relative phase, the population dynamics for 
$\ket{\phi^-}$ remain entirely identical to $\ket{\phi^+}$, yielding the exact 
same concurrence landscape as shown in Figure~\ref{fig:ad_id_surf}.

\item \textbf{For a noisy $\ket{\psi^+}$ mixed with a pure $\ket{\psi^+}$:}

Applying the AD Kraus operators to $\ket{\psi^+}$, the noise causes the 
$\ket{01}$ and $\ket{10}$ states to decay into the ground state $\ket{00}$. 
The noisy density matrix $\rho'$ is:
\[
\rho' = \begin{bmatrix}
    p & 0 & 0 & 0 \\
    0 & \frac{1-p}{2} & \frac{1-p}{2} & 0 \\
    0 & \frac{1-p}{2} & \frac{1-p}{2} & 0 \\
    0 & 0 & 0 & 0
\end{bmatrix}.
\]
Mixing with the pure $\ket{\psi^+}$ reference state with probability $r$:
\begin{equation}
\rho'' = \begin{bmatrix}
    p(1-r) & 0 & 0 & 0 \\
    0 & \frac{(1-r)(1-p)+r}{2} & \frac{(1-r)(1-p)+r}{2} & 0 \\
    0 & \frac{(1-r)(1-p)+r}{2} & \frac{(1-r)(1-p)+r}{2} & 0 \\
    0 & 0 & 0 & 0
\end{bmatrix}.
\label{eq:ad_psi_mix}
\end{equation}
The concurrence simplifies 
to:
\begin{equation}
    C = \max\!\left[0,\,1-x\right].
    \label{eq:ad_identical_psi_C}
\end{equation}
By symmetry, mixing a noisy $\ket{\psi^-}$ with a pure 
$\ket{\psi^-}$ introduces a negative sign to $\rho''_{23}$ and $\rho''_{32}$, which is 
subsequently eliminated by the absolute value in 
Eq.~(\ref{eq:x_state_concurrence}). Both identical mixtures within the 
$\ket{\psi}$ subspace therefore share this concurrence relation, shown in 
Figure~\ref{fig:ad_id_psi_surf}.

\begin{figure}[htbp]
    \centering
    \includegraphics[width=0.7\textwidth]{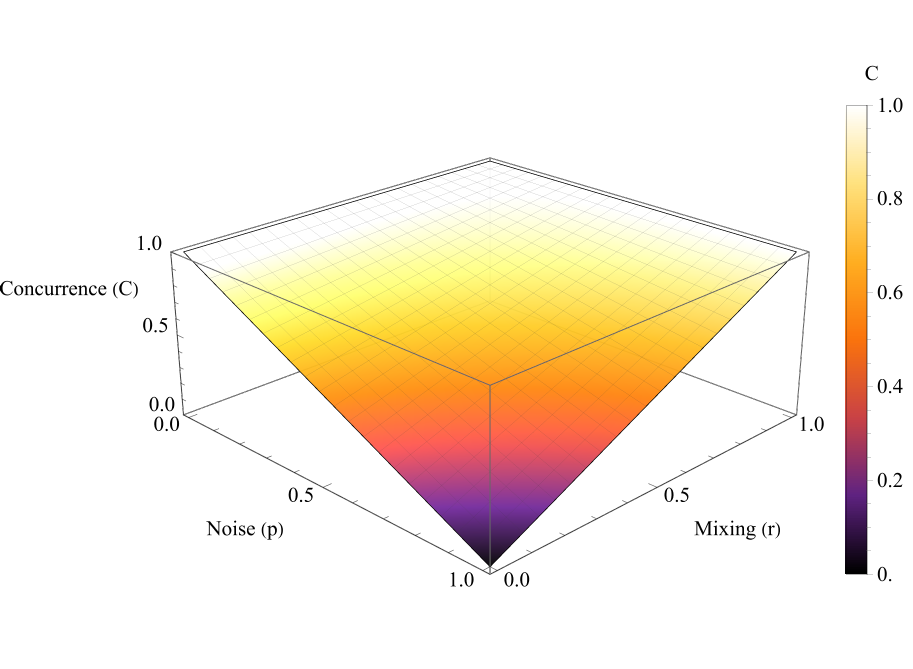}
    \caption{Concurrence $C$ as a function of mixing parameter $r$ for a noisy 
    $\ket{\psi^+}$ parametrized with $p$ mixed with a pure $\ket{\psi^+}$.}
    \label{fig:ad_id_psi_surf}
\end{figure}

\FloatBarrier

\textbf{In the non-identical Bell state cases:}

\item \textbf{For a noisy $\ket{\phi^-}$ mixed with a pure $\ket{\phi^+}$:}

We now analyze the mixture of a noisy $\ket{\phi^-}$ state with a pure 
$\ket{\phi^+}$ reference state.

Applying the AD Kraus operators to $\ket{\phi^-} = \frac{1}{\sqrt{2}}(\ket{00} 
- \ket{11})$, the noisy density matrix $\rho'$ is:
\[
\rho' = \begin{bmatrix}
    \frac{1+p^2}{2} & 0 & 0 & -\frac{1-p}{2} \\
    0 & \frac{p(1-p)}{2} & 0 & 0 \\
    0 & 0 & \frac{p(1-p)}{2} & 0 \\
    -\frac{1-p}{2} & 0 & 0 & \frac{(1-p)^2}{2}
\end{bmatrix}.
\]
We mix this noisy ensemble with the pure $\ket{\phi^+}$ state with probability 
$r$:
\[
\rho'' = (1-r)\rho' + r\rho_{\ket{\phi^+}}.
\]
Expanding the mixture:
\begin{equation}
\rho'' = \begin{bmatrix}
    \frac{(1-r)(1+p^2)+r}{2} & 0 & 0 & \frac{-(1-r)(1-p)+r}{2} \\
    0 & \frac{(1-r)p(1-p)}{2} & 0 & 0 \\
    0 & 0 & \frac{(1-r)p(1-p)}{2} & 0 \\
    \frac{-(1-r)(1-p)+r}{2} & 0 & 0 & \frac{(1-r)(1-p)^2+r}{2}
\end{bmatrix}.
\label{eq:ad_phimix_rho}
\end{equation}
The concurrence is given as:
\begin{equation}
    C = \max\!\left[0,\,|1-2r-x| - p(1-p)(1-r)\right].
    \label{eq:ad_phimix_C}
\end{equation}
The corresponding concurrence surface is shown in 
Figure~\ref{fig:ad_phiminus_phiplus}.

\begin{figure}[htbp]
    \centering
    \includegraphics[width=0.7\textwidth]{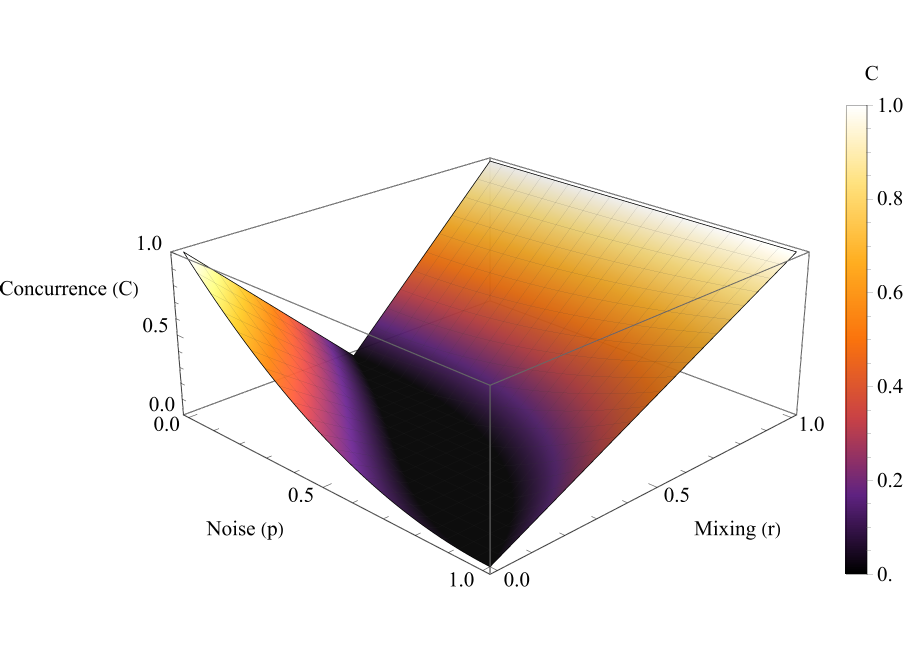}
    \caption{Concurrence $C$ as a function of mixing parameter $r$ for a noisy 
    $\ket{\phi^-}$ parametrized with $p$ mixed with a pure $\ket{\phi^+}$.}
    \label{fig:ad_phiminus_phiplus}
\end{figure}

\item \textbf{For a noisy $\ket{\psi^+}$ mixed with a pure $\ket{\phi^+}$:}

We now evaluate the mixture of a noisy $\ket{\psi^+}$ state with a pure 
$\ket{\phi^+}$ reference state (states from different bases of the Bell basis).

Applying the AD Kraus operators to $\ket{\psi^+}$, the noisy density matrix 
$\rho'$ decays into the $\ket{00}$ ground state as derived previously:
\[
\rho' = \begin{bmatrix}
    p & 0 & 0 & 0 \\
    0 & \frac{1-p}{2} & \frac{1-p}{2} & 0 \\
    0 & \frac{1-p}{2} & \frac{1-p}{2} & 0 \\
    0 & 0 & 0 & 0
\end{bmatrix}.
\]
Mixing with the pure $\ket{\phi^+}$ state with probability $r$:
\[
\rho'' = (1-r)\rho' + r\rho_{\ket{\phi^+}}.
\]
The reconstructed density matrix populates both the inner and outer diagonal 
blocks:
\begin{equation}
\rho'' = \begin{bmatrix}
    (1-r)p + \frac{r}{2} & 0 & 0 & \frac{r}{2} \\
    0 & \frac{(1-r)(1-p)}{2} & \frac{(1-r)(1-p)}{2} & 0 \\
    0 & \frac{(1-r)(1-p)}{2} & \frac{(1-r)(1-p)}{2} & 0 \\
    \frac{r}{2} & 0 & 0 & \frac{r}{2}
\end{bmatrix}.
\label{eq:ad_cross_rho}
\end{equation}
The concurrence is:
\begin{equation}
    C = \max\!\left[0,\,r-(1-r)(1-p),\,
    (1-r)(1-p)-\sqrt{r(r+2x)}\right].
    \label{eq:ad_cross_C}
\end{equation}
The corresponding concurrence surface is shown in 
Figure~\ref{fig:ad_psiplus_phiplus}.

\begin{figure}[htbp]
    \centering
    \includegraphics[width=0.7\textwidth]{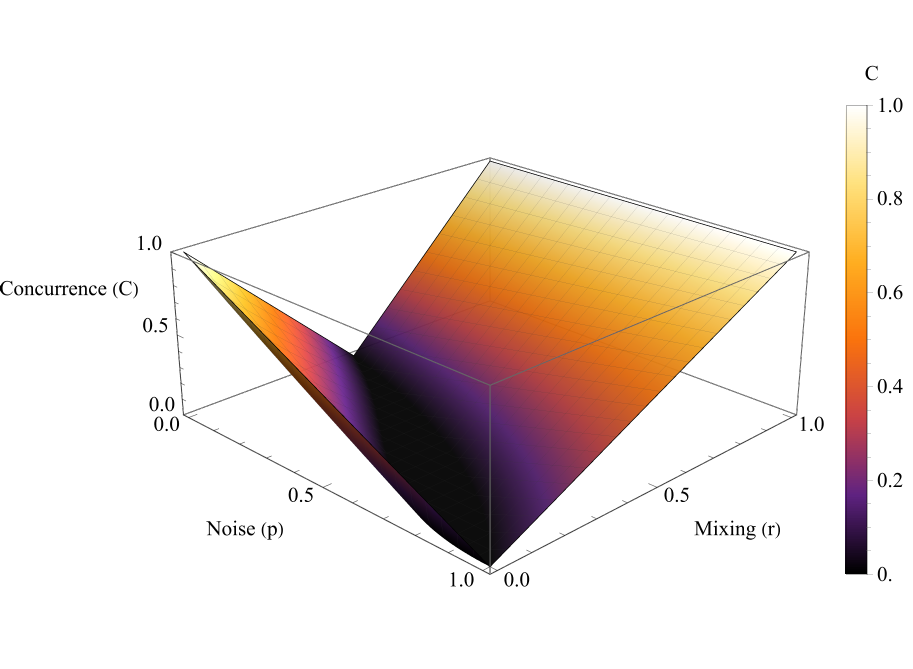}
    \caption{Concurrence $C$ as a function of mixing parameter $r$ for a noisy 
    $\ket{\psi^+}$ parametrized with $p$ mixed with a pure $\ket{\phi^+}$.}
    \label{fig:ad_psiplus_phiplus}
\end{figure}

\FloatBarrier

\item \textbf{For a noisy $\ket{\psi^-}$ mixed with a pure $\ket{\phi^+}$:}

To complete the analysis of mixtures between different bases, we consider a noisy 
$\ket{\psi^-}$ ensemble mixed with a pure $\ket{\phi^+}$ reference state. Under 
the AD channel, the population decay dynamics for $\ket{\psi^-}$ are exactly 
identical to those of $\ket{\psi^+}$, as both states symmetrically decay their 
excitations into the $\ket{00}$ ground state.

The reconstructed density matrix for this mixture differs from the previous 
$\ket{\psi^+}$ case only by a negative sign on the inner off-diagonal coherences:
\[
\rho''_{23} = \rho''_{32} = -\frac{(1-r)(1-p)}{2}.
\]
Because the X-state concurrence formula relies on the absolute magnitude 
$|\rho''_{23}|$, this negative sign is completely nullified during the calculation, 
and the concurrence is identical to the $\ket{\psi^+}$ case:
\begin{equation}
    C = \max\!\left[0,\,r-(1-r)(1-p),\,
    (1-r)(1-p)-\sqrt{r(r+2x)}\right].
    \label{eq:ad_cross_C2}
\end{equation}

\vspace{0.2cm}
\noindent \textbf{Note on the remaining mixtures between different bases:}

For the remaining mixtures of this type (specifically a noisy $\ket{\psi^\pm}$ 
ensemble with a pure $\ket{\phi^-}$ reference state), the results are 
analytically equivalent to the cases derived above. The phase difference of the 
pure reference state (from $\ket{\phi^+}$ to $\ket{\phi^-}$) alters only the 
sign of the outer off-diagonal components $\rho''_{14}$ and $\rho''_{41}$. 
Because the concurrence formula relies on $|\rho''_{14}|$, this sign is 
eliminated during the calculation. Consequently, all four mixtures of a noisy 
$\ket{\psi^\pm}$ with a pure $\ket{\phi^\pm}$ collapse into the same relation 
as Eq.~(\ref{eq:ad_cross_C2})). Duplicate figures and matrix derivations for 
these symmetric cases are therefore omitted.

\item \textbf{For a noisy $\ket{\psi^-}$ mixed with a pure $\ket{\psi^+}$:}

We now consider the non-identical mixture within the $\ket{\psi}$ basis: a noisy 
$\ket{\psi^-}$ ensemble mixed with a pure $\ket{\psi^+}$ reference state.

Applying the AD Kraus operators to $\ket{\psi^-} = \frac{1}{\sqrt{2}}(\ket{01} 
- \ket{10})$, the state decays into the $\ket{00}$ ground state just as 
$\ket{\psi^+}$ did, but the initial negative phase is preserved in the decaying 
off-diagonal coherences. The noisy density matrix $\rho'$ is:
\[
\rho' = \begin{bmatrix}
    p & 0 & 0 & 0 \\
    0 & \frac{1-p}{2} & -\frac{1-p}{2} & 0 \\
    0 & -\frac{1-p}{2} & \frac{1-p}{2} & 0 \\
    0 & 0 & 0 & 0
\end{bmatrix}.
\]
Mixing with the pure $\ket{\psi^+}$ state with probability $r$:
\[
\rho'' = (1-r)\rho' + r\rho_{\ket{\psi^+}}.
\]
\begin{equation}
\rho'' = \begin{bmatrix}
    (1-r)p & 0 & 0 & 0 \\
    0 & \frac{(1-r)(1-p)+r}{2} & \frac{-(1-r)(1-p)+r}{2} & 0 \\
    0 & \frac{-(1-r)(1-p)+r}{2} & \frac{(1-r)(1-p)+r}{2} & 0 \\
    0 & 0 & 0 & 0
\end{bmatrix}.
\label{eq:ad_nonid_psi_rho}
\end{equation}
The concurrence simplifies to:
\begin{equation}
    C = \max\!\left[0,\,|r-(1-r)(1-p)|\right].
    \label{eq:ad_nonid_psi_C}
\end{equation}
The corresponding concurrence surface is shown in 
Figure~\ref{fig:ad_nonid_psi}.

\begin{figure}[htbp]
    \centering
    \includegraphics[width=0.7\textwidth]{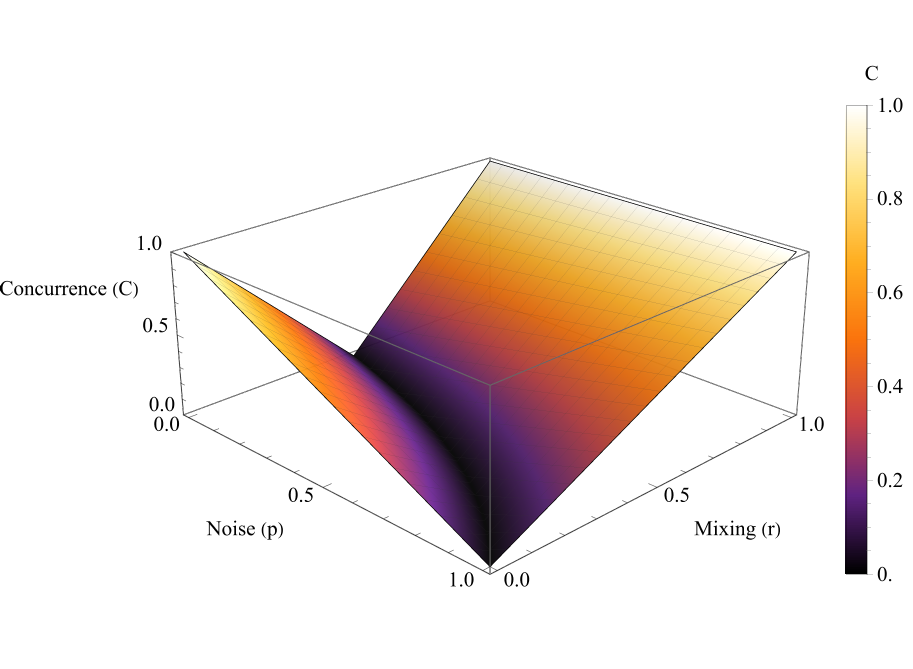}
    \caption{Concurrence $C$ as a function of mixing parameter $r$ for a noisy 
    $\ket{\psi^-}$ parametrized with $p$ mixed with a pure $\ket{\psi^+}$.}
    \label{fig:ad_nonid_psi}
\end{figure}

\end{itemize}
The five graphs Figs.~\ref{fig:ad_id_surf}--\ref{fig:ad_nonid_psi} illustrate all possible combinations. Fig~\ref{fig:ad_id_surf} (\ref{fig:ad_id_psi_surf}) for identical $\phi(\psi)$ states, showing nonlinear and linear dependence on the noise parameter. Fig~\ref{fig:ad_phiminus_phiplus}, \ref{fig:ad_psiplus_phiplus}, \ref{fig:ad_nonid_psi} illustrate the mixing of different $(\phi,\phi), (\phi,\psi)$, and $(\psi,\psi)$ states.
\FloatBarrier

\section{Entanglement versus Non-Locality Thresholds Under Noise}
\label{sec:thresholds}
While the concurrence $C$ provides a robust mathematical quantification of the 
degree of entanglement, not all entangled states are capable of demonstrating 
true quantum non-locality. In the presence of noise, a quantum state can enter 
a regime of ``Bell-Local Entanglement'' (or hidden non-locality), wherein the 
state is inseparable ($C > 0$) yet completely fails to violate the CHSH inequality via projective measurements. To characterize the practical utility 
of our mixed noisy states, we now map the boundary between mathematical 
entanglement and quantum non-locality using the Horodecki criterion and the 
diagonal $T$-matrix elements of Eqs.~(\ref{eq:t1})--(\ref{eq:t3}), together 
with the effective noise parameter $x = p(1-r)$ of Eq.~(\ref{eq:effective_noise}). 
A state exhibits true non-locality if and only if $M(T) > 1$; if $C > 0$ but 
$M(T) \leq 1$, the state is Bell-local.

\subsection{Depolarizing Channel Thresholds}
\label{sec:dp_thresh}

\subsubsection{Identical Bell States:} 
For the mixture of identical states under 
Depolarizing (DP) noise, the concurrence simplifies to 
$C = \max[0,\, 1 - \tfrac{3}{2}x]$. Evaluating the Horodecki elements for 
this reconstructed density matrix yields $t_1 = 1-x$, $t_2 = -(1-x)$, and 
$t_3 = 1-x$. Therefore, $M(T) = 2(1-x)^2$. Setting the thresholds for 
entanglement ($C = 0$) and non-locality ($M(T) = 1$), we find that the state 
is entangled for $x < 2/3$, but only exhibits non-locality for 
$x < 1 - 1/\sqrt{2}$.

\subsubsection{Non-Identical Bell States:}
Here whether mixing two $\ket{\phi}$-basis states 
with opposite phases (e.g.\ $\ket{\phi^+}$ with $\ket{\phi^-}$), two 
$\ket{\psi}$-basis states with opposite phases (e.g.\ $\ket{\psi^+}$ with 
$\ket{\psi^-}$), or states from different bases (e.g.\ $\ket{\phi^+}$ with 
$\ket{\psi^\pm}$), the reconstructed density matrices yield an identical set 
of squared $T$-matrix elements: $(1-x)^2$, $(1-2r-x)^2$, and $(1-2r-x)^2$. 
Thus, all 12 non-identical permutations collapse into a single universal 
non-locality relation:
\begin{equation}
    M(T) = (1-x)^2 + (1-2r-x)^2.
    \label{eq:dp_nonid_M}
\end{equation}
This boundary competes with the universal non-identical concurrence relation 
$C = \max[0,\, |1-2r-x| - \tfrac{1}{2}x]$.

\begin{figure}[htbp]
    \centering
    \begin{subfigure}[b]{0.48\textwidth}
        \centering
        \includegraphics[width=\linewidth]{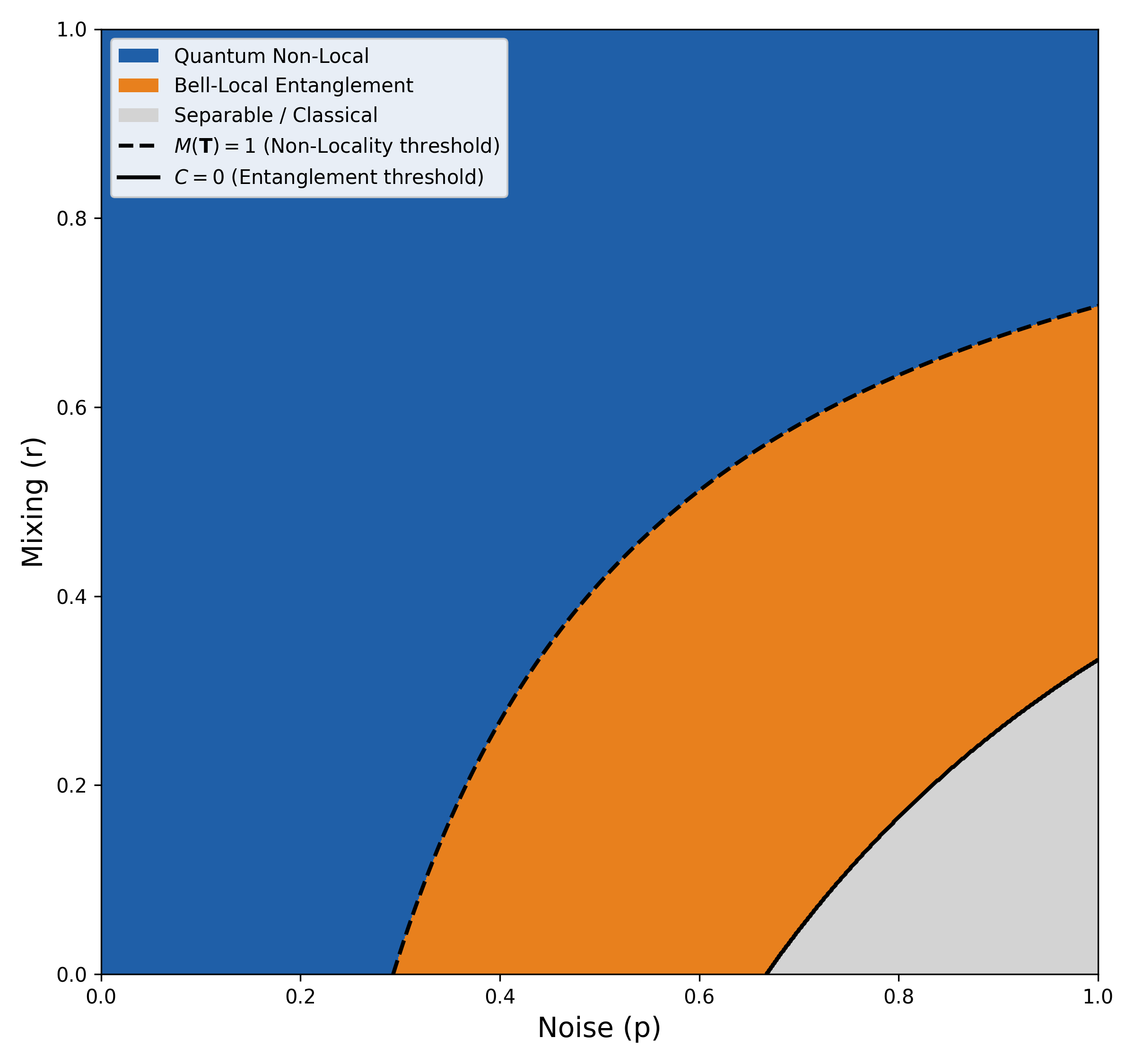}
        \caption{Identical States}
        \label{fig:dp_phase_id}
    \end{subfigure}\hfill
    \begin{subfigure}[b]{0.48\textwidth}
        \centering
        \includegraphics[width=\linewidth]{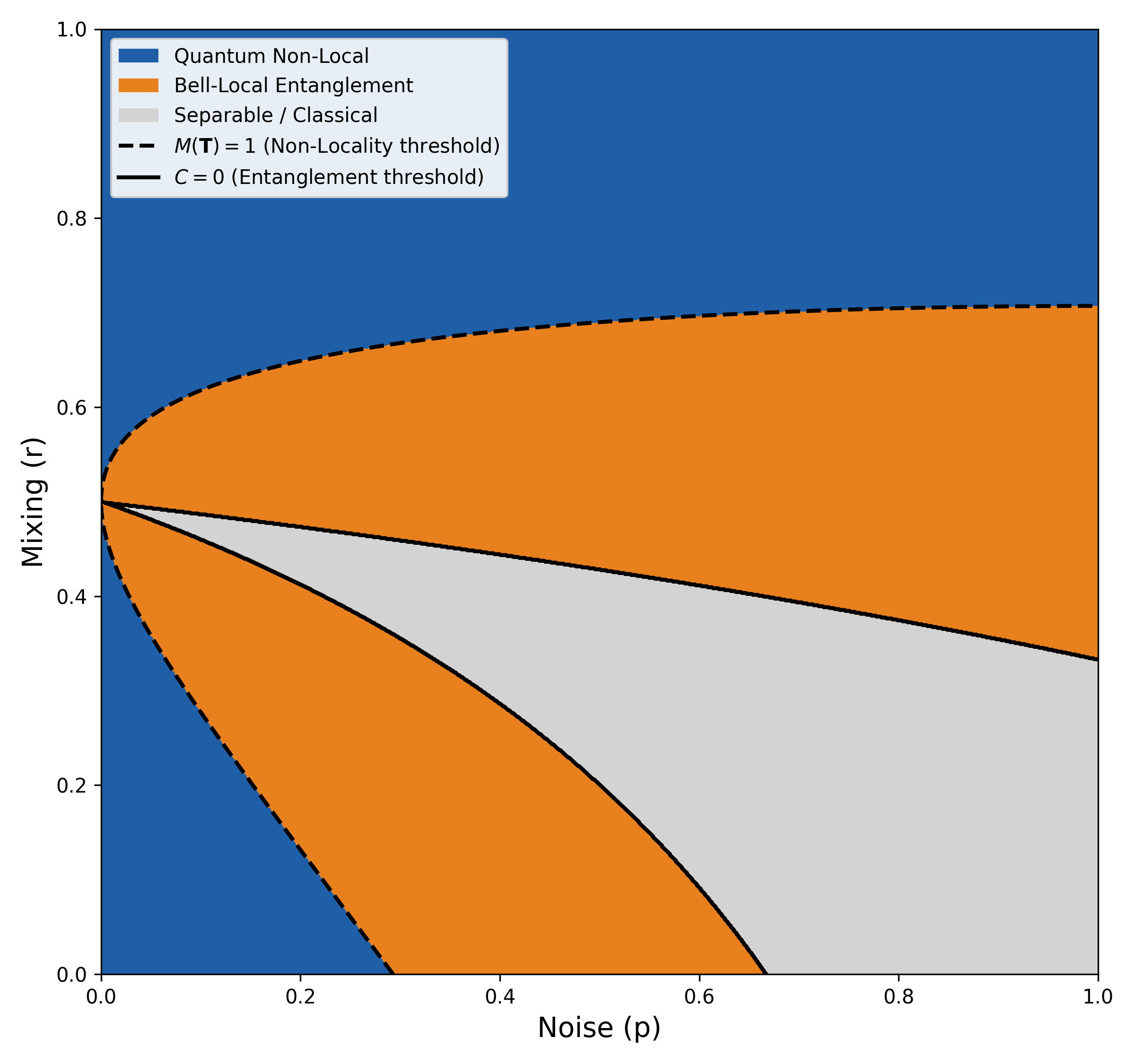}
        \caption{Non-Identical States}
        \label{fig:dp_phase_nonid}
    \end{subfigure}
    \caption{Phase diagrams mapping the entanglement and non-locality boundaries 
    under the Depolarizing channel. The orange region represents Bell-local 
    entanglement, where the state is mathematically inseparable ($C > 0$) but 
    fails to violate the CHSH inequality ($M(T) \leq 1$).}
    \label{fig:DP_Phase}
\end{figure}

To visually capture these dynamics, we map the corresponding phase diagrams 
in Figure~\ref{fig:DP_Phase}.

\FloatBarrier

\textbf{Physical Interpretation of Depolarizing Phase Diagrams:} In Panel~(a), the massive 
Bell-local region (orange) demonstrates that while injecting a pure state 
(increasing $r$) can successfully increase the degree of entanglement, the recovered entanglement is mostly classical-like 
in the sense that it fails to violate the CHSH inequality. For instance, at 
maximum noise ($p = 1$), one must inject a $>70\%$ pure state 
composition ($r > 1/\sqrt{2}$) to recover actionable non-locality.

Panel~(b) illustrates the destructive nature of mixing non-identical Bell states. 
At $p = 0$, an equal mixture ($r = 0.5$) of any two non-identical Bell states 
perfectly cancels all coherence, falling the state directly into the separable 
regime. 
\subsection{Phase Damping Channel Thresholds}
\label{sec:pd_thresh}

Unlike the Depolarizing channel, Phase Damping (PD) strictly affects the 
off-diagonal coherences of the density matrix while leaving the populations 
(diagonal elements) completely invariant. This unique physical property 
fundamentally alters the Horodecki $T$-matrix and the survival of quantum 
non-locality. Using the same effective noise parameter $x = p(1-r)$, we map 
the thresholds for the three distinct mixing cases.

\subsubsection{ Identical Bell States}
For the mixture of identical states 
(e.g., $\ket{\phi^+}$ mixed with pure $\ket{\phi^+}$) under PD noise, the 
concurrence simplifies to $C = \max[0,\, 1-x]$. Because the diagonal 
populations are untouched by the noise, evaluating the $T$-matrix yields 
$t_1 = 1-x$, $t_2 = -(1-x)$, and $t_3 = 1$. The sum of the two largest 
squared elements is always $M(T) = 1 + (1-x)^2$. Setting the thresholds 
for entanglement ($C > 0$) and non-locality ($M(T) > 1$), we find that they are 
mathematically identical: both survive for $x < 1$. Under this 
specific mixture, the Bell-local regime vanishes entirely; any existing 
entanglement guarantees a violation of the CHSH inequality.

\subsubsection{Non-Identical States ($\ket{\phi^+}$ with $\ket{\phi^-}$, same 
basis with opposite phases):}
When mixing two $\ket{\phi}$-basis states with 
opposite phases, the concurrence is governed by: 
$C = |1-2r-x|$. The corresponding $T$-matrix elements are 
$t_1 = 1-2r-x$, $t_2 = -(1-2r-x)$, and $t_3 = 1$. This yields a universal 
non-locality relation of $M(T) = 1 + (1-2r-x)^2$. Since $C = |1-2r-x|$, 
these two expressions satisfy $M(T) = 1 + C^2$ exactly, meaning $M(T) > 1$ 
whenever $C > 0$ and $M(T) = 1$ precisely when $C = 0$. The entanglement 
boundary and the CHSH non-locality boundary therefore coincide exactly, and 
the Bell-local entanglement regime (orange region) is entirely absent: any 
state in this mixture that is entangled is guaranteed to violate the CHSH 
inequality. The separable boundary itself (the curve $|1-2r-x| = 0$, 
equivalently $r = (1-p)/(2-p)$) is present in the phase diagram as the 
boundary between the blue and grey regions.

\subsubsection{ Non-Identical States ($\ket{\phi^+}$ with $\ket{\psi^\pm}$, 
different bases):} 
\begin{figure}[htbp]
    \centering
    \begin{subfigure}[b]{0.32\textwidth}
        \centering
        \includegraphics[width=\textwidth]{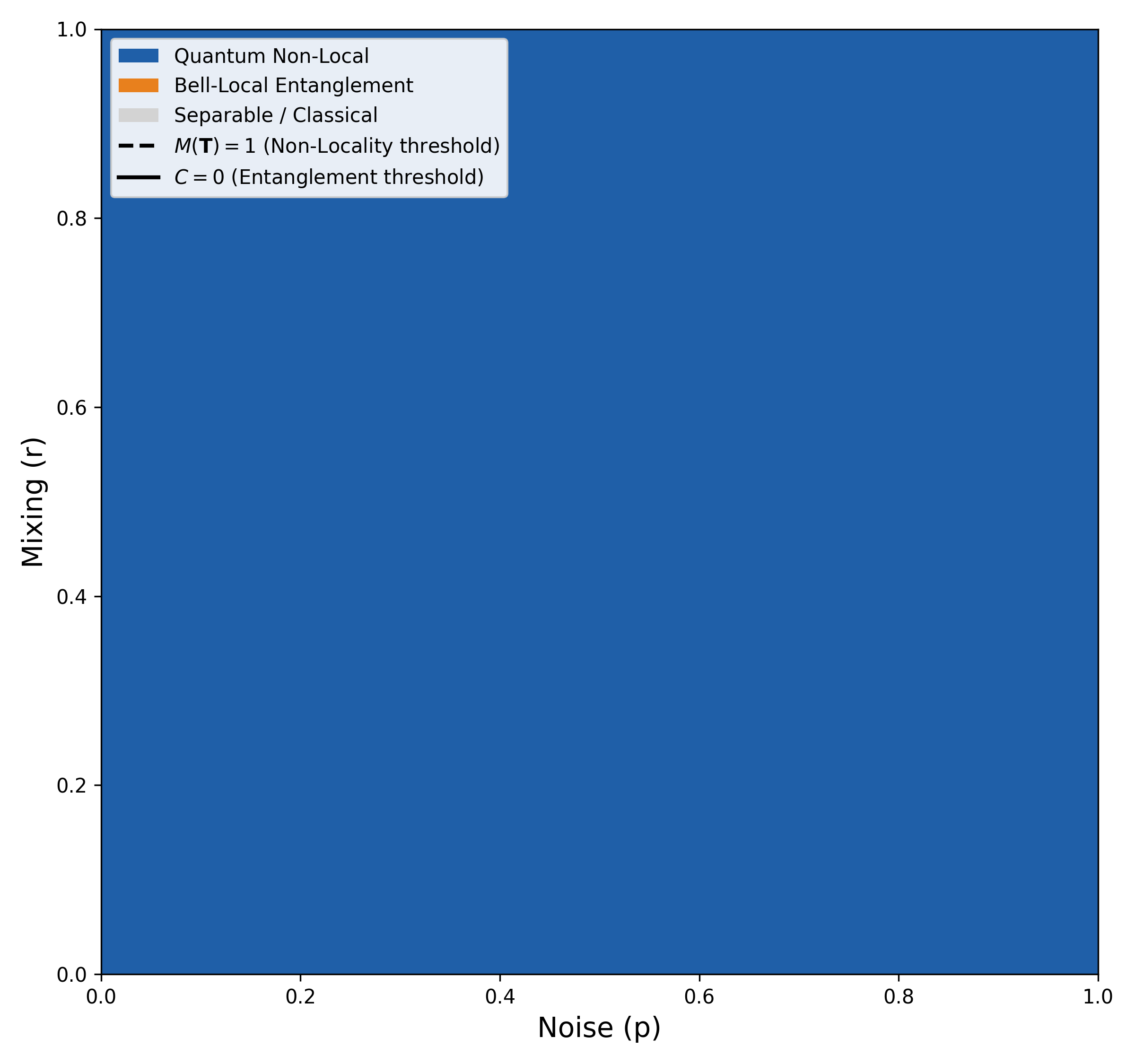}
        \caption{Identical states\\($\ket{\phi^+}$ with $\ket{\phi^+}$)}
        \label{fig:pd_identical}
    \end{subfigure}
    \hfill
    \begin{subfigure}[b]{0.32\textwidth}
        \centering
        \includegraphics[width=\textwidth]{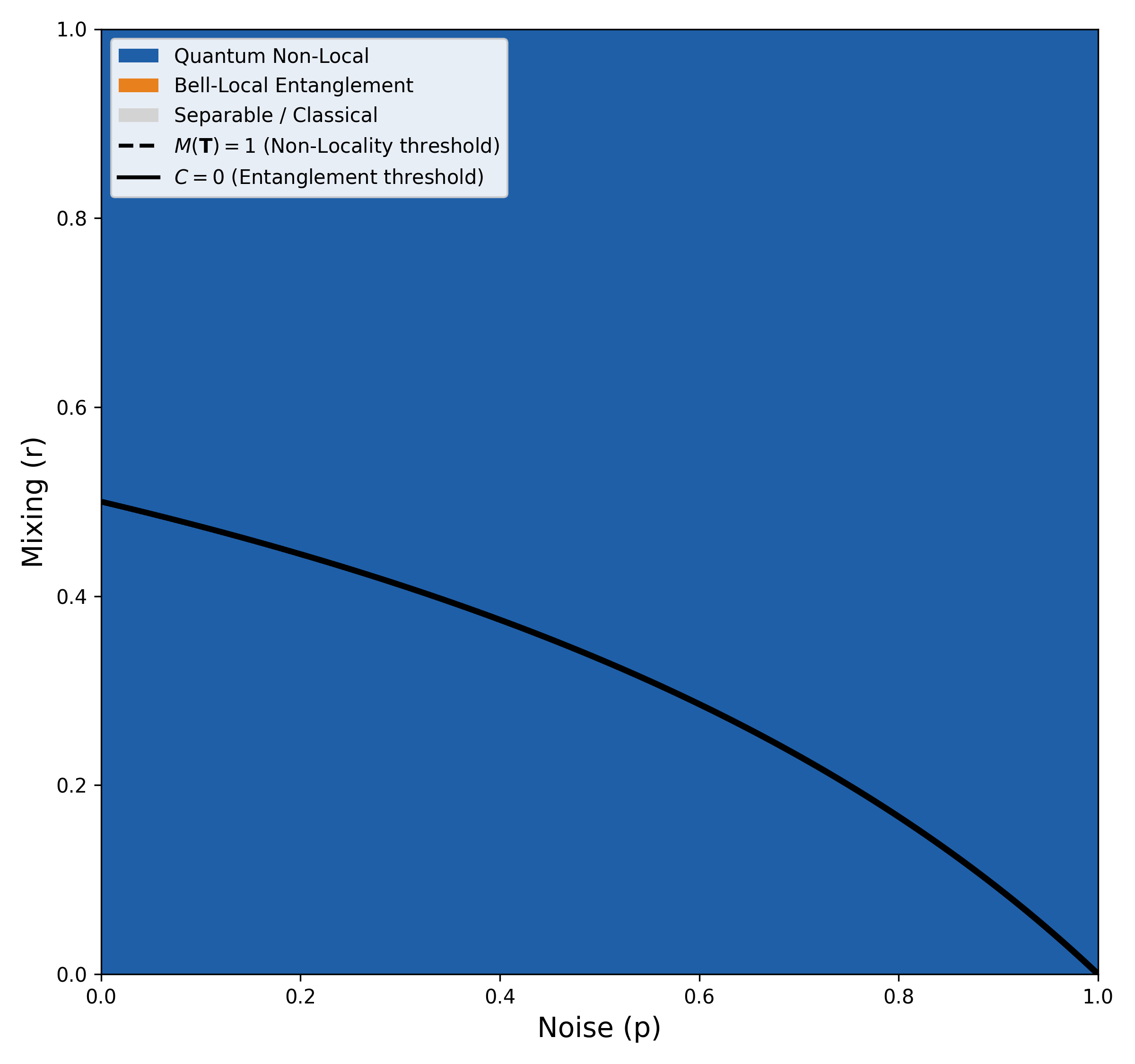}
        \caption{Same basis, opposite phases\\($\ket{\phi^+}$ with $\ket{\phi^-}$)}
        \label{fig:pd_same}
    \end{subfigure}
    \hfill
    \begin{subfigure}[b]{0.32\textwidth}
        \centering
        \includegraphics[width=\textwidth]{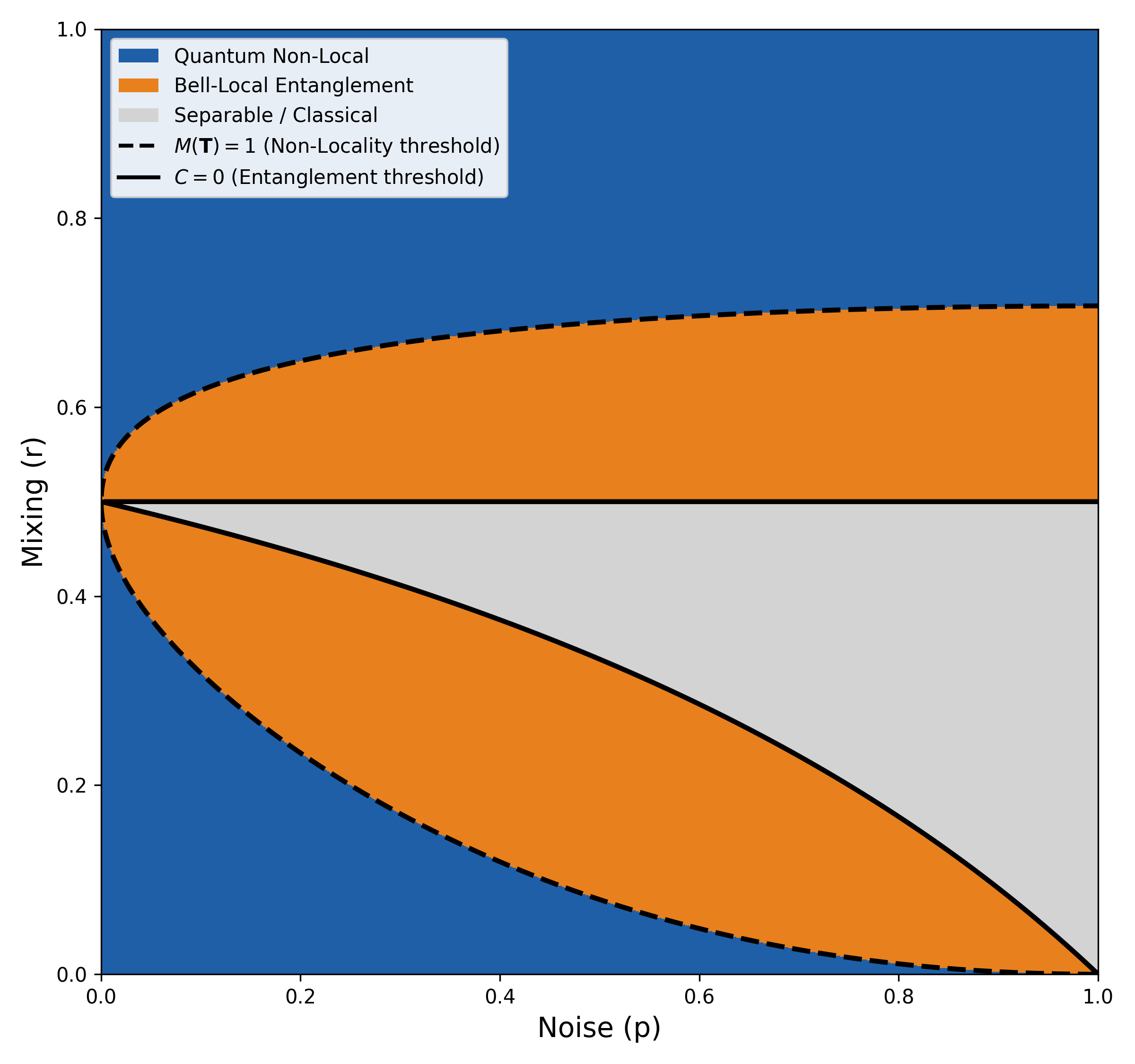}
        \caption{Different bases\\($\ket{\phi^+}$ with $\ket{\psi^\pm}$)}
        \label{fig:pd_cross}
    \end{subfigure}
    \caption{Phase diagrams for Bell state mixtures under the Phase Damping 
    (PD) channel. The orange region represents Bell-local entanglement 
    ($C > 0$ but $M(T) \leq 1$), which is absent for cases (a) and (b) 
    but re-emerges in case (c).}
    \label{fig:pd_all_cases}
\end{figure}

When mixing states from different bases of the Bell basis, 
we have the concurrence:
$C = \max[0,\, 1-2r-x,\, 2r-1]$ and the $T$ matrix is given as:
\begin{align}
    t_1 &= 1 - x, \\
    t_2 &= -(1 - 2r - x), \\
    t_3 &= 1 - 2r.
\end{align}
For these mixtures, $M(T)$ is determined by the sum of the two largest squares 
among $(1-x)^2$, $(1-2r-x)^2$, and $(1-2r)^2$. The phase diagrams for 
all three PD mixing cases are collected in Figure~\ref{fig:pd_all_cases}.

\FloatBarrier
\subsection{Amplitude Damping Channel Thresholds}
\label{sec:ad_thresh}

Unlike the Depolarizing (DP) and Phase Damping (PD) channels, the Amplitude 
Damping (AD) channel models the asymmetric, directional loss of energy from a 
quantum system to its environment through spontaneous emission. 
Physically, this means the excited state $\ket{1}$ decays to the ground state 
$\ket{0}$ with probability $p$, while the ground state gets populated with probability $p$.

Under AD noise, the mathematical symmetry between the 
$\ket{\phi}$-basis states and the 
$\ket{\psi}$-basis states is broken. The AD channel therefore requires us to analyze each 
mixing case separately, as identical mixtures and mixtures between different 
bases behave radically differently depending on which specific Bell states are 
used.

\subsubsection{Identical $\ket{\phi}$-Basis States ($\ket{\phi^+}$ with $\ket{\phi^+}$, and $\ket{\phi^-}$ with $\ket{\phi^-}$)} To illustrate the unique mechanics 
of the AD channel, we first analyze the baseline case of mixing a noisy 
$\ket{\phi^+}$ state (subject to AD noise $p$) with a pure $\ket{\phi^+}$ 
state (with probability $r$). Since the $\ket{\phi^-}$ states yield the same 
density matrix under AD noise, this result accounts for both identical 
$\ket{\phi}$ mixtures. As derived in Section~\ref{sec:ad_mixing}, the 
reconstructed density matrix is:
\begin{equation}
\rho'' =
\begin{bmatrix}
\frac{(1-r)(1+p^2)+r}{2} & 0 & 0 & \frac{1-p(1-r)}{2} \\
0 & \frac{(1-r)p(1-p)}{2} & 0 & 0 \\
0 & 0 & \frac{(1-r)p(1-p)}{2} & 0 \\
\frac{1-p(1-r)}{2} & 0 & 0 & \frac{(1-r)(1-p)^2+r}{2}
\end{bmatrix}.
\end{equation}
The concurrence is:
\begin{equation}
    C = \max[0,\,(1-p)^2(1-r)+r].
\end{equation}
Extracting the $T$-matrix elements:
\begin{align}
    t_1 &= 1 - p(1-r), \\
    t_2 &= -[1 - p(1-r)], \\
    t_3 &= (1-r)(1-2p+2p^2)+r,
\end{align}
\begin{equation}
    M(T) = [1-p(1-r)]^2 + \max\!\left([1-p(1-r)]^2,\,
    [(1-r)(1-2p+2p^2)+r]^2\right).
\end{equation}
This is illustrated in Fig~\ref{fig:ad_all_cases}(\subref{fig:ad_identical_phi}).
\subsubsection{Identical $\ket{\psi}$-Basis States ($\ket{\psi^+}$ with $\ket{\psi^+}$, and $\ket{\psi^-}$ with $\ket{\psi^-}$)} From Eq.~(\ref{eq:ad_psi_mix}), 
the reconstructed density matrix is:
\begin{equation}
\rho'' = \begin{bmatrix}
    p(1-r) & 0 & 0 & 0 \\
    0 & \frac{(1-r)(1-p)+r}{2} & \frac{(1-r)(1-p)+r}{2} & 0 \\
    0 & \frac{(1-r)(1-p)+r}{2} & \frac{(1-r)(1-p)+r}{2} & 0 \\
    0 & 0 & 0 & 0
\end{bmatrix}.
\end{equation}
The concurrence is:
\begin{equation}
    C = 1 - p(1-r).
\end{equation}
Using $x = p(1-r)$, the $T$-matrix elements are:
\begin{align}
    t_1 &= 2(\rho_{14}+\rho_{23}) = 1-x, \\
    t_2 &= 2(\rho_{23}-\rho_{14}) = 1-x, \\
    t_3 &= \rho_{11}+\rho_{44}-\rho_{22}-\rho_{33} = 2x-1.
\end{align}
Because $t_1^2 = t_2^2 = (1-x)^2$:
\begin{equation}
    M(T) = (1-x)^2 + \max\!\left[(1-x)^2,\,(2x-1)^2\right]
    = \begin{cases}
    2(1-x)^2 & x \leq \tfrac{2}{3}, \\
    5x^2 - 6x + 2 & x > \tfrac{2}{3}.
    \end{cases}
\end{equation}
The non-locality threshold $M(T) = 1$ in the regime $x \leq 2/3$ gives:
\begin{equation}
    2(1-x)^2 = 1 \implies x_{\mathrm{crit}} = 1 - \frac{1}{\sqrt{2}} 
    \approx 0.293,
\end{equation}
with the phase boundary $r(p) = 1 - 0.293/p$. At this threshold the 
concurrence is:
\begin{equation}
    C = 1 - x_{\mathrm{crit}} = \frac{1}{\sqrt{2}} \approx 0.707,
\end{equation}
demonstrating that when the state loses its capacity to violate the CHSH 
inequality, it still retains substantial entanglement ($C \approx 0.707$). This is illustrated in Figure~\ref{fig:ad_all_cases}(\subref{fig:ad_identical_psi}).

\subsubsection{Same-Basis, Opposite-Phase States ($\ket{\phi^+}$ mixed with $\ket{\phi^-}$)}The reconstructed density matrix for this mciture is:
\begin{equation}
\rho'' =
\begin{bmatrix}
\frac{(1-r)(1+p^2)+r}{2} & 0 & 0 & \frac{-(1-r)(1-p)+r}{2} \\
0 & \frac{(1-r)p(1-p)}{2} & 0 & 0 \\
0 & 0 & \frac{(1-r)p(1-p)}{2} & 0 \\
\frac{-(1-r)(1-p)+r}{2} & 0 & 0 & \frac{(1-r)(1-p)^2+r}{2}
\end{bmatrix}.
\end{equation}
The concurrence is:
\begin{equation}
    C = \max[0,\,|r-(1-r)(1-p)|-p(1-p)(1-r)].
\end{equation}
Extracting the $T$-matrix elements:
\begin{align}
    t_1 &= r-(1-r)(1-p), \\
    t_2 &= -[r-(1-r)(1-p)], \\
    t_3 &= (1-r)(1-2p+2p^2)+r,
\end{align}
\begin{equation}
    M(T) = [r-(1-r)(1-p)]^2 + \max\!\left([r-(1-r)(1-p)]^2,\,
    [(1-r)(1-2p+2p^2)+r]^2\right).
\end{equation}
This is illustrated in Fig~\ref{fig:ad_all_cases}(\subref{fig:ad_non_identical_phi}).
\subsubsection{Different-Basis States ($\ket{\phi^+}$ mixed with $\ket{\psi^\pm}$)} We now examine mixtures between states from 
different bases of the Bell basis. Because the noisy state is from the 
$\ket{\psi}$ basis while the pure reference state is from the $\ket{\phi}$ 
basis, the density matrix exhibits non-zero coherences on both off-diagonals. 
The results are invariant whether the noisy state is $\ket{\psi^+}$ or 
$\ket{\psi^-}$, as the phase difference ($\pm$) is irrelevant by the absolute 
magnitudes in both the concurrence and Horodecki calculations. The reconstructed 
density matrix is:
\begin{equation}
\rho'' =
\begin{bmatrix}
\frac{2p(1-r)+r}{2} & 0 & 0 & \frac{r}{2} \\
0 & \frac{(1-r)(1-p)}{2} & \pm\frac{(1-r)(1-p)}{2} & 0 \\
0 & \pm\frac{(1-r)(1-p)}{2} & \frac{(1-r)(1-p)}{2} & 0 \\
\frac{r}{2} & 0 & 0 & \frac{r}{2}
\end{bmatrix}.
\end{equation}
We then find the concurrence as:
\begin{equation}
    C = \max\!\left[0,\,r-(1-r)(1-p),\,
    (1-r)(1-p)-\sqrt{r(r+2x)}\right].
\end{equation}
The $T$-matrix elements are:
\begin{align}
    t_1 &= r\pm(1-r)(1-p), \\
    t_2 &= \pm(1-r)(1-p)-r, \\
    t_3 &= (1-r)(2p-1)+r.
\end{align}
Squaring these elements eliminates the phase dependency ($\pm$), giving a 
result valid for both $\ket{\psi^+}$ and $\ket{\psi^-}$:
\begin{equation}
    M(T) = [r+(1-r)(1-p)]^2 + \max\!\left([r-(1-r)(1-p)]^2,\,
    [(1-r)(2p-1)+r]^2\right).
\end{equation}
This is illustrated in Fig~\ref{fig:ad_all_cases}(\subref{fig:ad_different_bases}).
\subsubsection{Same-Basis, Opposite-Phase States ($\ket{\psi^+}$ mixed with $\ket{\psi^-}$} 
\begin{figure}[htbp]
    \centering
    \begin{subfigure}[b]{0.32\textwidth}
        \centering
        \includegraphics[width=\textwidth]{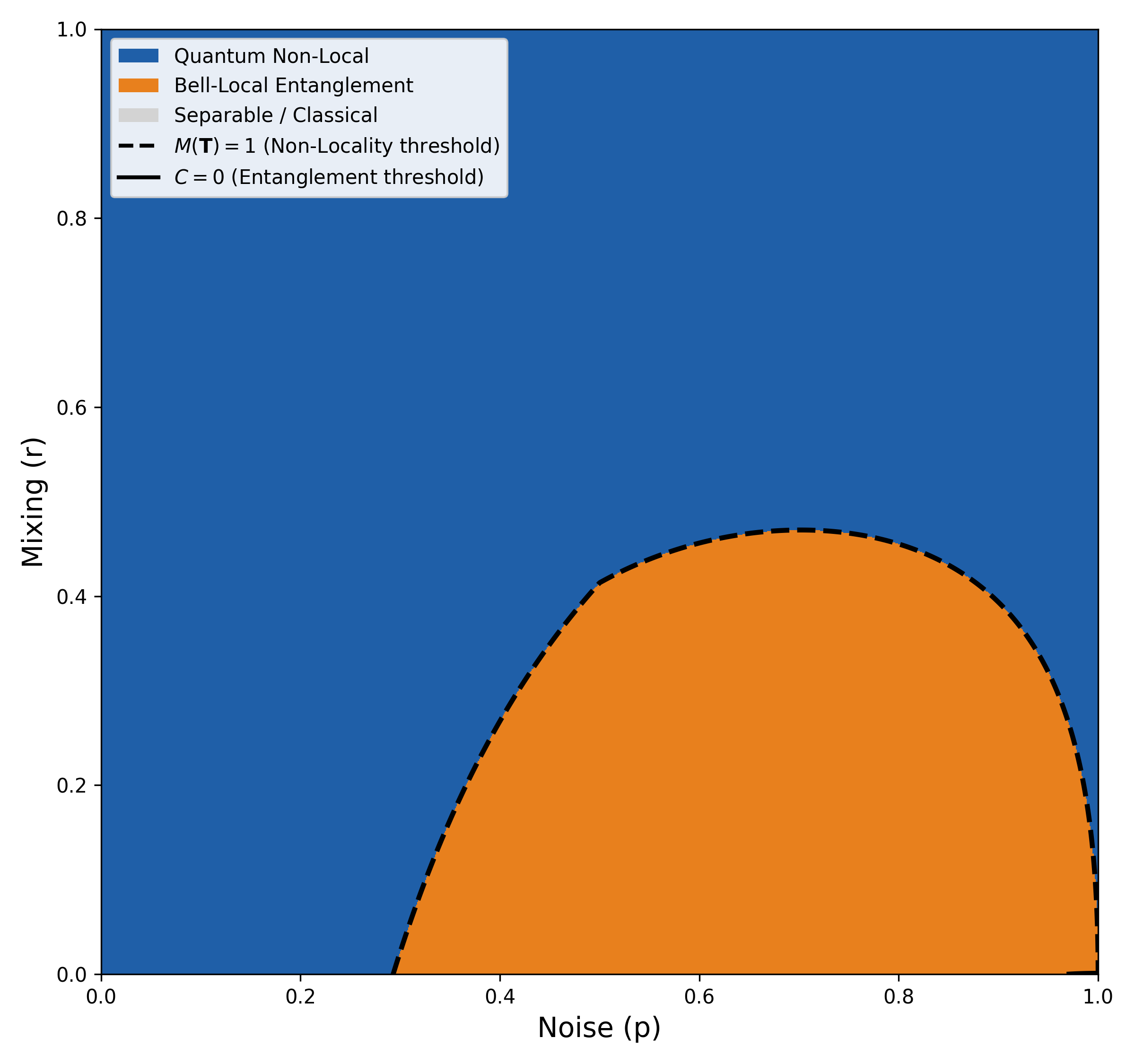}
        \caption{Identical $\ket{\phi}$ states\\($\ket{\phi^+}$ with 
        $\ket{\phi^+}$)}
        \label{fig:ad_identical_phi}
    \end{subfigure}
    \hfill
    \begin{subfigure}[b]{0.32\textwidth}
        \centering
        \includegraphics[width=\textwidth]{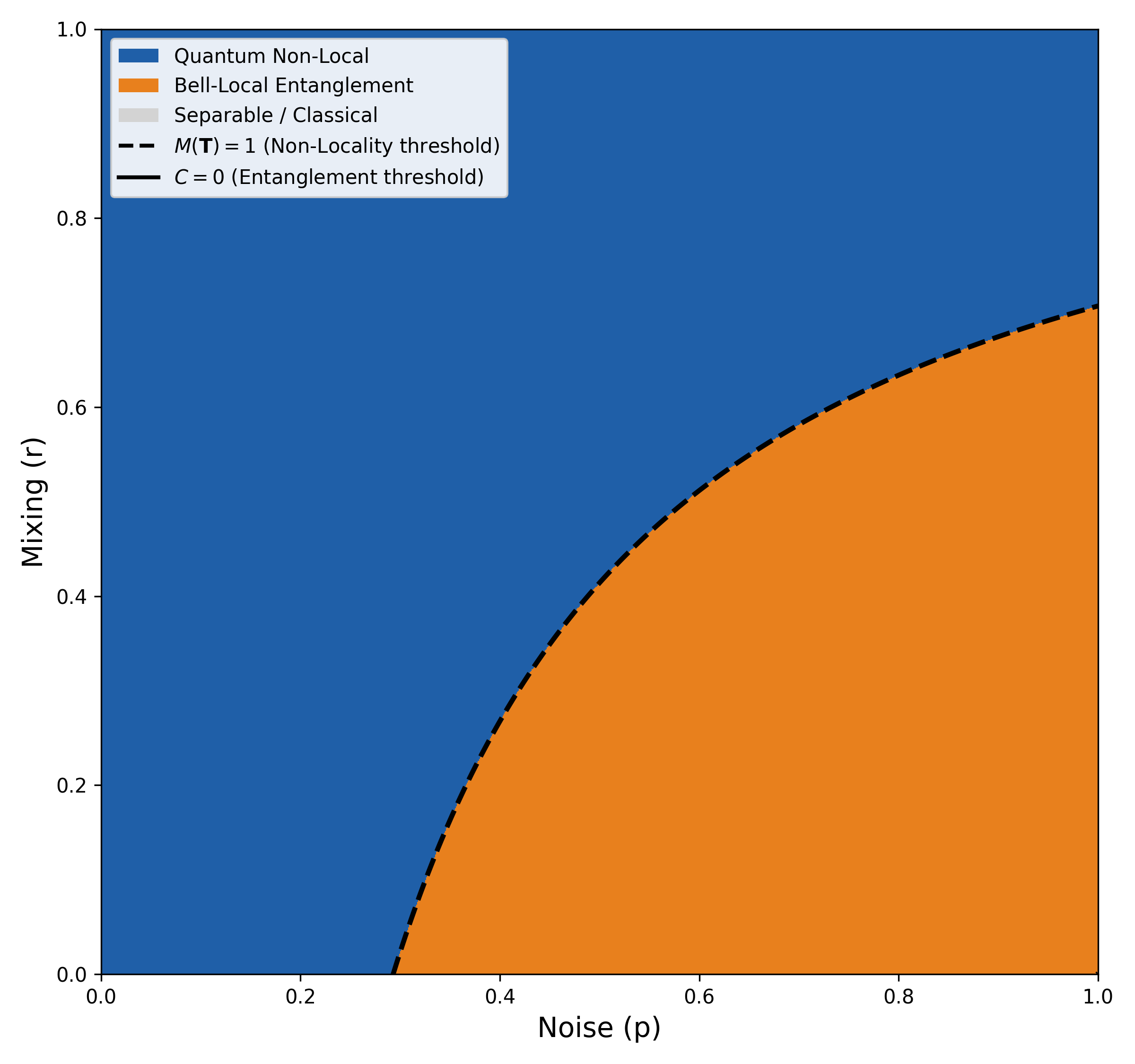}
        \caption{Identical $\ket{\psi}$ states\\($\ket{\psi^+}$ with 
        $\ket{\psi^+}$)}
        \label{fig:ad_identical_psi}
    \end{subfigure}
    \hfill
    \begin{subfigure}[b]{0.32\textwidth}
        \centering
        \includegraphics[width=\textwidth]{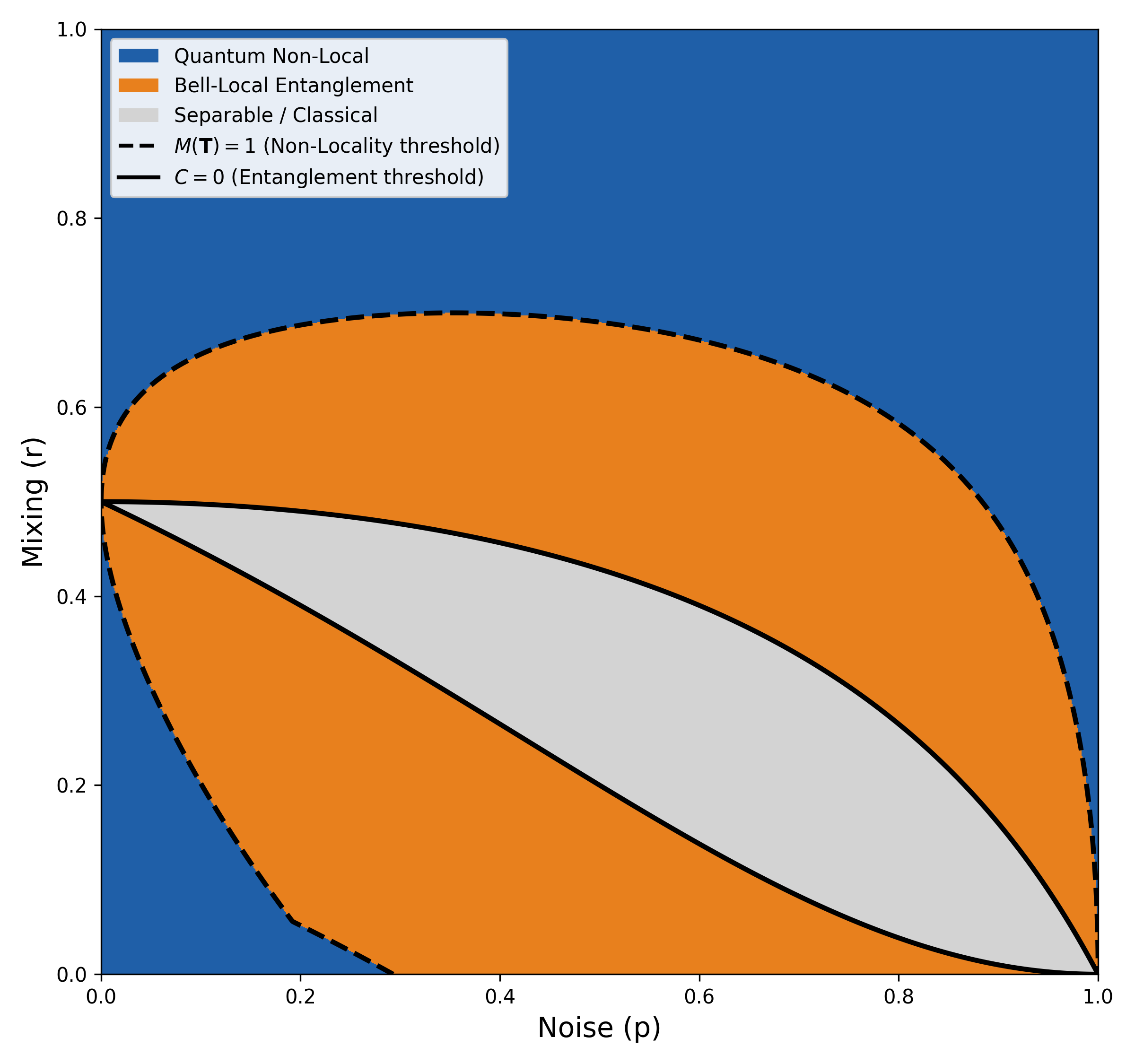}
        \caption{Same basis, opposite phases\\($\ket{\phi^+}$ with 
        $\ket{\phi^-}$)}
        \label{fig:ad_non_identical_phi}
    \end{subfigure}

    \vspace{1.5em}

    \begin{subfigure}[b]{0.32\textwidth}
        \centering
        \includegraphics[width=\textwidth]{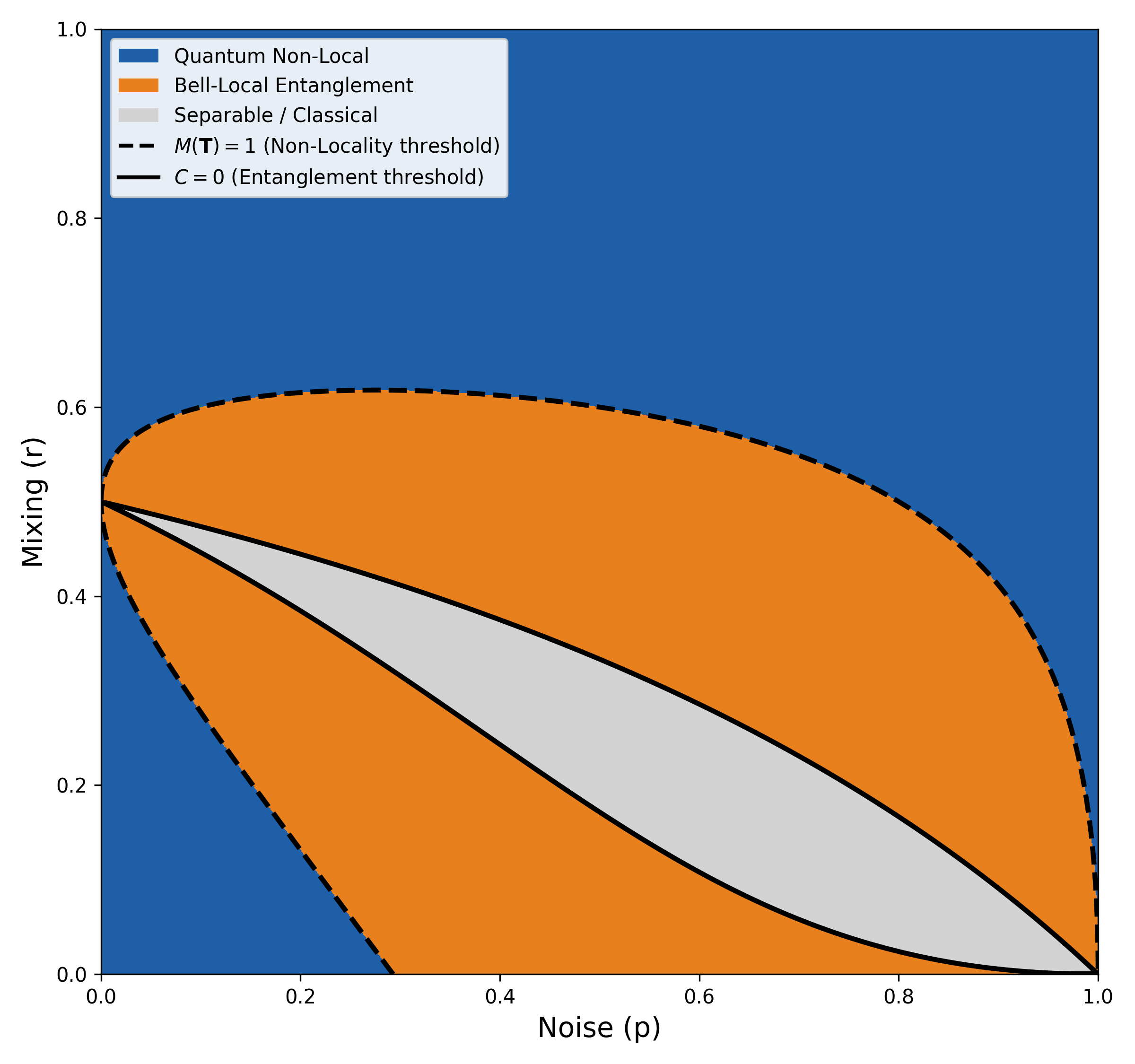}
        \caption{Different bases\\($\ket{\phi^+}$ with $\ket{\psi^\pm}$)}
        \label{fig:ad_different_bases}
    \end{subfigure}
    \hspace{0.05\textwidth}
    \begin{subfigure}[b]{0.32\textwidth}
        \centering
        \includegraphics[width=\textwidth]{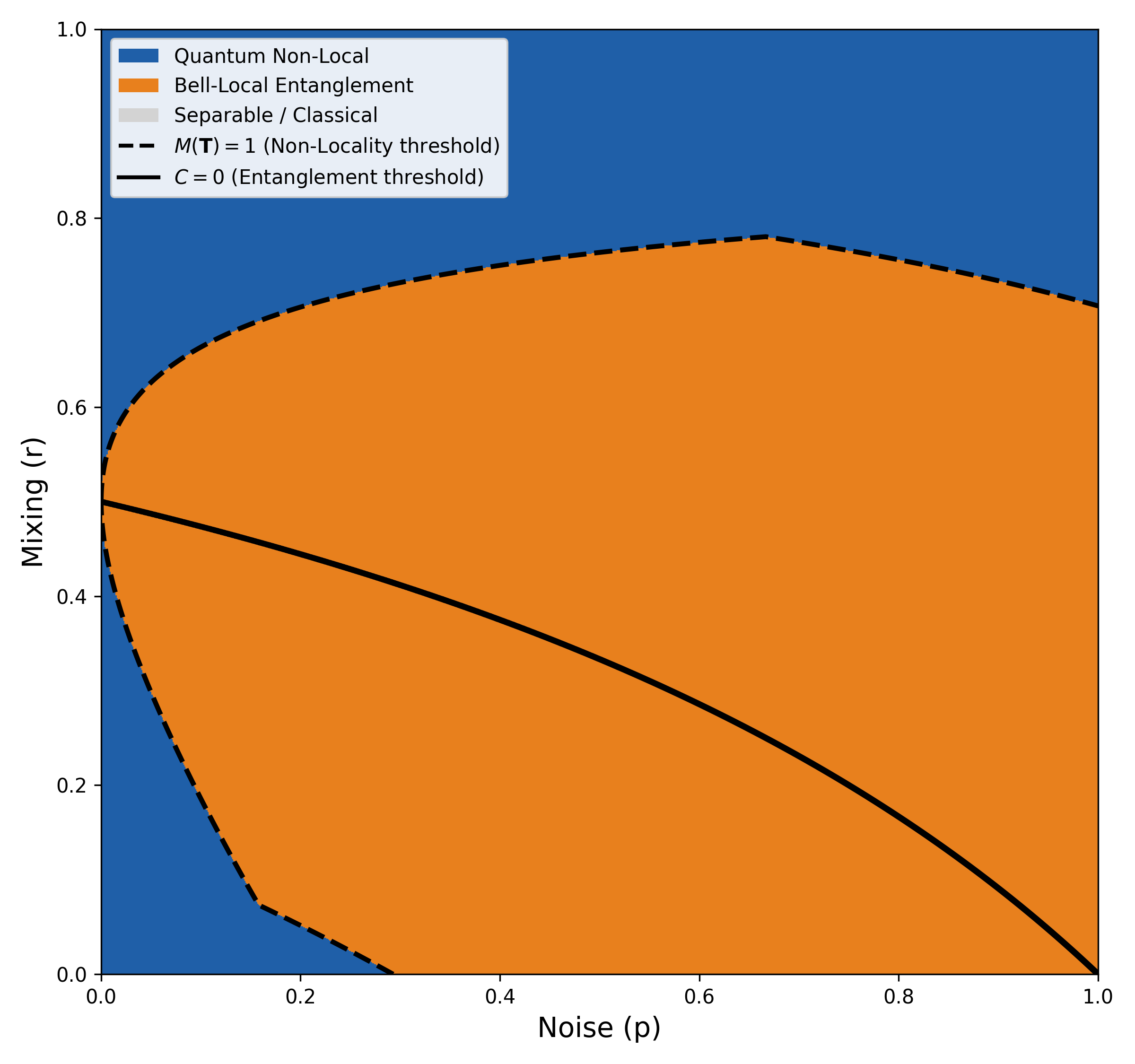}
        \caption{Same basis, opposite phases\\($\ket{\psi^+}$ with 
        $\ket{\psi^-}$)}
        \label{fig:ad_non_identical_psi}
    \end{subfigure}
    \caption{Phase diagrams for Bell state mixtures under the Amplitude 
    Damping (AD) channel, illustrating the boundary dynamics between the 
    quantum non-local, Bell-local, and separable regimes across the five 
    distinct mixing cases.}
    \label{fig:ad_all_cases}
\end{figure}

We conclude by analyzing the 
mixture of two $\ket{\psi}$-basis states with opposite phases. The reconstructed 
density matrix is:
\begin{equation}
\rho'' =
\begin{bmatrix}
p(1-r) & 0 & 0 & 0 \\
0 & \frac{(1-r)(1-p)+r}{2} & \frac{-(1-r)(1-p)+r}{2} & 0 \\
0 & \frac{-(1-r)(1-p)+r}{2} & \frac{(1-r)(1-p)+r}{2} & 0 \\
0 & 0 & 0 & 0
\end{bmatrix}.
\end{equation}
The concurrence for this mixture is:
\begin{equation}
    C = \max[0,\,|r-(1-r)(1-p)|].
\end{equation}
The elements for $T$ is given as:
\begin{align}
    t_1 &= r-(1-r)(1-p), \\
    t_2 &= r-(1-r)(1-p), \\
    t_3 &= 2p(1-r)-1,
\end{align}
\begin{equation}
    M(T) = [r-(1-r)(1-p)]^2 + \max\!\left([r-(1-r)(1-p)]^2,\,
    [2p(1-r)-1]^2\right).
\end{equation}
This is illustrated in Fig~\ref{fig:ad_all_cases}(\subref{fig:ad_non_identical_psi}).

\subsubsection*{Physical Interpretation of Amplitude Damping Thresholds}

In Fig~\ref{fig:ad_all_cases}(\subref{fig:ad_identical_phi}), the state never becomes separable for any $r>0$ and so the diagram is entirely blue and orange. As $p$ tends to 1, the noisy branch completely decays to $\ket{00}$ which again restores a clean correlation that is strong enough to violate CHSH, entering into the blue region. In Fig~\ref{fig:ad_all_cases}(\subref{fig:ad_identical_psi}), the CHSH violation is lost once $x>1-\frac{1}{\sqrt{2}}\approx0.293$, even though the state remains entangled (orange). Since we are using the $\ket{\psi}$ states, the maximal AD noise doesn't restore correlation structure unlike for the situation in Fig~\ref{fig:ad_all_cases}(\subref{fig:ad_identical_phi}). 

In Fig~\ref{fig:ad_all_cases}(\subref{fig:ad_non_identical_phi}) and Fig~\ref{fig:ad_all_cases}(\subref{fig:ad_different_bases}), we have all three regions. At $p=0$ and $r=0.5$, we have $C=0$ and as noise increases, the grey region expands. At high noise near $p=1$, the noisy branch fully decays to $\ket{00}$ and entanglement re-emerges. The grey region is smaller and the orange region is larger for Fig~\ref{fig:ad_all_cases}(\subref{fig:ad_different_bases}) compared to Fig~\ref{fig:ad_all_cases}(\subref{fig:ad_non_identical_phi}) since their coherences are on different off-diagonals of the density matrix. Also, the recovery of entanglement at high noise is stronger and faster in Fig~\ref{fig:ad_all_cases}(\subref{fig:ad_different_bases}) compared to Fig~\ref{fig:ad_all_cases}(\subref{fig:ad_non_identical_phi}). For Fig~\ref{fig:ad_all_cases}(\subref{fig:ad_non_identical_psi}), the gray region has a much smaller area and the orange region is also larger. 
\FloatBarrier
\section{Conclusion}
\label{sec:conclusion}

In this work, we have presented a systematic and comprehensive characterization 
of quantum entanglement in two-qubit Bell states subjected to three physically 
motivated noise channels (Depolarizing (DP), Phase Damping (PD), and 
Amplitude Damping (AD)) across three distinct operational regimes: pure 
superposition, noiseless mixing, and asymmetric mixing of one noisy and one 
pure Bell state. Throughout, we used von Neumann entropy to quantify 
entanglement in pure superposed states and Wootters' concurrence to track 
entanglement in mixed-state configurations including noise. We further extended our analysis 
to map the boundary between mathematical entanglement and quantum non-locality 
using the Horodecki criterion, revealing the phase structure across the 
noise-mixing parameter space.

\subsection*{Superposition and Noiseless Mixing}
Our analysis of pure Bell state superpositions revealed that entanglement 
depends critically on the relative phase $\phi$ between the superposed states. 
The entropy reaches its minimum at equal superposition weight $s = 0.5$, with 
complete separability ($S = 0$) emerging only at specific values of $\phi$ 
determined by the combinations: $\phi = 0$ or $\phi = \pi$ for either
same bases but opposite phases (\ $\ket{\phi^+}$ with $\ket{\phi^-}$) or different basis but same phases ($\ket{\phi^+}$ with $\ket{\psi^+}$) and 
$\phi = \pi/2$ for different bases and different phases (e.g.\ $\ket{\phi^+}$ with 
$\ket{\psi^-}$). Notably, each Bell state has exactly two partners for which 
the superposition remains maximally entangled ($S = 1$) for all $s$ and $\phi=0$: 
itself (e.g. $\ket{\phi^+}$ with $\ket{\phi^+}$, and its opposite phase and different basis partner  (e.g. $\ket{\phi^+}$ with $\ket{\psi^-}$). This complete 
structure is captured in Table~\ref{tab:superposition}.

In the noiseless mixing regime, we found a universal result: mixing any Bell 
state with itself preserves maximal entanglement ($C = 1$) for all $r$, while 
mixing any two non-identical Bell states produces a characteristic V-shaped 
concurrence $C = |2r - 1|$ that vanishes at equal mixing weights $r = 0.5$, 
regardless of which specific pair is chosen. This universality holds across 
all twelve non-identical combinations and contrasts sharply with the 
superposition case, where the specific pair and relative phase $\phi$ 
determined the entanglement structure.
\subsection*{Depolarizing Channel}

Under Depolarizing noise, we derived the concurrence for 
all sixteen Bell state mixing combinations. For identical state mixtures, the 
concurrence follows $C = \max\left[0,\, 1 - \tfrac{3}{2}p(1-r)\right]$, 
showing monotonic decrease with noise; we note that this case reduces 
exactly to the Werner state $\rho = (1-x)\ket{\phi^+}\bra{\phi^+} + xI/4$ 
with $x = p(1-r)$, and our thresholds recover the known separability ($x < 2/3$) 
and CHSH ($x < 1-1/\sqrt{2}$) boundaries~\cite{werner1989,horodecki1995} as a 
consistency check. For non-identical mixtures, we get the
concurrence $C = \max\left[0,\, |1-2r-p(1-r)| - \tfrac{1}{2}p(1-r)\right]$, 
which gives a counterintuitive feature: whenever $x > 1-2r$, equivalently 
$p(1-r) > 1-2r$, increasing noise \emph{raises} the concurrence rather than 
degrading it. This condition holds for all $p \in (0,1)$ when $r \geq 0.5$, 
and extends into smaller $r$ values as $p$ increases. 
The mechanism is that the Depolarizing channel reduces the magnitude of the 
noisy branch's off-diagonal element, which was canceling against the reference 
contribution; as this cancellation weakens, the net coherence and hence 
concurrence rises.

The non-locality analysis revealed a large Bell-local entanglement region for 
DP noise, confirming that while mixing with a pure reference state can increase entanglement, all of the recovered entanglement fails to 
violate the CHSH inequality. For non-identical states, all twelve permutations 
collapse to a single universal non-locality boundary, reflecting the symmetry 
of the Bell basis under the Depolarizing channel.

\subsection*{Phase Damping Channel}

Phase Damping, which destroys coherence (off diagonals), produces 
a notable result. In PD, while the X state structure is unchanged, the off diagonal elements are changed.
For identical Bell state mixtures, the concurrence is given by 
$C = \max[0,\, 1-p(1-r)]$, which is less sensitive to noise than the DP case. 
For two states from the same basis with opposite phases (e.g.\ $\ket{\phi^+}$ 
with $\ket{\phi^-}$), we find $C = |1-2r-p(1-r)|$. For states from different 
bases (e.g.\ $\ket{\phi^+}$ with $\ket{\psi^\pm}$), the concurrence is 
governed by the relation between the outer off-diagonal coherence and the 
inner diagonal elements from the pure reference state.

We find several interesting results in the phase diagrams of the PD channel. 
For identical states, the resulting state always violates the CHSH inequality 
unless $p = 1$ and $r = 0$, the unique point of complete separability. Even 
under large noise with small but nonzero $r$, the state retains positive 
concurrence and violates CHSH. For $\ket{\phi^+}$ with $\ket{\phi^-}$ 
(same basis, opposite phases), noise increases concurrence and maintains a 
CHSH violation everywhere except on the separable boundary curve 
$r = (1-p)/(2-p)$, with no Bell-local entanglement region present. For states 
from different bases ($\ket{\phi}$ with $\ket{\psi}$), the concurrence 
increase with noise lies entirely within the Bell-local regime; the state 
fails to violate the CHSH inequality in that region.

\subsection*{Amplitude Damping Channel}

Amplitude Damping, which models the decay of excited states into 
the ground state, introduces a fundamental asymmetry between the 
$\ket{\phi}$-basis states (basis at $\ket{00}$ and $\ket{11}$) and the 
$\ket{\psi}$-basis states (basis at $\ket{01}$ and $\ket{10}$). Because of this asymmetry unlike DP and PD, 
we get qualitatively different entanglement dynamics for the two 
bases, even under identical noise conditions.

For a single Bell state, the $\ket{\psi}$ states exhibit linear concurrence 
decay $C = 1-p$ due to their symmetric structure, while the $\ket{\phi}$ 
states show nonlinear decay $C = (1-p)^2$ because the AD channel acts on the 
doubly-excited $\ket{11}$ component at rate proportional to $p^2$ since both 
qubits must decay. When mixing is introduced, the identical $\ket{\phi}$ 
mixture yields $C = \max\left[0,\,(1-p)^2(1-r)+r\right]$. The identical 
$\ket{\psi}$ mixture, by contrast, exhibits a cleaner $C = 1-p(1-r)$ 
structure.

For identical $\ket{\phi}$ mixtures, the concurrence decreases monotonically 
with the noise parameter, but the Horodecki parameter $M(\mathbf{T})$ is 
non-monotonic: it dips below the non-locality threshold in the intermediate 
noise regime and then rises back above it as $p \to 1$, provided $r > 0$. The 
physical reason is that as $p \to 1$ the AD channel reduces the noisy branch onto $\ket{00}$, so the full mixture approaches 
$(1-r)\ket{00}\bra{00} + r\ket{\phi^+}\bra{\phi^+}$, for which 
$M(\mathbf{T}) = 1 + r^2 > 1$. Non-locality is therefore restored not by noise 
acting constructively in a general sense, but because maximal damping 
repurifies the noisy branch (decaying to $\ket{00}$). 

For identical $\ket{\psi}$ mixtures, $M(\mathbf{T})$ is also non-monotonic;
it reaches a minimum near $x = 2/3$ before rising back toward 1, but 
unlike the $\ket{\phi}$ case it never re-crosses the non-locality threshold. 
The contrast between the two bases is therefore that both show non-monotonic $M(\mathbf{T})$, 
but only the $\ket{\phi}$ basis recovers a CHSH violation.

For non-identical mixtures within the $\ket{\phi}$ basis ($\ket{\phi^+}$ with 
$\ket{\phi^-}$) and across different bases ($\ket{\phi}$ with $\ket{\psi}$), 
the concurrence increases with noise and the CHSH inequality is violated, with 
the $\ket{\psi}$-basis mixtures showing a smaller separable region. For the 
non-identical $\ket{\psi}$-basis mixture ($\ket{\psi^+}$ with $\ket{\psi^-}$), 
since there is no $\ket{11}$ component, the entanglement boundary is purely by the phase 
 changes between these two states rather than noise-induced decay,
producing a qualitatively distinct phase diagram structure among all five 
AD cases.

\subsection*{Broader Significance and Future Directions}
In this paper, we provide a comprehensive analysis of entanglement and non-locality in the combinations of two qubit maximally entangled Bell state under three decoherence channels. Throughout this work, the noise acts on one Bell state of the mixture where one state undergoes decoherence with some probability. As our results show, noise can enhance entanglement under particular mixing with a pure reference Bell state.

The distinction between mathematical entanglement ($C > 0$) and operational 
non-locality ($M(\mathbf{T}) > 1$) proved to be strongly channel-dependent in 
ways not captured by concurrence alone. For NISQ applications where Bell 
inequality violation is required for many protocols, our phase diagrams provide direct guidance on 
which noise regimes and mixing strategies yield actionable quantum correlations.

We mention here two extensions of this work, that we are interested in: First, generalizing the analysis 
to multi-qubit entangled states such as GHZ and W states for the three qubit states and verifying whether 
entanglement enhancement persists or is specific to 
the two-qubit Bell basis. Second, connecting the entanglement-recovery 
conditions identified here to concrete entanglement distillation 
protocols~\cite{ref-nielsen} would clarify their practical relevance for 
near-term quantum hardware.
\section*{Acknowledgements}
This research began as a part of the Fletcher Research Internship program at Brigham Young University, Provo, Utah. We would like to thank the Department of Physics and Astronomy at Brigham Young University and the Department of Physics and Astronomy at the University of Southern Mississippi for their support.

\appendix
\section{Derivation of the Amplitude Damping Kraus operators for single- and two-qubit systems}
\label{app:ad_kraus}

Amplitude Damping is a special type of noise that corresponds to the decay of an atom from an excited state to a ground state with emission of a photon to the environment. For a system, we have two states: $\ket{0}_s$ representing the ground state and $\ket{1}_s$ representing the excited state. A full amplitude-damping error corresponds to the jump of $\ket{1}_s$ to $\ket{0}_s$ with no coherence between them. The resulting density matrix is given as:
\[
\begin{bmatrix}
    |\alpha|^2&\alpha\beta^*\\
    \beta\alpha^*&|\beta|^2
\end{bmatrix}\xrightarrow{100\%  \;AD}\begin{bmatrix}
    1&0\\
    0&0
\end{bmatrix}.
\]
To fully understand how amplitude damping affects both coherence and the release of energy from the excited state, we need to understand the evolution of the Kraus operators for this noise.
For the setting, we have:\\
System (Qubit): $\ket{0}_s$ (ground) and $\ket{1}_s$ (excited)\\
Environment (Field): $\ket{0}_E$ (no photon) and $\ket{1}_E$ (one photon)\\
The interaction between the system and environment is governed by a Jaynes--Cummings-type Hamiltonian~\cite{jaynes1963}. The interaction unitary $U$ does two things over a time $t$:
\begin{enumerate}
    \item If the atom is in $\ket{0}$: it cannot decay further,
    \[
    U\ket{0}_S\ket{0}_E=\ket{0}_S\ket{0}_E.
    \]
    \item  If the atom is in $\ket{1}$: it has a probability $p$ of decaying to $\ket{0}$ and emitting a photon into the environment, and a probability $(1-p)$ of staying excited:
    \[
    U\ket{1}_S\ket{0}_E=\sqrt{1-p}\,\ket{1}_S\ket{0}_E+\sqrt{p}\,\ket{0}_S\ket{1}_E.
    \]
Square roots are used because these are probability amplitudes in the wavefunction.
\end{enumerate}

\subsection{Deriving the Kraus operators}
Kraus operators are simply `slices' of this unitary matrix taken with respect to the environment. We find them by sandwiching the unitary between the environment's basis states:
\[E_k=\bra{k}_E \;U\;\ket{0}_E.\]
Operator $E_0$ (environment stays in $\ket{0}$): this operator represents the `no decay happened' branch.
\begin{itemize}
    \item Input $\ket{0}_S$: output is $\ket{0}_S$ (coefficient is 1)
    \item Input $\ket{1}_S$: output is $\sqrt{1-p}\;\ket{1}_S$
    \[
    E_0=\begin{bmatrix}
        1&0\\
        0&\sqrt{1-p}
    \end{bmatrix}
    \]
\end{itemize}
Operator $E_1$ (environment flips to $\ket{1}$): this operator represents `decay happened'. The environment detected a photon.
\begin{itemize}
    \item Input $\ket{0}_S$: output is 0
    \item Input $\ket{1}_S$: output is $\sqrt{p}\;\ket{0}_S$
    \[
    E_1=\begin{bmatrix}
        0&\sqrt{p}\\
        0&0
    \end{bmatrix}.
    \]
\end{itemize}

\subsection{Applying the same process for two-qubit states}
System (Qubit):
\begin{itemize}
    \item $\ket{00}_S$ (ground state)
    \item $\ket{01}_S$ (first qubit ground, second qubit excited)
    \item $\ket{10}_S$ (first qubit excited, second qubit ground)
    \item $\ket{11}_S$ (excited state)
\end{itemize}
\textit{Note: for the environment (field), we can use the same logic.}
By solving the probability for the various input systems, we get
\[
K_0=\begin{bmatrix}
    1&0&0&0\\
    0&\sqrt{1-p}&0&0\\
    0&0&\sqrt{1-p}&0\\
    0&0&0&(1-p)
\end{bmatrix},
\quad K_1=\begin{bmatrix}
    0&\sqrt{p}&0&0\\
    0&0&0&0\\
    0&0&0&\sqrt{p(1-p)}\\
    0&0&0&0\\
\end{bmatrix},
\]
\[
K_2=\begin{bmatrix}
    0&0&\sqrt{p}&0\\
    0&0&0&\sqrt{p(1-p)}\\
    0&0&0&0\\
    0&0&0&0\\
\end{bmatrix}, \quad K_3=\begin{bmatrix}
    0&0&0&p\\
    0&0&0&0\\
    0&0&0&0\\
    0&0&0&0
\end{bmatrix}.
\]
\newpage
\bibliographystyle{unsrt}

\begin{thebibliography}{999}

\bibitem[Bennett \textit{et al}. (1996)]{bennett1996}
Bennett, C. H.; Bernstein, H. J.; Popescu, S.; Schumacher, B.
Concentrating Partial Entanglement by Local Operations.
\textit{Phys. Rev. A} \textbf{1996}, \textit{53}, 2046--2052.


\bibitem[Clauser \textit{et al}. (1969)]{clauser1969}
Clauser, J. F.; Horne, M. A.; Shimony, A.; Holt, R. A. Proposed Experiment to
Test Local Hidden-Variable Theories. \textit{Phys. Rev. Lett.} \textbf{1969},
\textit{23}, 880--884.

\bibitem[Gisin (1991)]{gisin1991}
Gisin, N. Bell's Inequality Holds for All Non-Product States.
\textit{Phys. Lett. A} \textbf{1991}, \textit{154}, 201--202.

\bibitem[Horodecki \textit{et al}. (1995)]{horodecki1995}
Horodecki, R.; Horodecki, P.; Horodecki, M. Violating Bell Inequality by Mixed
Spin-$\tfrac{1}{2}$ States: Necessary and Sufficient Condition.
\textit{Phys. Lett. A} \textbf{1995}, \textit{200}, 340--344.

\bibitem[Jaynes and Cummings (1963)]{jaynes1963}
Jaynes, E. T.; Cummings, F. W. Comparison of Quantum and Semiclassical Radiation
Theories with Application to the Beam Maser. \textit{Proc. IEEE} \textbf{1963},
\textit{51}, 89--109.

\bibitem[Kandala \textit{et al.} (2019)]{ref-kandala}
Kandala, A.; Temme, K.; C\'orcoles, A. D.; Mezzacapo, A.; Chow, J. M.; Gambetta, J. M.
Error Mitigation Extends the Computational Reach of a Noisy Quantum Processor.
\textit{Nature} \textbf{2019}, \textit{567}, 491--495.

\bibitem[Nielsen and Chuang (2010)]{ref-nielsen}
Nielsen, M. A.; Chuang, I. L. \textit{Quantum Computation and Quantum Information};
Cambridge University Press: Cambridge, United Kingdom, 2010.

\bibitem[Pironio \textit{et al}. (2010)]{pironio2010}
Pironio, S.; Ac\'in, A.; Massar, S.; Boyer de la Giroday, A.; Matsukevich, D. N.;
Maunz, P.; Olmschenk, S.; Hayes, D.; Luo, L.; Manning, T. A.; Monroe, C.
Random Numbers Certified by Bell's Theorem.
\textit{Nature} \textbf{2010}, \textit{464}, 1021--1024.

\bibitem[Preskill (2018)]{preskill2018}
Preskill, J. Quantum Computing in the NISQ Era and Beyond.
\textit{Quantum} \textbf{2018}, \textit{2}, 79.

\bibitem[Schr\"odinger (1935)]{schrodinger1935}
Schr\"odinger, E. Discussion of Probability Relations Between Separated Systems.
\textit{Math. Proc. Camb. Phil. Soc.} \textbf{1935}, \textit{31}, 555--563.

\bibitem[Steeb and Hardy (2006)]{ref-steeb}
Steeb, Willi-Hans; Hardy, Yorick. \textit{Problems and Solutions in Quantum Computing
and Quantum Information}; World Scientific: Singapore, 2006; p. 80.

\bibitem[von Neumann (1955)]{ref-vonneumann}
von Neumann, J. \textit{Mathematical Foundations of Quantum Mechanics};
Princeton University Press: Princeton, NJ, 1955.

\bibitem[Werner (1989)]{werner1989}
Werner, R. F. Quantum states with Einstein-Podolsky-Rosen correlations admitting a
hidden-variable model. \textit{Phys. Rev. A} \textbf{1989}, \textit{40}, 4277.

\bibitem[Wootters (1998)]{wootters1998}
Wootters, W. K. Entanglement of Formation of an Arbitrary State of Two Qubits.
\textit{Phys. Rev. Lett.} \textbf{1998}, \textit{80}, 2245--2248.

\bibitem[Wootters (2001)]{ref-wootters}
Wootters, W. K. Entanglement of Formation and Concurrence.
\textit{Quantum Inf. Comput.} \textbf{2001}, \textit{1}, 27--44.

\bibitem[Yu and Eberly (2004)]{yu2004sudden}
Yu, Ting; Eberly, J. H. Finite-Time Disentanglement via Spontaneous Emission.
\textit{Phys. Rev. Lett.} \textbf{2004}, \textit{93}, 140404.

\bibitem[Yu and Eberly (2007)]{yu2007evolution}
Yu, Ting; Eberly, J. H. Evolution from entanglement to decoherence of bipartite
mixed ``X'' states. \textit{Quantum Inf. Comput.} \textbf{2007}, \textit{7}, 459--468.

\end{thebibliography}
\isAPAandChicago{}{%

}

\end{document}